\documentclass{aa}
\usepackage{graphicx}
\usepackage{txfonts}
\usepackage{natbib,twoopt}
\usepackage[breaklinks=true]{hyperref} 

\newcommand{\angstrom}{\mbox{\normalfont\AA}}

\begin{document}

   \title{How do wavelength correlations affect transmission spectra? Application of a new fast and flexible 2D Gaussian process framework to transiting exoplanet spectroscopy}
   
   \author{Mark Fortune \orcid{0000-0002-8938-9715} \inst{1}\fnmsep\thanks{E-mail: fortunma@tcd.ie},
           Neale P. Gibson \orcid{0000-0002-9308-2353} \inst{1},
           Daniel Foreman-Mackey \orcid{0000-0002-9328-5652} \inst{2},
            Thomas M. Evans-Soma \orcid{0000-0001-5442-1300} \inst{3,4},
            Cathal Maguire \orcid{0000-0002-9061-780X} \inst{1},
            Swaetha Ramkumar \orcid{0000-0003-0815-8366} \inst{1}
            }

   \institute{School of Physics, Trinity College Dublin, University of Dublin, Dublin 2, Ireland
        \and
             Center for Computational Astrophysics, Flatiron Institute, New York, NY, USA
        \and
            School of Information and Physical Sciences, University of Newcastle, Callaghan, NSW, Australia
        \and
            Max Planck Institute for Astronomy, K\"{o}nigstuhl 17, D-69117 Heidelberg, Germany
             }

    \titlerunning{A 2D Gaussian Process Method for Transmission Spectroscopy}
    \authorrunning{M. Fortune et al.}
   \date{Received XXX; accepted XXX}

 
  \abstract
   {The use of Gaussian processes (GPs) is a common approach to account for correlated noise in exoplanet time series, particularly for transmission and emission spectroscopy. This analysis has typically been performed for each wavelength channel separately, with the retrieved uncertainties in the transmission spectrum assumed to be independent. However, the presence of noise correlated in wavelength could cause these uncertainties to be correlated, which could significantly affect the results of atmospheric retrievals. We present a method that uses a GP to model noise correlated in both wavelength and time simultaneously for the full spectroscopic dataset. To make this analysis computationally tractable, we introduce a new fast and flexible GP method that can analyse 2D datasets when the input points lie on a (potentially non-uniform) 2D grid - in our case a time by wavelength grid - and the kernel function has a Kronecker product structure. This simultaneously fits all light curves and enables the retrieval of the full covariance matrix of the transmission spectrum. Our new method can avoid the use of a `common-mode' correction, which is known to produce an offset to the transmission spectrum. Through testing on synthetic datasets, we demonstrate that our new approach can reliably recover atmospheric features contaminated by noise correlated in time and wavelength. In contrast, fitting each spectroscopic light curve separately performed poorly when wavelength-correlated noise was present. It frequently underestimated the uncertainty of the scattering slope and overestimated the uncertainty in the strength of sharp absorption peaks in transmission spectra. Two archival VLT/FORS2 transit observations of WASP-31b were used to compare these approaches on real observations. Our method strongly constrained the presence of wavelength-correlated noise in both datasets, and significantly different constraints on atmospheric features such as the scattering slope and strength of sodium and potassium features were recovered.
    }


   \maketitle
%

\section{Introduction}
\label{sec:1}

   Low-resolution transmission spectroscopy has been a powerful technique for probing the atmospheric composition of exoplanets ever since the first detection of an exoplanet atmosphere \citep{Charbonneau2002}. The technique relies upon observing an exoplanet transit - when an exoplanet appears to pass in front of its host star - and analysing the decrease in flux during the transit as a function of wavelength. The resulting transmission spectrum contains information about the planetary atmosphere \citep{Seager2000, Brown2001}.

   The field of exoplanet atmospheres has entered a new era due to the recent launch of JWST. Early Release Science observations of WASP-39b using JWST NIRSpec's PRISM mode have produced a 33$\sigma$ detection of H$_2$O in addition to strong detections of Na, CO, and CO$_2$ \citep{PRISM2022} - far exceeding what had been achieved with previous ground-based and space-based observations \citep{Nikolov2016, Sing2016}. As the exoplanet community pushes JWST further to its limits towards smaller terrestrial planets (e.g. \citealt{Greene2023, Zieba2023}), the importance of careful treatment of systematics - astrophysical or instrumental - will become of increasing importance.
   
   Gaussian processes (GPs) were introduced in \citet{Gibson2012} to account for the uncertainty that correlated noise - also referred to as systematics - produces in the resulting transmission spectrum in a statistically robust way. GPs have been shown to provide more reliable estimates of uncertainties when compared to other common techniques such as linear basis models \citep{Gibson2014}. This difference may help explain contradictory results in the field. For example, WASP-31b was reported to have a strong potassium signal at 4.2$\sigma$ using data from the Hubble Space Telescope (HST) \citep{Sing2015}, which were analysed using linear basis models. However, follow-up measurements in \citet{Gibson2017} with the FOcal Reducer and low dispersion Spectrograph (FORS2) on the Very Large Telescope (VLT) \citep{Appenzeller1998} found no evidence of potassium. High-resolution observations in \citet{Gibson2019} using the Ultraviolet and Visual Echelle Spectrograph (UVES) on the VLT as well as low-resolution observations from the Inamori-Magellan Areal Camera and Spectrograph (IMACS) on Magellan \citep{McGruder2020} also failed to reproduce this detection. These results are more consistent with the re-analysis of the HST data using GPs which reduced the significance of the potassium signal to $2.5\sigma$ \citep{Gibson2017}, demonstrating the importance of careful treatment of systematics.

    In addition to inconsistent detections of species, conflicting measurements of the slope of the transmission spectrum are also common (e.g. \citealt{Sedaghati2017, Sedaghati2021, Espinoza2019}). A slope in the transmission spectrum can be caused by Rayleigh scattering or from scattering by aerosols including clouds or hazes in the atmosphere \citep{Etangs2008} and it is therefore often referred to as a scattering slope. However, stellar activity can also produce an apparent scattering slope, which is typically used to explain these contradictory results \citep{Espinoza2019, McCullough2014, Rackham2018}. It is worth considering whether these effects could be caused by systematics because it is difficult to obtain direct evidence that stellar activity is the cause of these contradictions. This could be particularly relevant for measurements of extreme scattering slopes such as in \citet{May2020}, where the authors note that a combination of atmospheric scattering and stellar activity still struggle to explain the observed slope.
    
    One potential issue with current data analyses is that each transit depth in the transmission spectrum is fit separately and is assumed to have an independent uncertainty. \citet{Ih2021} studied the effect on retrieved atmospheric parameters if this assumption is incorrect and if transit depth uncertainties are correlated. The authors suggest that both instrumental and stellar systematics could cause correlations in transmission spectra uncertainties, which they showed could significantly impact atmospheric retrievals, but they did not provide a method for retrieving these correlations. \citet{Holmberg2023} report the presence of wavelength-correlated systematics in observations of WASP-39b and WASP-96b using the Single Object Slitless Spectroscopy (SOSS) mode of JWST's Near Infrared Imager and Slitless Spectrograph (NIRISS). They demonstrate why these systematics should result in correlated uncertainties in the transmission spectrum and made multiple simplifying assumptions to derive an approximate covariance matrix of the transmission spectrum, although their method cannot not account for both time and wavelength-correlated systematics.
    
    In this work, we demonstrate a statistically robust way to account for the presence of both time-correlated and wavelength-correlated systematics and its use on both simulated datasets and to real transit observations of WASP-31b from VLT/FORS2 (originally analysed in \citealt{Gibson2017}). We model the noise present across the full dataset as a Gaussian process. By simultaneously fitting all spectroscopic light curves for transit depth, we also explored the joint posterior of all transit depths using Markov chain Monte Carlo (MCMC) and recovered the covariance matrix of the transmission spectrum. We present an efficient optimisation that can dramatically speed up the required log-likelihood calculation based on work in \citet{Saatchi2011} and \citet{Rakitsch2013}. We assume that the inputs to the kernel function of the GP lie on a 2D grid - such as a time by wavelength grid - and refer to this as a 2D GP. Intuitively this assumption is valid for many datasets that can be neatly arranged in a 2D grid, for example an image typically has a 2D grid structure where the inputs describing correlations could be chosen as the x and y coordinates of each pixel. It is not required that the grid is uniform; that is to say for transmission spectroscopy neither the time points nor wavelengths need to be uniformly separated. In contrast, we refer to GPs where the input(s) all lie along a single dimension such as time as 1D GPs. This includes the standard approach of fitting individual transit light curves using a GP.
    
    One example where 2D GPs have already been used would be in radial velocity analysis where multiple parallel time series observations may be jointly fit with a GP \citep{Rajpaul2015}. In this case, time is one of the dimensions and we can consider the other dimension to consist of a small number ($\sim$3) of different observables stacked together such as radial velocity and any suitable activity indicators. This is in contrast to transmission spectroscopy where dozens or even hundreds of parallel time series may be observed simultaneously in the form of different wavelength channels and the number of observations in time may also be greater. Unfortunately, the algorithms developed for radial velocity either do not scale to the significantly larger datasets encountered in transmission spectroscopy (as noted in \citealt{PyanetiII}), or else they make strong assumptions about the form of correlations in both dimensions \citep{Gordon2020, sleaf} which were not sufficient for our analysis of real observations.
    
    In particular, \citet{Gordon2020} introduced a 2D GP method that could be applied to transmission spectroscopy and scales better than the method introduced here but with a much stronger assumption that the shape of the correlated noise is identical at different wavelengths but can change in amplitude. With weaker assumptions about the correlation in wavelength their method has worse scaling in the number of wavelength channels. The covariance matrix in the time dimension is also limited to celerite kernels as introduced in \citet{Foreman_Mackey_2017}. Similarly, the extension of the \citet{Gordon2020} method derived in \citet{sleaf} is still too limited to apply the kernel function used to fit the VLT/FORS2 data in this work. Our method has significant advantages for transmission spectroscopy as it is computationally tractable even when the length of both dimensions is of the order of hundreds of points and can be used with any general covariance matrices describing the noise in each dimension.

    This paper is laid out as follows; Sect.~\ref{sec:2} gives a brief introduction to Gaussian processes, the challenges of scaling them to two dimensional datasets and outlines the mathematics of the 2D GP method used in this paper, Sect.~\ref{sec:3} compares our method to standard approaches on simulated datasets containing wavelength-correlated noise. Finally, Sect.~\ref{sec:4} re-analyses archival VLT/FORS2 observations using this method and the discussion and conclusions are presented in Sects.~\ref{sec:5} and \ref{sec:6}.
    
    The code developed for this work has been made available as a Python package called \textsc{luas}\footnote{\emph{luas} is the word for speed in the Irish language, pronounced like `Lewis'}. It is available on GitHub\footnote{\href{https://github.com/markfortune/luas}{https://github.com/markfortune/luas}} along with documentation and tutorials.
    

    \section{2D Gaussian processes for transmission spectroscopy}
    \label{sec:2}


    \subsection{Introduction to 2D Gaussian processes}
    \label{sec:GPs}
    To fit our transit observations we first model the transit using a deterministic function. This function is typically referred to as the `mean function' in the context of GPs and denoted by $\vec{\mu}$. This light curve model includes a transit depth parameter for each spectroscopic light curve, which we fit for to obtain our transmission spectrum (see Sect.~\ref{sec:light curve} for more details on the light curve model). When using 2D GPs, we combine light curve models for each spectroscopic light curve to form our 2D mean function $\vec{\mu}(\vec{\lambda}, \vec{t}, \vec{\theta})$, which is a function of model parameters $\vec{\theta}$ as well as the wavelength channels $\vec{\lambda}$ and the times of our flux observations $\vec{t}$. 
    
    We model the noise in our observations as a Gaussian process. This means that if we take any two arbitrary input locations $\vec{x}_i$ and $\vec{x}_j$ out of the collection of points in a dataset $D$, we assume that the noise observed at these points follows a multivariate Gaussian distribution with the covariance given by a kernel function of our choosing. In the case of 1D GPs fitting a single transit light curve, the covariance might be described by the commonly used squared-exponential kernel as a function of the time separation of each data point, giving covariance matrix:
    \begin{equation}
        \mathbf{K}_{ij} = h^2 \exp\left(-\frac{|t_i - t_j|^2}{2 l_{t}^2}\right) + \sigma^2 \delta_{ij},
    	\label{eq:kernel}
    \end{equation}
    where $h$ is the height scale or amplitude of the correlated noise, $t_i$ and $t_j$ are the times of $\vec{x}_i$ and $\vec{x}_j$, $l_{t}$ is the length scale of the correlated noise in time and we have also included a white noise term with variance $\sigma^2$ added to the diagonal ($\delta_{ij}$ is the Kronecker delta). The choice of kernel function is problem-specific with a range of kernel functions to choose from (see \citealt{Rasmussen}).
    
    If we now consider fitting multiple spectroscopic light curves with a GP then any two flux measurements have both a time separation and a wavelength separation. We can combine correlations in both time and wavelength into a single kernel. If we choose the squared-exponential kernel for both dimensions and introduce a wavelength length scale $l_\lambda$ we get:
    \begin{equation}
        \mathbf{K}_{ij} = h^2 \exp\left(-\frac{|\lambda_i - \lambda_j|^2}{2 l_{\lambda}^2}\right) \exp\left(-\frac{|t_i - t_j|^2}{2 l_{t}^2}\right) + \sigma^2 \delta_{ij},
    	\label{eq:kernel2D}
    \end{equation}
    where $\lambda_i$ and $\lambda_j$ are the wavelength values at locations $\vec{x}_i$ and $\vec{x}_j$. We can also mix kernel functions, that is we could model time with the Matérn 3/2 kernel and wavelength with a squared-exponential kernel.
    
    In general, the covariance matrix describing the noise in the observations $\mathbf{K}$ is a function of inputs $\vec{\lambda}$ and $\vec{t}$ and is a function of parameters such as $h, l_{t}, l_{\lambda}$ and $\sigma$ which are typically referred to as hyperparameters. We include both hyperparameters and light curve parameters in $\vec{\theta}$.
    
    Overall, our dataset of flux observations $\vec{y}$ are modelled as a multivariate Gaussian distribution, with the likelihood (the probability of the data given our model) given by:
    \begin{align}
        p(D|\Vec{\theta}) &= \mathcal{N}(\vec{y}|\vec{\mu}(\vec{\lambda}, \vec{t}, \vec{\theta}),\mathbf{K}(\vec{\lambda}, \vec{t}, \vec{\theta}))
        \label{eq:likelihood}
    \end{align}
    Taking the logarithm of this (to avoid numerical errors) we get the log-likelihood of our model:
    \begin{equation}
        \log(p(D|\vec{\theta})) = -\frac{1}{2}\vec{r}^T \mathbf{K}^{-1} \vec{r} - \frac{1}{2}\log|\mathbf{K}| - \frac{N}{2}\log(2\pi),
        \label{eq:loglikelihood}
    \end{equation}
    where we have defined our residuals vector $\vec{r} = \vec{y} - \vec{\mu}$.

    In accordance with Bayes' theorem, we can add the logarithm of any prior probability to our log-likelihood to get the log-posterior of all the parameters\footnote{ignoring the constant normalisation term which is not required for MCMC inference}. This equation can then be used by inference methods such as Markov Chain Monte Carlo (MCMC) to explore the marginal probability distributions of each parameter.

    \subsection{The computational cost of 2D Gaussian processes}
    \label{sec:complexity}
    
    The log-likelihood function in Eq.~(\ref{eq:loglikelihood}) can be used for analysing any general dataset including 1D or 2D datasets. However, the computational cost becomes prohibitive for large datasets. Assume we have flux observations measured at $N$ different times and binned into $M$ different wavelength channels. If we move from analysing individual light curves to all spectroscopic light curves simultaneously then the covariance matrix describing the correlation will go from an $N \times N$ matrix to an $MN \times MN$ matrix to describe correlations in both time and wavelength. Studying Eq.~(\ref{eq:loglikelihood}), it should be noted that it is necessary to invert the covariance matrix as well as calculate its determinant. The runtime of both of these computations scales as the cube of the number of data points in the most general case. We can write this scaling of runtime in `big $\mathcal{O}$' notation as $\mathcal{O}(M^3 N^3)$. The covariance matrix $\mathbf{K}$ also requires storing $M^2 N^2$ entries in memory. The scaling in memory is therefore $\mathcal{O}(M^2 N^2)$. This poor scaling in both runtime and memory would make most transmission spectroscopy datasets unfeasible to analyse and is the motivation for introducing a new log-likelihood optimisation.

    The optimised GP method introduced in this work is based on \citet{Saatchi2011} and \citet{Rakitsch2013} which both present exact methods of optimising the calculation of Eq.~(\ref{eq:loglikelihood}). Both methods assume that all inputs for the GP lie on a (potentially non-uniform) 2D grid. In the context of transmission spectroscopy this simply means that if wavelength and time are used as the two inputs then the same set of wavelength bands must be chosen for all points in time (but the wavelength bands do not need to be evenly separated). The grid must also be complete with no missing data, which means outliers at a specific time and wavelength cannot simply be removed from the dataset without either removing all other data points at that specific time or that particular wavelength. However, a common approach for dealing with non-Gaussian outliers has just been to replace them using an interpolating function (e.g. \citealt{Gibson2017}). This approach keeps the inputs on a complete grid and so this grid assumption likely satisfies most low-resolution transmission spectroscopy datasets.
    
    Both methods differ in the range of kernel functions they can be used with. The \citet{Saatchi2011} method is simpler and is described first, with the \citet{Rakitsch2013} method being a more general extension with a similar computational scaling.
    
    \subsection{Kronecker products}
    \label{sec:kronecker products}
    
    Some basics of Kronecker product algebra are useful to explain the optimisations used in this work. See \citet{Saatchi2011} for more details, proofs of the results and how these results generalise to more than two dimensions.
    
    The Kronecker product between two matrices may be written using the $\otimes$ symbol. It can be thought of as multiplying every element of the first matrix by every element in the second matrix with each multiplication producing its own term in the resulting matrix. For example:
    \begin{equation}
    \mathbf{A} \otimes \mathbf{B} =
    \begin{bmatrix}
    a_{11} & a_{12} & a_{13}\\
    a_{21} & a_{22} & a_{23}\\
    a_{31} & a_{32} & a_{33}
    \end{bmatrix} \otimes \mathbf{B} = \begin{bmatrix}
    a_{11}\mathbf{B} & a_{12}\mathbf{B} & a_{13}\mathbf{B}\\
    a_{21}\mathbf{B} & a_{22}\mathbf{B} & a_{23}\mathbf{B}\\
    a_{31}\mathbf{B} & a_{32}\mathbf{B} & a_{33}\mathbf{B}
    \end{bmatrix}.
    \end{equation}
    If $\mathbf{A}$ is $M \times M$ and $\mathbf{B}$ is $N \times N$, then $\mathbf{A} \otimes \mathbf{B}$ is $MN \times MN$.
    
    The following equations hold:
    \begin{align}
        &(\mathbf{A} \otimes \mathbf{B})(\mathbf{C} \otimes \mathbf{D}) = \mathbf{A}\mathbf{C} \otimes \mathbf{B}\mathbf{D}, \\
        &\mathbf{A} \otimes (\mathbf{B} + \mathbf{D}) = \mathbf{A} \otimes \mathbf{B} + \mathbf{A} \otimes \mathbf{D}, \\
        &(\mathbf{A} \otimes \mathbf{B})^{-1} = \mathbf{A}^{-1} \otimes \mathbf{B}^{-1}.
        \label{eq:kronecker inverse}
    \end{align}

    Both methods presented assume the covariance matrix of the noise can be expressed using Kronecker products. The \citet{Saatchi2011} method assumes the covariance matrix can be expressed as:
    \begin{equation}
    \mathbf{K} = \mathbf{K}_{\lambda} \otimes \mathbf{K}_{t} + \sigma^2 \mathbb{I},
    \label{eq:Saatchi_cov}
    \end{equation}
    where we have written the covariance matrix $\mathbf{K}$ as a Kronecker product of separate covariance matrices: $\mathbf{K}_{\lambda}$ constructed using a wavelength kernel function $k_{\lambda}(\lambda, \lambda')$; and $\mathbf{K}_{t}$ constructed using the time kernel function $k_{t}(t, t')$. It also assumes that white noise is constant across the full dataset with variance $\sigma^2$ - a limitation of the \citet{Saatchi2011} method. An equivalent way of stating this restriction is that the kernel function can be written as:
    \begin{equation}
        k(\lambda_i, \lambda_j, t_i, t_j) = k_{\lambda}(\lambda_i, \lambda_j) k_{t}(t_i, t_j) + \sigma^2 \delta_{ij}.
        \label{eq:kronecker kernel}
    \end{equation}

    In most transmission spectroscopy datasets, the amplitude of white noise varies significantly in wavelength across the dataset, making the \citet{Saatchi2011} method highly restrictive. In contrast, the \citet{Rakitsch2013} method is valid for any covariance matrix that is the sum of two independent Kronecker products:
    \begin{equation}\label{eq:kronsum}
    \mathbf{K} = \mathbf{K}_{\lambda} \otimes \mathbf{K}_{t} + \mathbf{\Sigma}_{\lambda} \otimes \mathbf{\Sigma}_{t},
    \end{equation}
    or equivalently any kernel function of the form:
    \begin{equation}
        k(\lambda_i, \lambda_j, t_i, t_j) = k_{\lambda_1}(\lambda_i, \lambda_j) k_{t_1}(t_i, t_j) + k_{\lambda_2}(\lambda_i, \lambda_j) k_{t_2}(t_i, t_j).
        \label{eq:kronecker kernel rakitsch}
    \end{equation}
    This method is much more general with the second set of covariance matrices $\Sigma_\lambda$ and $\Sigma_t$ being able to account for white noise that varies in amplitude across the dataset. We can choose $\mathbf{K}_{\lambda}$ and $\mathbf{K}_{t}$ to include our correlated noise terms and $\mathbf{\Sigma}_{\lambda}$ and $\mathbf{\Sigma}_{t}$ to be diagonal matrices containing the white noise terms. This allows us to account for any correlated noise that is separable in wavelength and time as well as any white noise that is separable in wavelength and time. The white noise as a function of time and wavelength $\sigma(\lambda, t)$ must be able to be written as
    \begin{equation}
        \sigma(\lambda, t) = \sigma(\lambda)\sigma(t).
    \end{equation}
    However, there is no requirement that $\mathbf{\Sigma}_{\lambda}$ and $\mathbf{\Sigma}_{t}$ be diagonal and the method works for any valid covariance matrices $\mathbf{K}_{\lambda}, \mathbf{K}_{t}, \mathbf{\Sigma}_{\lambda}$ and $\mathbf{\Sigma}_{t}$.
    
    Unlike the method introduced in \citet{Gordon2020} that uses a celerite kernel for the time covariance matrix and is limited to a single regression variable within that matrix, multiple regression variables may be used within any of these covariance matrices. Multiple regression variables are often used in transmission spectroscopy as it is argued they can provide extra information about how strongly correlated the noise may be for different flux observations. For example, the width of the spectral trace on the detector can change in time and a kernel function could be chosen where the correlation in the noise between two points is an explicit function of both the time-separation and difference in trace widths (e.g. \citealt{Gibson2012b, Lowe2020}) which is not possible using a celerite kernel \citep{Foreman_Mackey_2017}. Due to the kronecker product structure, any regression variables used must also lie on a 2D grid, which will be true if they vary in time but are constant in wavelength (or vice versa). One caveat is that multiple regression variables may be used with the \citet{Gordon2020} method by treating them as additional time series to be fit by the GP (see our discussion of \citealt{Rajpaul2015} and \citealt{sleaf} in Sect.~\ref{sec:1}).
    
    \subsection{Log-likelihood calculation for uniform white noise}
    \label{sec:saatchi}

    The algorithms described in Chapter 5 of \citet{Saatchi2011} allow for the calculation of Eq.~(\ref{eq:loglikelihood}) in $\mathcal{O}((M+N)MN)$ time after an initial matrix decomposition step that scales as $\mathcal{O}(M^3 + N^3)$. This holds for any covariance matrix expressible in the form of Eq.~(\ref{eq:Saatchi_cov}).
    
    To understand this method, first note that it takes advantage of Kronecker product structure to speed up matrix-vector products $[\mathbf{A} \otimes \mathbf{B}] \vec{c}$ for an arbitrary $MN$ long vector $\vec{c}$ using
    \begin{equation}\label{eq:Algo14}
        [\mathbf{A} \otimes \mathbf{B}] \vec{c} = \text{vec}(\mathbf{A} \mathbf{C} \mathbf{B}^T),
    \end{equation}
    where we reshape $\vec{c}$ into an $M \times N$ matrix with $\mathbf{C}_{ij} = c_{(i-1)M + j}$ and the $\text{vec}()$ operator reverses this such that $\text{vec}(\mathbf{C}) = \vec{c}$.
    
    As $\mathbf{A} \otimes \mathbf{B}$ is an $MN \times MN$ matrix, we would expect multiplication by a vector $\vec{c}$ to take $\mathcal{O}(M^2N^2)$ operations to calculate in general. However, Eq.~(\ref{eq:Algo14}) takes $\mathcal{O}((M+N)MN)$ operations. We also do not need to store the full matrix $\mathbf{A} \otimes \mathbf{B}$ in memory but instead just $\mathbf{A}$ and $\mathbf{B}$ separately, reducing the memory requirement from $\mathcal{O}(M^2N^2)$ to $\mathcal{O}(M^2+N^2)$.
    
    We combine this with Eq.~(\ref{eq:kronecker inverse}) and find that once we perform the $\mathcal{O}(M^3 + N^3)$ operations of computing $\mathbf{K}_{\lambda}^{-1}$ and $\mathbf{K}_{t}^{-1}$ then
    \begin{equation}\label{eq:invAlgo14}
        \mathbf{K}^{-1} \vec{r} = [\mathbf{K}_{\lambda}^{-1} \otimes \mathbf{K}_{t}^{-1}] \vec{r}
    \end{equation}
    can be computed in $\mathcal{O}((M+N)MN)$ operations and using $\mathcal{O}(M^2+N^2)$ memory.
    
    While it would be sufficient to use any method such as Cholesky factorisation to compute $\mathbf{K}_{\lambda}^{-1}$ and $\mathbf{K}_{t}^{-1}$, using the eigendecomposition of each matrix permits white noise to be easily accounted for. We exploit a particular property of eigendecomposition that the addition of a constant to the diagonal of a matrix shifts the eigenvalues but leaves the eigenvectors unchanged. We denote the eigendecomposition of some matrix $\mathbf{A}$ as:
    \begin{equation}
        \mathbf{A} = \mathbf{Q}_\mathbf{A} \mathbf{\Lambda}_\mathbf{A} \mathbf{Q}_\mathbf{A}^T,
    \end{equation}
    after which the inverse can be easily computed as:
    \begin{equation}
        \mathbf{A}^{-1} = \mathbf{Q}_\mathbf{A} \mathbf{\Lambda}_\mathbf{A}^{-1} \mathbf{Q}_\mathbf{A}^T.
    \end{equation}
    
    For two matrices $\mathbf{A}$ and $\mathbf{B}$ it can be shown that
    \begin{equation}
        (\mathbf{A} \otimes \mathbf{B})^{-1} = [\mathbf{Q}_\mathbf{A} \otimes \mathbf{Q}_\mathbf{B}] [\mathbf{\Lambda}_\mathbf{A} \otimes \mathbf{\Lambda}_\mathbf{B}]^{-1} [\mathbf{Q}_\mathbf{A}^T \otimes \mathbf{Q}_\mathbf{B}^T].
        \label{eq:ABinv}
    \end{equation}
    
    We can then add a white noise term to Eq.~(\ref{eq:ABinv}) simply by shifting the eigenvalues:
    \begin{align}
    \mathbf{K}^{-1} \vec{r} &=  (\mathbf{K}_{\lambda} \otimes \mathbf{K}_{t} + \sigma^2 \mathbb{I}_{MN})^{-1} \vec{r} \nonumber \\
    &= [\mathbf{Q}_{\lambda} \otimes \mathbf{Q}_{t}] [\mathbf{\Lambda}_{\lambda} \otimes \mathbf{\Lambda}_{t} + \sigma^2 \mathbb{I}_{MN}]^{-1} [\mathbf{Q}_{\lambda}^T \otimes \mathbf{Q}_{t}^T] \vec{r}, \label{eq:K_sph_inv}
    \end{align}
    where $\mathbb{I}_{MN}$ represents the $MN \times MN$ identity matrix. We use Eq.~(\ref{eq:Algo14}) to multiply $\vec{r}$ by the eigenvector matrices $\mathbf{Q}_{\lambda}$ and $\mathbf{Q}_{t}$. The middle term containing the product of eigenvalues is diagonal and so the inverse is easily calculated.
    
    Calculating $\log|\mathbf{K}|$ can be performed in $\mathcal{O}(MN)$ time using:
    \begin{equation}\label{eq:logdet}
    \log|\mathbf{K}| =  \sum_{i=1}^{N} \log((\mathbf{\Lambda}_{\lambda} \otimes \mathbf{\Lambda}_{t})_{ii} + \sigma^2).
    \end{equation}
    
    \subsection{Accounting for non-uniform white noise}
    \label{sec:rakitsch}
    
    The assumption of uniform white noise can be avoided by using a linear transformation presented in \citet{Rakitsch2013}. Using this transformation, we can efficiently solve for the log-likelihood for any covariance matrix that is the sum of two Kronecker products - as described by Eq.~(\ref{eq:kronsum}).
    
    First, we solve for the eigendecomposition of $\mathbf{\Sigma}_{\lambda}$ and $\mathbf{\Sigma}_{t}$, notating this as follows:
    \begin{align}
        \mathbf{\Sigma}_{\lambda} &= \mathbf{Q}_{\mathbf{\Sigma}_{\lambda}} \mathbf{\Lambda}_{\mathbf{\Sigma}_{\lambda}} \mathbf{Q}_{\mathbf{\Sigma}_{\lambda}}^T, \\
        \mathbf{\Sigma}_{t} &= \mathbf{Q}_{\mathbf{\Sigma}_{t}} \mathbf{\Lambda}_{\mathbf{\Sigma}_{t}} \mathbf{Q}_{\mathbf{\Sigma}_{t}}^T.
    \end{align}
    In the case that $\mathbf{\Sigma}_{\lambda}$ is diagonal then we simply have $\mathbf{\Sigma}_{\lambda} = \mathbf{\Lambda}_{\lambda}$ and $\mathbf{Q}_{\mathbf{\Sigma}_{\lambda}} = \mathbb{I}_N$ (and similarly for $\mathbf{\Sigma}_{t}$).
    
    We transform $\mathbf{K}_{\lambda}$ and $\mathbf{K}_{t}$ using these eigendecompositions:
    \begin{align}
        \tilde{\mathbf{K}}_{\lambda} &= \mathbf{\Lambda}_{\mathbf{\Sigma}_{\lambda}}^{-\frac{1}{2}} \mathbf{Q}_{\mathbf{\Sigma}_{\lambda}}^T \mathbf{K}_{\lambda} \mathbf{Q}_{\mathbf{\Sigma}_{\lambda}} \mathbf{\Lambda}_{\mathbf{\Sigma}_{\lambda}}^{-\frac{1}{2}}, \\
        \tilde{\mathbf{K}}_{t} &= \mathbf{\Lambda}_{\mathbf{\Sigma}_{t}}^{-\frac{1}{2}} \mathbf{Q}_{\mathbf{\Sigma}_{t}}^T \mathbf{K}_{t} \mathbf{Q}_{\mathbf{\Sigma}_{t}} \mathbf{\Lambda}_{\mathbf{\Sigma}_{t}}^{-\frac{1}{2}}.
    \end{align}
    Doing this allows us to write our covariance matrix $\mathbf{K}$ as a product of three terms:
    \begin{equation}
    \mathbf{K} = [\mathbf{Q}_{\mathbf{\Sigma}_{\lambda}} \mathbf{\Lambda}_{\mathbf{\Sigma}_{\lambda}}^{\frac{1}{2}} \otimes \mathbf{Q}_{\mathbf{\Sigma}_{t}} \mathbf{\Lambda}_{\mathbf{\Sigma}_{t}}^{\frac{1}{2}}] [\tilde{\mathbf{K}}_{\lambda} \otimes \tilde{\mathbf{K}}_{t} + \mathbb{I}_\mathrm{MN}] [\mathbf{\Lambda}_{\mathbf{\Sigma}_{\lambda}}^{\frac{1}{2}} \mathbf{Q}_{\mathbf{\Sigma}_{\lambda}}^T \otimes \mathbf{\Lambda}_{\mathbf{\Sigma}_{t}}^{\frac{1}{2}} \mathbf{Q}_{\mathbf{\Sigma}_{t}}^T].
    \label{eq:K_rakitsch}
    \end{equation}
    Multiplying this out reproduces Eq.~(\ref{eq:kronsum}).
    
    The inverse of this equation can be calculated to be:
    \begin{align}
    \mathbf{K}^{-1} &= [\mathbf{Q}_{\mathbf{\Sigma}_{\lambda}} \mathbf{\Lambda}_{\mathbf{\Sigma}_{\lambda}}^{-\frac{1}{2}} \otimes \mathbf{Q}_{\mathbf{\Sigma}_{t}} \mathbf{\Lambda}_{\mathbf{\Sigma}_{t}}^{-\frac{1}{2}}] [\tilde{\mathbf{K}}_{\lambda} \otimes \tilde{\mathbf{K}}_{t} + \mathbb{I}_\mathrm{MN}]^{-1} \nonumber \\
    &\times [\mathbf{\Lambda}_{\mathbf{\Sigma}_{\lambda}}^{-\frac{1}{2}} \mathbf{Q}_{\mathbf{\Sigma}_{\lambda}}^T \otimes \mathbf{\Lambda}_{\mathbf{\Sigma}_{t}}^{-\frac{1}{2}} \mathbf{Q}_{\mathbf{\Sigma}_{t}}^T].
    \label{eq:K_inv_R_rakitsch}
    \end{align}
    We can now solve $\mathbf{K}^{-1} \vec{r}$ in three steps. $\mathbf{K}^{-1}$ is the product of three matrices, two of which are of the form $\mathbf{A} \otimes \mathbf{B}$ and so we can use Eq.~(\ref{eq:Algo14}). The middle matrix in this product is equivalent to Eq.~(\ref{eq:K_sph_inv}) because we have a Kronecker product plus a constant added to the diagonal being multiplied by a vector. In this case, we need the eigendecomposition of the transformed covariance matrices $\tilde{\mathbf{K}}_{\lambda}$ and $\tilde{\mathbf{K}}_{t}$ and the `white noise' term is set to $\sigma=1$. The scaling has not changed significantly compared to the method with constant white noise although the eigendecomposition of four matrices instead of two is now required (although this is trivial if $\mathbf{\Sigma}_{\lambda}$ and $\mathbf{\Sigma}_{t}$ are diagonal).
    
    $\log|\mathbf{K}|$ is now computed as follows:
    \begin{equation}\label{eq:logdet2}
        \log|\mathbf{K}| =  \sum_{i=1}^{MN} \log((\mathbf{\Lambda}_\mathrm{\tilde{\mathbf{K}}_{\lambda}} \otimes \mathbf{\Lambda}_{\mathrm{\tilde{\mathbf{K}}}_{t}})_{ii} + 1) + \log((\mathbf{\Lambda}_{\mathbf{\Sigma}_{\lambda}} \otimes \mathbf{\Lambda}_{{\mathbf{\Sigma}}_{t}})_{ii}).
    \end{equation}
    
    The log-likelihood can now be solved for any covariance matrix that can be written in the form of Eq.~(\ref{eq:kronsum}) with an overall scaling of $\mathcal{O}(2M^3 + 2N^3 + MN(M+N))$ operations and $\mathcal{O}(M^2 + N^2)$ memory. Compared to a more general approach of performing Cholesky factorisation on the full $MN \times MN$ covariance matrix $\mathbf{K}$, our optimised method provides greater than order of magnitude improvements in runtime for typical transmission spectroscopy datasets (shown in Sect.~\ref{sec:benchmarking}).

    \subsection{Efficient inference with large numbers of parameters}
    
    As multiple parameters are typically fit for each spectroscopic light curve, parameter inference can become computationally expensive when simultaneously fitting multiple light curves and a good choice of inference method may be required. Hamiltonian Monte Carlo (HMC) was chosen as it can scale significantly better when fitting large numbers of parameters compared to other more traditional MCMC methods such as Metropolis-Hastings or Affine-Invariant MCMC \citep{Neal2011}. HMC can make use of the gradient of the log-likelihood to take longer trajectories through parameter space for each step of the MCMC compared to these other methods. Specifically, we use No U-Turn Sampling (NUTS) because it eliminates the need to hand-tune parameters while achieving a similar sampling efficiency to HMC \citep{Hoffman2011}.

    To provide the gradients of the log-likelihood calculation for NUTS, the above algorithms were implemented in the Python package \textsc{JAX} \citep{JAX}. \textsc{JAX} has the capability of providing the values as well as the gradients of functions, often with minimal additional runtime cost. It has an implementation of \textsc{NumPy} that allows \textsc{NumPy} code to be converted into \textsc{JAX} code with limited alterations. It also allows the same code to be compiled to run on either a CPU or GPU - where GPUs may provide significant computational advantages. Novel analytic expressions (that were used in combination with \textsc{JAX}) for the gradients and hessian of the log-likelihood were developed to aid numerical stability - as discussed in Appendices~\ref{app:gradients} and \ref{app:hessian} - with numerical stability further discussed in Appendix~\ref{app:stability}.
    

\section{Testing on simulated data}
\label{sec:3}


    As we do not know {\it a priori} if wavelength-correlated systematics are present in a real dataset, this could lead to challenging model selection problems deciding whether to fit the datasets with wavelength-independent 1D GPs or a 2D GP that can account for wavelength-correlated systematics. We will demonstrate using simulated data that 2D GPs can accurately account for both wavelength-independent and wavelength-correlated systematics. This avoids the need for model selection - which could become computationally intractable for large numbers of parameters and is also known to place a heavy reliance on the choice of priors\footnote{Model selection may still be required for the choice of kernel function (e.g. squared exponential) but this is an issue with GPs in general.} \citep{model_selection}. For example, the 2D GP method can fit for a wavelength length scale $l_\lambda$ - with the log-likelihood varying significantly with $l_\lambda$ - while the 1D GP method does not. This would result in the favoured model (e.g. measured using the Bayesian evidence) being strongly dependent on the arbitrary choice of prior on $l_\lambda$. This dependence on the choice of priors can be significantly reduced by avoiding model selection and instead marginalising over both wavelength-independent 1D GPs and wavelength-correlated 2D GPs. This can be performed simply by fitting with a 2D GP using a kernel such as from Eq.~(\ref{eq:kernel2D}) and choosing a prior where $l_\lambda$ can go to small values where the 2D GP becomes equivalent to joint-fitting wavelength-independent 1D GPs. While the choice of prior on $l_\lambda$ will still affect how strongly the 1D and 2D GP methods are weighted, both methods are still marginalised over, reducing the importance of the choice of priors.

    We aim to validate this approach by testing both methods across different sets of simulated data containing either wavelength-independent or wavelength-correlated systematics. We will show that 2D GPs can have an added benefit of sharing hyperparameters between light curves to better constrain systematics and improve accuracy. We also use these datasets to characterise how wavelength-correlated systematics may contaminate transmission spectra when fitting with 1D GPs.
    
    Finally, we study how we can account for common-mode systematics - systematics that are constant in wavelength - using 2D GPs. Typically, these are accounted for using a separate common-mode correction step before fitting for the transmission spectrum (to be described in Sect.~\ref{sec:common_mode}), we show that these systematics can be accounted for using a 2D GP while simultaneously retrieving the transmission spectrum.
    
    To summarise, we will demonstrate four major points:

    \begin{enumerate}[(i)]
    \item Sharing hyperparameters between light curves can improve the reliability of results compared to individual 1D GP fits.
    
    \item 2D GPs can accurately account for wavelength-independent systematics by fitting for a wavelength length scale parameter.
    
    \item When systematics are correlated in wavelength, 2D GPs can accurately retrieve transmission spectra while 1D GPs may retrieve erroneous results.

    \item Common-mode systematics can be accounted for using 2D GPs without requiring a separate common-mode correction step.
    
    \end{enumerate}

    Sections \ref{sec:light curve} - \ref{sec:metrics} will describe how the synthetic data were generated and analysed, with Sections \ref{sec:shared_params} - \ref{sec:common_mode} explaining how the results demonstrate points (i), (ii), (iii) and (iv).
    
    \subsection{Light curve model}
    \label{sec:light curve}
    In this work, our light curve model is calculated using the equations of \citet{mandelagol} using quadratic limb-darkening parameters $c_1$ and $c_2$. The other parameters describing the light curve are the central transit time ($T_0$), period ($P$), system scale ($a/R_\mathrm{*}$) and impact parameter ($b$) as well as the planet-to-star radius ratio ($\rho = R_\mathrm{p} / R_*$) we aim to measure. We chose a standard approach of using the transit depth ($\rho^2 = (R_\mathrm{p} / R_*)^2$) to fit each light curve (i.e. \citealt{PRISM2022, G3952022}). We can also fit a linear baseline function to each light curve, which introduces the flux out-of-transit parameter ($F_\mathrm{oot}$) and a fit to the slope of the baseline flux ($T_\mathrm{grad}$).
    
    Our mean function parameters therefore include five parameters that may be fit for each light curve independently ($\rho^2$, $c_1$, $c_2$, $F_\mathrm{oot}$, $T_\mathrm{grad}$) and four parameters that are shared across all light curves ($T_0$, $P$, $a/R_\mathrm{*}$, $b$). For the simulations, the transit depths $\rho^2$ were the only mean function parameters being fit for to reduce runtimes for the large number of simulations (the other parameters were kept fixed to their injected values).

    We note that the 1D GP method generated light curves using \textsc{batman} \citep{batman} to fit the data while the 2D GP method used \textsc{jaxoplanet} \citep{jaxoplanet}. Any differences in the light curves generated between the packages were orders of magnitude below the level of white noise being added. \textsc{jaxoplanet} is a Python package that can generate transit light curves similar to \textsc{batman} but is implemented in \textsc{JAX}. This allows for the calculation of gradients of the log-likelihood - as required to implement No U-Turn Sampling (NUTS).
    
    \subsection{Generating the exoplanet signal}
    \label{sec:mf}
    Each simulated dataset contained 100 time points and 16 wavelength channels. The parameter ranges used for generating the exoplanet signal for all sets of simulations are included in Table~\ref{tab:mf}. They are similar to the ranges chosen in \citet{Gibson2014} and are designed to represent typical hot Jupiter transits.
    
    The time range for these observations is [-0.1, 0.1] days sampled uniformly in time. For 100 time points this gives a cadence of approximately 3 minutes. This cadence is longer than many real observations - such as from VLT/FORS2 or JWST observations - but is still well-sampled. This was required to reduce runtimes and allow for the analysis of a large number of simulated datasets. The parameter ranges chosen resulted in the proportion of time spent in transit varying within the range [26\%, 65\%].
    
    The wavelength range considered was [3850\AA, 8650\AA] which for 16 evenly separated wavelength bins gives a bin width of 300\AA. This wide wavelength range ensures light curves with a range of limb darkening parameters were included. Limb darkening parameters were calculated by first uniformly sampling a confirmed transiting exoplanet listed on the NASA Exoplanet Archive\footnote{list downloaded on 2022 October 19} and selecting the listed host star parameters. These parameters were used to compute the limb darkening parameters using \textsc{PyLDTk}, allowing for a realistic range of stellar limb-darkening profiles \citep{LDTK}.
    
    \begin{table}
    	\centering
    	\caption{Mean function parameter ranges used for all simulations.}
    	\label{tab:mf}
    	\begin{tabular}{lll} 
    		\hline
    		Parameter & Symbol & Range\\
    		\hline
    		  Central transit time (days)  & $T_0$ & 0\\
    		Period (days) & $P$ & (2.75,3.25)\\
    		Scaled semi-major axis & $a/R_\mathrm{*}$ & (9,11)\\
            Impact parameter & $b$ & (0.05,0.85)\\
            Flux Out-of-transit & $F_\mathrm{oot}$ & 1\\
            Linear Baseline Flux & $T_\mathrm{grad}$ & 0\\
    		\hline
    	\end{tabular}
    \end{table}

    \begin{table}
    	\centering
    	\caption{Transmission spectrum parameter ranges for all simulations.}
    	\label{tab:trans_spec}
    	\begin{tabular}{lll} 
    		\hline
    		Parameter & Symbol & Range\\
    		\hline
    		  Radius Ratio & $R_\mathrm{p}/R_*$ & (0.08,0.12)\\
            Scattering Slope & $m$ & 0 or -0.005\\
            Radius Change at K & $\Delta \rho_\mathrm{K}$ & 0 or 0.005\\
    		\hline
    	\end{tabular}
    \end{table}
    
    The synthetic transmission spectra were generated using a simple model with the aim of testing the retrieval of typical sharp and broad features within transmission spectra. The planet-to-star radius ratio was randomly drawn as described in Table~\ref{tab:trans_spec}. To study transmission spectra that may have a scattering slope or flat `grey' cloud deck, there was a 50\% chance of a Rayleigh-scattering slope being included and a 50\% chance the spectrum was flat. The Rayleigh-scattering was implemented using Eq.~(\ref{eq:scattering}) from \citet{Etangs2008}:
    \begin{align}
         m = \frac{d R_\mathrm{p}/R_*}{d (\ln \lambda)} = \frac{\alpha H}{R_*},
        \label{eq:scattering}
    \end{align}
    where $m$ is the slope of the transmission spectrum, $H$ is the scale height and $\alpha$ describes the strength of scattering ($\alpha=-4$ for Rayleigh scattering).
    
    For simplicity, $m=0$ was used for a flat `cloudy' model and $m=-0.005$ for Rayleigh scattering. For context, WASP-31b has a reported height scale of 1220 km and stellar radius of $R_* = 1.12 M_\odot$ \citep{Sing2015, Anderson2011} which results in a predicted slope of $m = -0.0063$ for Rayleigh scattering.
    
\begin{figure}
    	\includegraphics[width=\columnwidth]{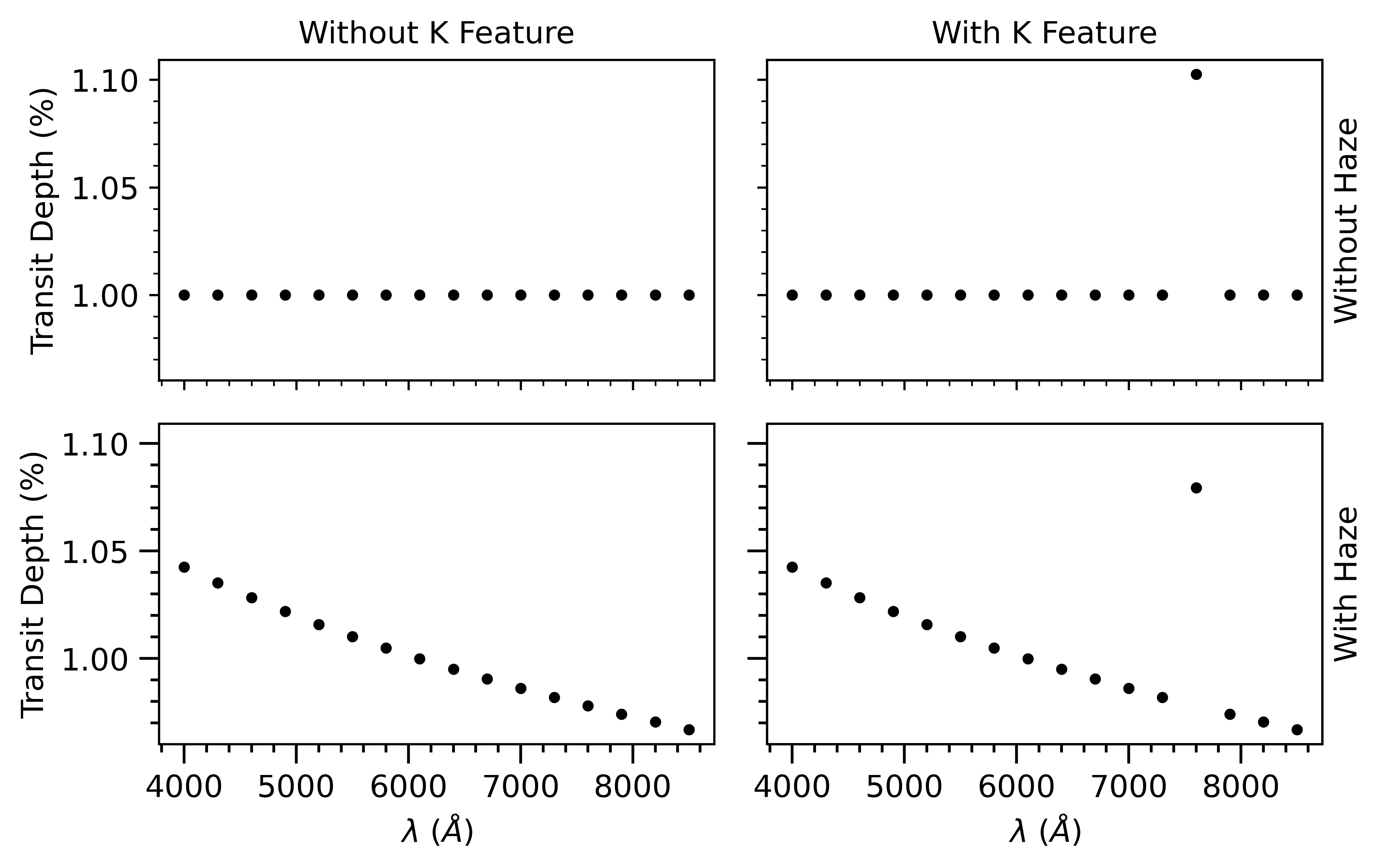}
        \caption{Four equally likely types of transmission spectra generated for a radius ratio of $\rho = 0.1$, showing the spectrum in transit depths.}
        \label{fig:example spectra}
    \end{figure}

    In order to compare how the different methods may constrain a sharp spectral feature such as a K feature (a sharp potassium absorption doublet at $\sim$7681$\angstrom$), each dataset had a 50\% probability of adding a change in radius ratio of $\Delta \rho_\mathrm{K} = 0.005$ to the single wavelength bin which covers the wavelength range $7450\angstrom$ to $7750\angstrom$. Overall this resulted in an even 25\% probability of the spectrum being flat and featureless, hazy and featureless, flat with a K feature, or hazy with a K feature (see Fig.~\ref{fig:example spectra}).
    
    \subsection{Generating systematics}
    \label{sec:systematics}

    500 synthetic transmission spectroscopy datasets (each containing 16 light curves) were generated for each of the four sets of simulations. All sets of simulations used the same transit signal model as described in Sect.~\ref{sec:mf}. Parameter ranges used to generate systematics for Simulations 1-4 are included in Table~\ref{tab:sim_1-4_hp}. Systematics generated were multiplied by each of the transit models generated. We chose to multiply by the transit signal because many sources of systematics would be expected to be proportional to flux (i.e. from varying instrumental or atmospheric throughput).
    
    For Simulation 1, noise was generated using Eq.~(\ref{eq:kernel}) to build a covariance matrix and by taking random draws from it independently for each wavelength. These datasets therefore contained wavelength-independent systematics (equivalently this is the limit of Eq.~(\ref{eq:kernel2D}) when $l_\lambda \to 0$). For Simulations 2 and 3, noise was generated using the kernel in Eq.~(\ref{eq:kernel2D}) and both represent scenarios where wavelength-correlated systematics are present but with different ranges of wavelength length scales. Simulation 4 represents contamination from common-mode systematics - which are often encountered in real observations and are relevant to the VLT/FORS2 analysis in Sect.~\ref{sec:4}. For each simulated dataset, a single draw of correlated noise was generated using the kernel in Eq.~(\ref{eq:kernel}) (with negligible white noise $\sigma = 10^{-6}$ for numerical stability). All wavelength channels used this same draw of correlated noise so that it was constant in wavelength (which is the limit of Eq.~\ref{eq:kernel2D} approached as $l_\lambda \to \infty$). White noise was then added to the full dataset from the parameter range in Table~\ref{tab:sim_1-4_hp}.

    For examples of noise-contaminated light curves generated, three synthetic light curves from a Simulation 1 dataset are plotted in the left plot of Fig.~\ref{fig:example recovery}, while all light curves from a Simulation 3 dataset are shown in the left plot of Fig.~\ref{fig:sim III light curves}. Simulation 2 datasets would appear similar to Simulation 3 but with the systematics changing shape more significantly between different light curves. Simulation 4 datasets would have the same shape systematics in all light curves (but with different white noise values).

     \begin{table}
    	\centering
    	\caption{Parameter ranges used to generate systematics for Simulations 1-4.}
    	\label{tab:sim_1-4_hp}
    	\begin{tabular}{lll} 
    		\hline
    		Parameter & Range\\
    		\hline
            $h$: Height scale & $(0.0005,0.0010)$\\
            $l_{t}$: Time length scale (days) & $(0.004,0.100)$\\
            $\sigma$: White noise amplitude & $(0.0001,0.0010)$\\
            $l_{\lambda}$: Sim. 1 wavelength length scale (\angstrom) & $l_{\lambda} \to 0$\\
            $l_{\lambda}$: Sim. 2 wavelength length scale (\angstrom) & $(300,2250)$\\
            $l_{\lambda}$: Sim. 3 wavelength length scale (\angstrom) & $(2250,36000)$\\
            $l_{\lambda}$: Sim. 4 wavelength length scale (\angstrom) & $l_{\lambda} \to \infty$\\
    		\hline
    	\end{tabular}
    \end{table}

    \subsection{Inference methods}
    \label{sec:MCMC}
    
    Each of the fitting methods were initialised with the true values used in the simulations. A global optimiser was then used to locate the best-fit values for the parameters being fit. MCMCs were initialised by perturbing them from this best-fit value and run until convergence was reached - as measured by the Gelman-Rubin statistic \citep{GR} - or until a limit on the number of chains or samples was reached.
    
    The 1D GP fits were performed using the \textsc{GeePea} \citep{Gibson2012} and \textsc{inferno} \citep{Gibson2017} Python packages. \textsc{GeePea} was used for the implementation of the 1D GP with the Nelder-Mead simplex algorithm used to locate a best-fit value. Differential Evolution MCMC (DE-MC) was used to explore the posterior of each of the light curves using the implementation in \textsc{inferno}. Four parameters were fit for each MCMC: $\rho^2$, $h$, $l_{t}$ and $\sigma$. 100 independent chains were run in each MCMC with 200 burn-in steps and another 200 steps run for the chain. If any of the parameters had a GR statistic $>1.01$ then the chains were extended another 200 steps with this repeating up to a maximum chain length of 1000. This was sufficient to achieve formal convergence in over 99\% of simulations performed.
    
    In order to perform inference for the 2D GP fits, the log-likelihood calculation was implemented in \textsc{JAX} using the equations described in Sect.~\ref{sec:rakitsch} and Appendix~\ref{app:gradients}. The Python package \textsc{PyMC} was used for its implementation of Limited-Memory BFGS for best-fits and NUTS for inference \citep{Salvatier2015}. Limited-Memory BFGS is a gradient-based optimisation method that can be used for models with large numbers (potentially thousands) of parameters \citep{LBFGS}.
    
    MCMC inference for the 2D GP method initially used two chains with 1000 burn-in steps each followed by another 1000 steps used for inference. Convergence was checked and if all parameters had not yet converged then another two chains were run. This was repeated until a maximum of ten chains were run. This was also sufficient to achieve convergence in over 99\% of simulations performed.

    For many MCMC methods, an approximation of the covariance matrix of the parameters being fit can be used as a transformation to reduce correlations between parameters and improve sampling efficiency. In NUTS, the so-called mass matrix performs this function \citep{Hoffman2011}. A Laplace approximation was performed to find an approximate covariance matrix, which requires calculating the Hessian of the log-posterior at the location of best-fit (see Appendix~\ref{app:hessian}). This was found to be significantly more efficient than using the samples from the burn-in period to approximate the covariance matrix.
    
    To overcome a few other challenges in achieving convergence (see Appendix~\ref{app:sim MCMC} for details), blocked Gibbs sampling \citep{blocked_Gibbs} was used to update some groups of parameters together. Slice sampling \citep{Neal2003} was also used if the time or wavelength length scales were within certain ranges close to a prior limit, as this could sometimes sample these parameters more efficiently than NUTS. Priors used for both 1D and 2D GP methods are described in Appendix~\ref{app:priors} and were unchanged between Simulations 1-4.
    
    \subsection{Accuracy metrics}
    \label{sec:metrics}

\begin{figure*}
    	\includegraphics[width=\textwidth]{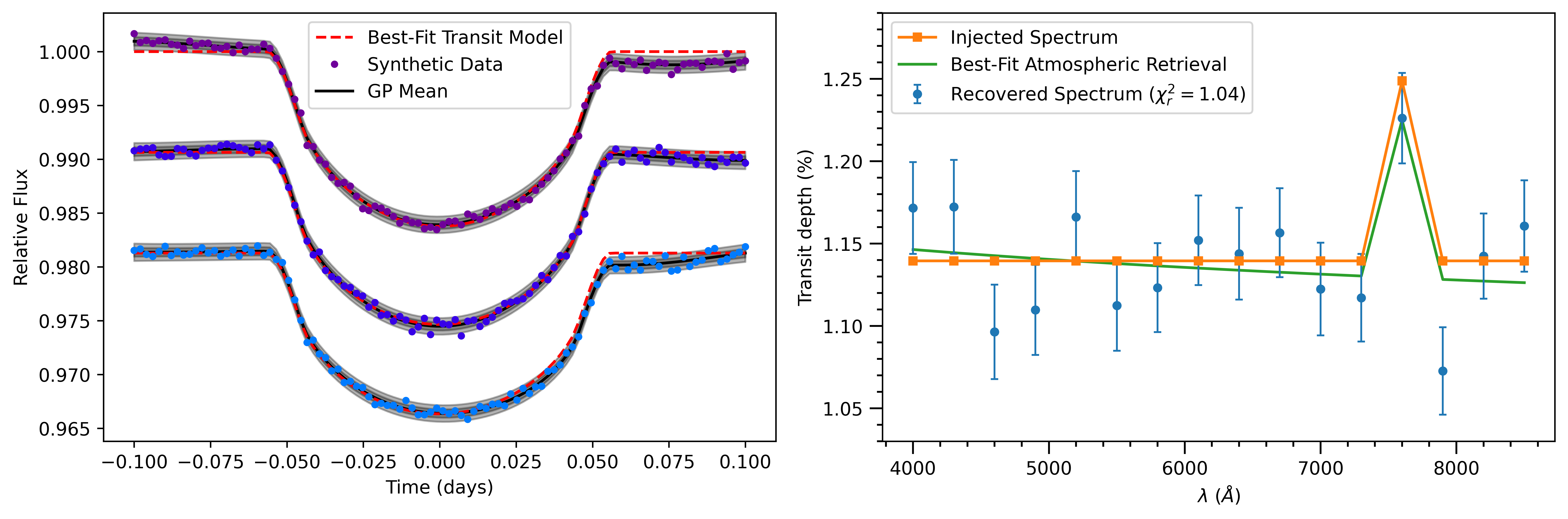}
        \caption{Example of synthetic dataset analysed in Simulation 1 (containing wavelength-independent systematics) showing some simulated light curves (left) and recovered transmission spectrum (right). Left: The three shortest wavelength light curves in the dataset being fit by the transit model combined with a 2D GP noise model (the other 13 light curves were simultaneously fit but not plotted). Right: Resulting transmission spectrum from the joint fit of all 16 light curves, with the three leftmost points corresponding to the light curves in the left plot. The recovered transmission spectrum was consistent with the injected spectrum ($\chi_r^2 = 1.04$). An atmospheric retrieval was performed on the recovered transmission spectrum (with the best-fit model plotted) and was consistent with the flat slope and strong K feature of the injected spectrum.}
        \label{fig:example recovery}
    \end{figure*}

    Multiple metrics were used to determine how accurately different methods could recover the injected transmission spectra. An example analysis of a single synthetic dataset from Simulation 1 is shown in Fig.~\ref{fig:example recovery}, demonstrating both the fitting of the light curves and how the resulting transmission spectrum and atmospheric retrieval compares to the injected spectrum. We will study how accurately each method can extract the transmission spectrum in addition to the accuracy of the retrieved atmospheric parameters.

    To measure the accuracy of the transmission spectra, we assumed that the uncertainty in the recovered transit depths was Gaussian. This implies that if we are accurately retrieving their mean and covariance, the reduced chi-squared $\chi_\mathrm{r}^2$ of the injected values should be distributed as a $\chi_\mathrm{r}^2$ distribution with 16 degrees of freedom. We compute the $\chi_\mathrm{r}^2$ of the injected transmission spectrum (accounting for covariance between transit depths) using:
    \begin{equation}
        \chi_\mathrm{r}^2 = (\bar{\vec{\rho}}_\mathrm{ret}^2 - \vec{\rho}_\mathrm{inj}^2)^T \mathbf{K}_\mathrm{\vec{\rho}^2; ret}^{-1} (\bar{\vec{\rho}}_\mathrm{ret}^2 - \vec{\rho}_\mathrm{inj}^2) / M,
        \label{eq:chi2}
    \end{equation}
    where $\bar{\vec{\rho}}_\mathrm{ret}^2$ and $\mathbf{K}_\mathrm{\rho^2; ret}$ represent the retrieved mean and covariance of the transmission spectrum respectively (with $\vec{\rho}^2$ representing the transit depths). $\vec{\rho}_\mathrm{inj}^2$ represents the injected transmission spectrum and $M$ is the number of light curves ($M = 16$ for all the simulations in this paper). As the 1D GP method fits each light curve independently, the covariance between all the transit depths is always assumed to be zero, resulting in the covariance matrices from the 1D GP method being diagonal. The 2D GP method will instead recover the full covariance matrix of transit depths from the MCMC.

    Since a single synthetic dataset is not sufficient to study the accuracy of each method, we chose to generate 500 synthetic datasets for each set of simulations to provide a sufficient sample to test each method. By comparing how the resulting 500 $\chi_\mathrm{r}^2$ values trace out the theoretical $\chi_\mathrm{r}^2$ distribution, we can determine if a method is accurately recovering the uncertainty in the transmission spectrum. For $M = 16$ degrees of freedom, $\chi_\mathrm{r}^2$ has mean $\mu = 1$ and variance $\sigma^2 = 2/M = 0.125$. The mean $\chi_\mathrm{r}^2$ being significantly larger than one suggests our uncertainties are too small on average, while the mean being smaller than one suggests our uncertainties are too large on average. We should expect the sample variance of 500 samples from a $\chi_\mathrm{r}^2$ distribution with 16 degrees of freedom to be distributed as $0.125 \pm 0.009$, if values diverge from this it may suggest the presence of outliers.
    
    We also performed one-sample Kolmogorov–Smirnov (K-S) tests on the distribution of recovered $\chi_\mathrm{r}^2$ values, which is a method for testing if a set of samples follows a given distribution - in this case a $\chi_\mathrm{r}^2$ distribution. It works by comparing the empirical cumulative distribution function (ECDF) of the samples to the cumulative distribution function (CDF) of the distribution chosen, returning the maximum deviation $D$ from the desired CDF. The K-S statistic $D$ for 500 samples and for a $\chi_\mathrm{r}^2$ distribution with 16 degrees of freedom should have $D$ < 0.060 at a p-value of 0.05 and $D$ < 0.072 at a p-value of 0.01.

    In addition to examining the $\chi_\mathrm{r}^2$ values, we can also examine how accurately we can retrieve atmospheric features. For our simulations, there are three features of the injected transmission spectra: the radius ratio $\rho = R_\mathrm{p}/R_*$, slope $m = d (R_\mathrm{p}/R_*)/d (\ln \lambda)$ and change in radius ratio in the wavelength bin centred on the K feature $\Delta \rho_\mathrm{K}$. For each simulation, these three features were determined by performing a simple atmospheric retrieval on the recovered transmission spectrum using the same atmospheric model which was used to generate the injected spectra. We determined the mean and uncertainty for each of these features using an MCMC (also using the NUTS algorithm). We measured the number of standard deviations - or the Z-score - each prediction was away from the true injected value for all 500 synthetic datasets (used previously in e.g. \citealt{Carter2009, Gibson2014, Ih2021}). If we are robustly retrieving a given atmospheric parameter, we expect these 500 Z-scores to follow a Gaussian distribution with mean $\mu = 0$ and variance $\sigma^2 = 1$.

    If the mean Z-score is zero, it would demonstrate that the parameter is not biased towards larger or smaller values than the true values. The Z-score values having unit variance shows that the uncertainty estimates are accurate. If the variance is less than one then that suggests the error bars are too large on average which might result in a missed opportunity to detect an atmospheric feature. If the variance is greater than one then the error bars are too small on average which could lead to false detections of signals.

    If two methods are both shown to produce reliable uncertainties based on these metrics, then we can also study which method gives smaller uncertainties on the transmission spectrum. Some methods described can produce tighter constraints than others but at the cost of making stronger assumptions about the systematics in the data.

    Tables of each accuracy metric for all methods tested are included in Appendix~\ref{app:sim_results}.

    \subsection{Result (i): Sharing hyperparameters can make constraints more reliable}
    \label{sec:shared_params}

    Before fitting for wavelength-correlations, we first demonstrate that simply by joint-fitting spectroscopic light curves and sharing hyperparameters between light curves, we can more reliably constrain transmission spectra. This technique has already been demonstrated on ground-based observations of WASP-94Ab \citep{Ahrer2022a}, but to examine reliability we use it on simulated datasets where the transmission spectrum is known.

    For this section, we used the Simulation 1 data containing only wavelength-independent systematics. We first fit each synthetic dataset with individual 1D GPs that fit a separate transit depth $\rho^2$, height scale $h$, time length scale $l_t$ and white noise amplitude $\sigma$ to each individual light curve. The kernel in Eq.~(\ref{eq:kernel}) was used for the GP (the same kernel used to generate the systematics). We compared this to joint-fitting all 16 spectroscopic light curves in each dataset, where we still fit for all 16 transit depths but used single shared values for $h$, $l_t$ and $\sigma$. This provides more information to constrain the values of the hyperparmeters and should produce better constraints than using wavelength-varying values. We note that this can be considered to be using a 2D GP as the kernel is a function of the two dimensions time and wavelength (the correlation in wavelength is given by the Kronecker delta $\delta_{\lambda_i \lambda_j}$). We therefore performed this joint-fit identically to the 2D GP method described in Sect.~\ref{sec:MCMC} and used the same kernel function but fixed the wavelength length scale $l_\lambda$ to be negligible\footnote{for numerical reasons we could not set $l_\lambda=0$ but instead set $l_\lambda=0.1\angstrom$}. Comparing Eq.~(\ref{eq:kernel}) and Eq.~(\ref{eq:kernel2D}), it can be seen that both kernel functions are mathematically equivalent in the limit as $l_\lambda \to 0$. We therefore refer to this method as the Hybrid method as it shares the same kernel function as the 1D GP method but joint-fits light curves and can benefit from shared hyperparameters like the 2D GP method.

    The Hybrid method had either equal or better performance compared to individual 1D GPs across every accuracy metric measured. For example, the distribution of $\chi_r^2$ values for the 1D GPs had a mean of $\bar{\chi}_r^2 = 1.35\pm0.04$, compared to $\bar{\chi}_r^2 = 1.00\pm0.02$. Similarly, the Z-scores of all three atmospheric parameters studied had sample variances that were constrained to be greater than one for the 1D GP fits but consistent with one for the joint-fits. 
    
    Our results suggest that 1D GPs underestimated the uncertainties of the transmission spectra on average, which may appear surprising given that the systematics in Simulation 1 were generated using 1D GPs with the exact same kernel function. The different results for the two methods appear to have been largely driven by a small number of outliers where individual light curves were fit to have much weaker systematics than was present. This did not happen with the joint-fits and since both methods have mathematically equivalent kernel functions, we interpret this result as the 1D GPs not having sufficient data to consistently fit the systematics. This could happen if an individual light curve happens to have systematics that have a similar shape to a transit dip, making it difficult to determine the amplitude of the systematics. The joint-fit can instead make use of more light curves to accurately constrain the systematics. We note that the 1D GP method had metrics consistent with ideal statistics for the 130 datasets with $l_t < 0.01$ days, which is much shorter than the minimum transit duration of 0.052 days and therefore less likely to mimic a transit dip. We conclude that sharing hyperparameters can improve the reliability of data analyses by utilising more data to account for the systematics, verifying Result (i).
    
    We note however that for real data we need to be careful with which parameters we assume are wavelength-independent. The amplitude of systematics and white noise could vary significantly across wavelength channels. For our VLT/FORS2 analysis in Sect.~\ref{sec:4}, we only assume that the length scales of correlated noise are the same across light curves, similar to \citet{Ahrer2022a}.

    \subsection{Result (ii): 2D GPs can account for wavelength-independent systematics}
    \label{sec:indep_systematics}

    As stated at the beginning of Sect.~\ref{sec:3}, 2D GPs (with the kernel given in Eq.~\ref{eq:kernel2D}) can be used to fit wavelength-independent systematics, meaning that there is no need to use model selection to determine whether to use 1D GPs or 2D GPs. We have already demonstrated that a 2D GP with a wavelength length scale fixed to negligible values (the Hybrid method) can accurately account for wavelength-independent systematics. For this test, we wanted to demonstrate that even if we fit for the wavelength length scale with a broad prior then we can retrieve similar results. This method has the benefit of avoiding the {\it a priori} assumption that systematics are wavelength-independent.

    Our results showed no statistically significant difference in any accuracy metric measured between the Hybrid method and fitting for $l_\lambda$ with a 2D GP. Both methods showed ideal $\chi_r^2$ statistics, based on both the mean and variance of the resulting $\chi_r^2$ distributions and the values of the K-S statistics. The Z-score of all retrieved atmospheric parameters were both consistent with the unit normal distribution. The average recovered transmission spectrum did have $2.4\%$ larger uncertainties however so there may have been a minor loss in precision from the more conservative approach of fitting for the wavelength length scale.
    
    We see that if systematics are wavelength-independent, a 2D GP can constrain the wavelength length scale to be sufficiently small to provide robust retrievals. This also has limited cost to the precision of the transmission spectrum compared to assuming the systematics are wavelength-independent {\it a priori}, confirming Result (ii).

    \subsection{Result (iii): 2D GPs can account for wavelength-correlated systematics}
    \label{sec:corr_systematics}

    \begin{figure}
    	\includegraphics[width=\columnwidth]{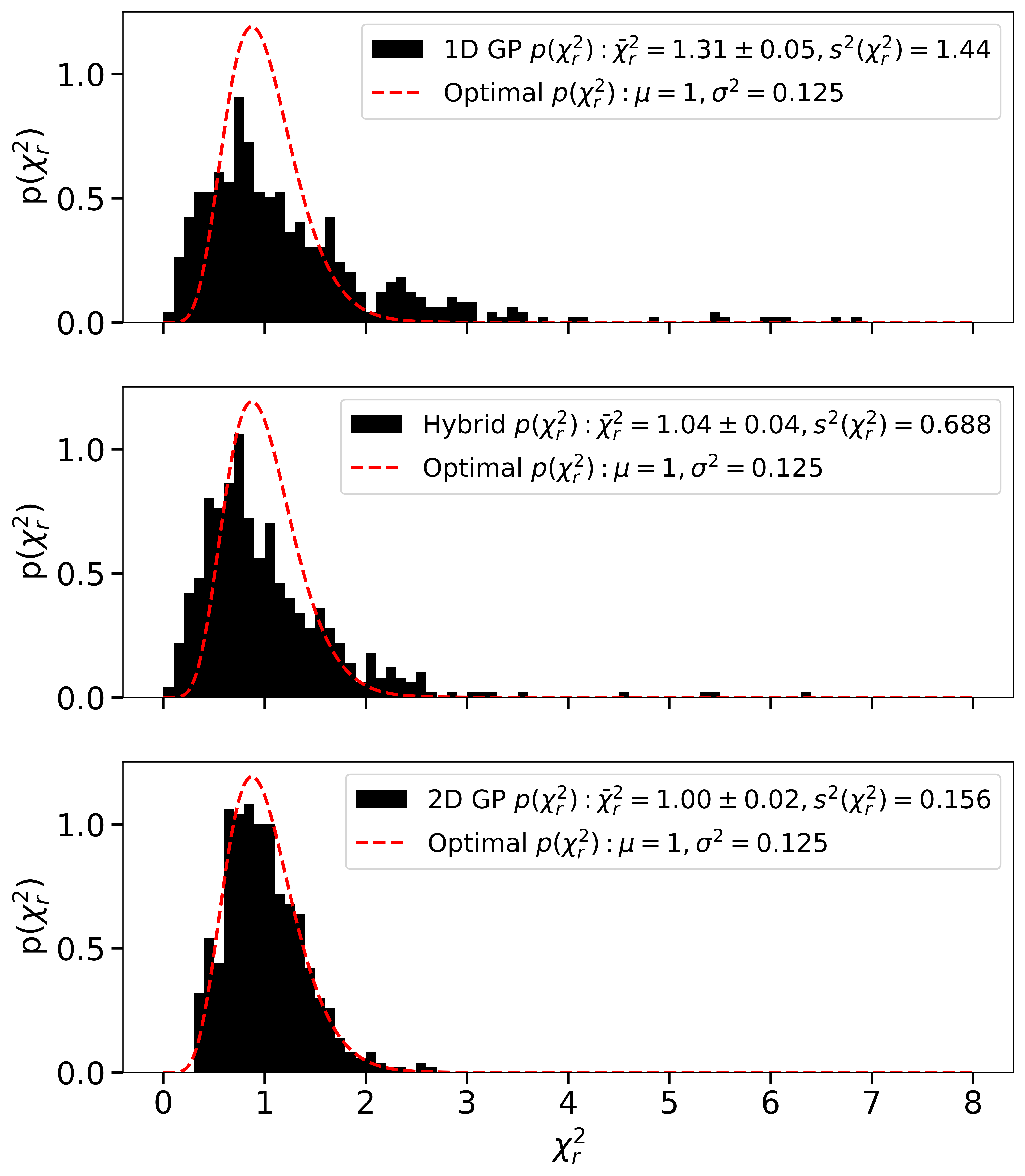}
        \caption{$\chi_\mathrm{r}^2$ histograms for all three methods with theoretical $\chi_\mathrm{r}^2$ distribution overplotted. Both the 1D GP and Hybrid methods have many outliers that increase the variance of the retrieved $\chi_\mathrm{r}^2$ values. The $\chi_\mathrm{r}^2$ values of the Hybrid method have the correct mean but with greater variance than the theoretical $\chi_\mathrm{r}^2$ distribution, while the 2D GP traces the theoretical distribution very closely.}
        \label{fig:sim II chi2}
    \end{figure}
    
    To study how accurately 1D GPs and 2D GPs can account for systematics correlated in both time and wavelength, we analysed Simulation 2 datasets that cover systematics correlated over wavelength length scales larger than one wavelength bin but smaller than half the wavelength range of the datasets. We fit all 500 datasets with 1D GPs and 2D GPs. However, in addition to having a different kernel to the 1D GP method, the 2D GP could also benefit from sharing hyperparameters (Result i). To isolate the effect of each of these differences, we also fit the Hybrid method to each dataset, which shares hyperparameters between light curves but uses a mathematically equivalent kernel to the 1D GP method.

    \begin{figure*}
    	\includegraphics[width=\textwidth]{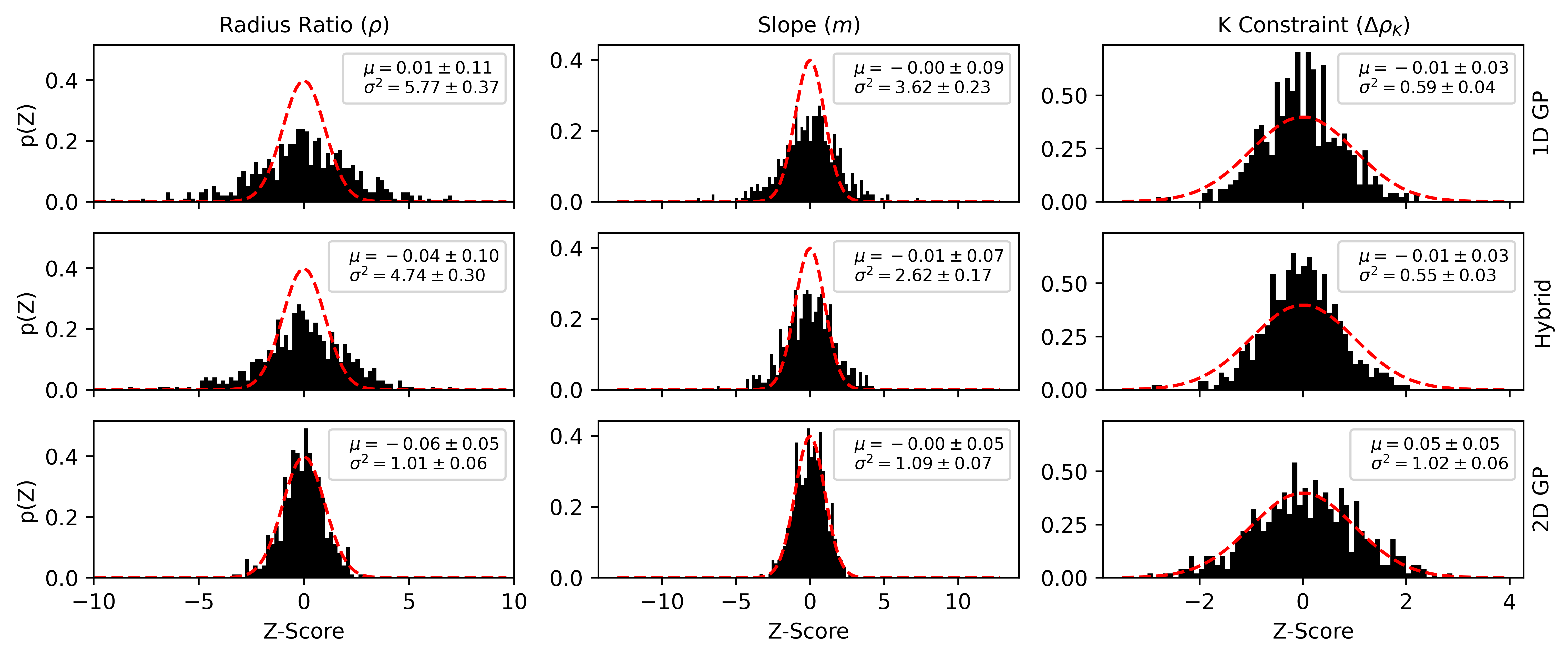}
        \caption{Z-scores of retrieved atmospheric features from Simulation 2, which had short wavelength length scale correlated noise. The three methods tested were fitting the light curves with 1D GPs (top row), the Hybrid method (middle row) and a 2D GP (bottom row). All methods have mean Z-scores consistent with zero, that is none of them biases the results towards measuring larger or smaller values of the features. The retrieved variance of Z-scores for the 2D GP method are all consistent with a variance of one (matching the Gaussian distribution plotted in red), but not for the other methods which indicates overestimation or underestimation of uncertainties.}
        \label{fig:sim II retrievals}
    \end{figure*}

    We found that only the 2D GP method had correctly distributed $\chi_r^2$ values and Z-scores of atmospheric parameters with unit variance. The $\chi_r^2$ distributions of each method are shown in Fig.~\ref{fig:sim II chi2}. The Hybrid method produces a distribution of $\chi_\mathrm{r}^2$ values with the correct mean but with greatly increased variance, while the 1D GP method performs similarly but with outliers producing a higher mean $\chi_r^2$ value (similar to Sect.~\ref{sec:shared_params}). We interpret the result of the Hybrid method as demonstrating that the systematics in each individual light curve are being accurately described by the kernel function of the Hybrid method (resulting in the correct mean $\chi_r^2$ value) but the correlation between light curves is not being accounted for (increasing the variance of the $\chi_r^2$ values). To understand this, consider fitting two light curves where it is incorrectly assumed that the systematics are independent, but the systematics are actually identical. If the systematics in one light curve are recovered with the correct systematics model but happen to result in a $2\sigma$ error in the measured transit depth, then the same measurement will occur in the other light curve. Both light curves analysed together would appear to result in two independent $2\sigma$ errors, resulting in a high $\chi_r^2$ value. However, when we account for the fact that the two light curves are perfectly correlated, this is really only a single $2\sigma$ error with a smaller $\chi_r^2$ value. Similarly, when the error in transit depth is very small in one light curve (i.e. 0.1$\sigma$), it will be small in both light curves, resulting in a $\chi_r^2$ value that is too low. The effect of this is that the $\chi_r^2$ distribution of two light curves can produce the correct mean but with increased variance when correlations between the light curves are not accounted for.
    
    The 1D GP and Hybrid methods were found to poorly retrieve atmospheric parameters: on average the radius ratio and slope uncertainties were underestimated and the uncertainty in the strength of a K feature was overestimated. This is shown in Fig.~\ref{fig:sim II retrievals}, where it can be seen that the Z-scores have variance that is either too high or too low for the different features using these two methods. In contrast, the Z-scores for the 2D GP method are consistent with unit normal distributions.
    
    To understand our interpretation of why the 1D GP kernel underestimates errors in broad features (such as the slope and radius ratio), but can overestimate errors for sharp features (such as the K feature), note that signals in data are most degenerate with systematics when both have similar length scales. For example, wavelength-independent systematics should be more degenerate with a sharp K feature compared to broader features. A slope in the transmission spectrum should be more degenerate with systematics which vary gradually in wavelength (i.e. with a long wavelength length scale). Both the 1D GP and Hybrid methods assume that the systematics are wavelength-independent, which could produce uncertainties that are too large when fitting the sharp K feature but too small for broader features. The 2D GP can constrain the systematics to have a longer wavelength length scale, which may increase the uncertainty in broader features while reducing uncertainty in sharp features. This result matches related work examining how different length scale correlations in exoplanet transmission or emission spectra affect atmospheric retrievals (\citealt{Ih2021, Evert2023}) but it had yet to be demonstrated how these correlations can arise from wavelength-correlated systematics in the spectroscopic light curves.

    \subsection{Result (iv): 2D GPs can account for common-mode systematics}
    \label{sec:common_mode}

    We first consider how to fit the Simulation 4 data containing common-mode systematics (i.e. systematics that are constant in wavelength) before analysing Simulation 3 containing long wavelength length scale systematics. Common-mode systematics are often encountered in transmission spectroscopy datasets, particularly from ground-based observations such as in the VLT/FORS2 observations analysed in Sect.~\ref{sec:4}. These systematics are often dealt with by performing a `common-mode correction'. First, a much broader wavelength range relative to the spectroscopic light curves is chosen to extract a `white' light curve. This was replicated for the simulations by averaging all spectroscopic light curves together. The systematics in this white light curve are then fit with a 1D GP similar to the individual spectroscopic light curve fits. The GP predictive mean is used to model the systematics in this white light curve, which is assumed to describe the common-mode systematic affecting all spectroscopic light curves. Each of the spectroscopic light curves is then divided by this recovered common-mode systematic, resulting in `common-mode corrected' spectroscopic light curves. Each of these corrected light curves is then fit separately to recover the transmission spectrum using the 1D GP approach already outlined, accounting for any wavelength-independent systematics in the light curves.

    There are multiple weaknesses of performing this separate common-mode correction. It is known that it can produce offsets to the resulting transmission spectrum, as the particular fit of the common-mode systematic chosen to correct the spectroscopic light curves can shift all transit depths equally. Only a single fit of the common-mode systematic is chosen to correct the light curves, so the transmission spectrum is conditioned on this particular fit instead of marginalising over all possible common-mode corrections. This likely has the effect of underestimating the uncertainty in the average radius ratio because different common-mode corrections offset the transmission spectrum differently. Finally, the assumption that a real dataset contains systematics that are constant in wavelength may be incorrect, as examined in Sect.~\ref{sec:4}. 2D GPs can avoid using a common-mode correction because they can account for common-mode systematics by using the kernel in Eq.~(\ref{eq:kernel2D}) with $l_\lambda$ set to be effectively infinite (in practise to very large values to maintain numerical stability).

    We tested three different approaches on each Simulation 4 dataset. First, all spectroscopic light curves were averaged together to perform a common-mode correction on each dataset, followed by fitting the corrected light curves with 1D GPs as performed for the other simulations. We also fit the data using a 2D GP with the wavelength length scale fixed to a very large value (50 times the wavelength range of the data) which effectively assumes {\it a priori} that the systematics present do not vary in wavelength. This is similar to the common-mode correction method but does not take into account remaining wavelength-independent systematics that were not simulated and also takes advantage of shared hyperparameters between light curves. Finally, we fit a 2D GP where the wavelength length scale was fit to the data (with a log-uniform prior from 1/4 of a wavelength bin to 50 times the total wavelength range of the dataset). This can be viewed as a more conservative approach that uses the data to constrain whether or not the systematics are constant in wavelength. Both 2D GPs used the kernel in Eq.~(\ref{eq:kernel2D}).

    We found as expected that the common-mode 1D GP method underestimated the uncertainty in the radius ratio. In contrast, both 2D GP methods were able to retrieve the radius ratio accurately with unit normal Z-score distributions, as they are not conditioned on a single fit of the common-mode systematics. All methods performed similarly at retrieving the strength of a K feature accurately.
    
    \begin{figure*}
    	\includegraphics[width=\textwidth]{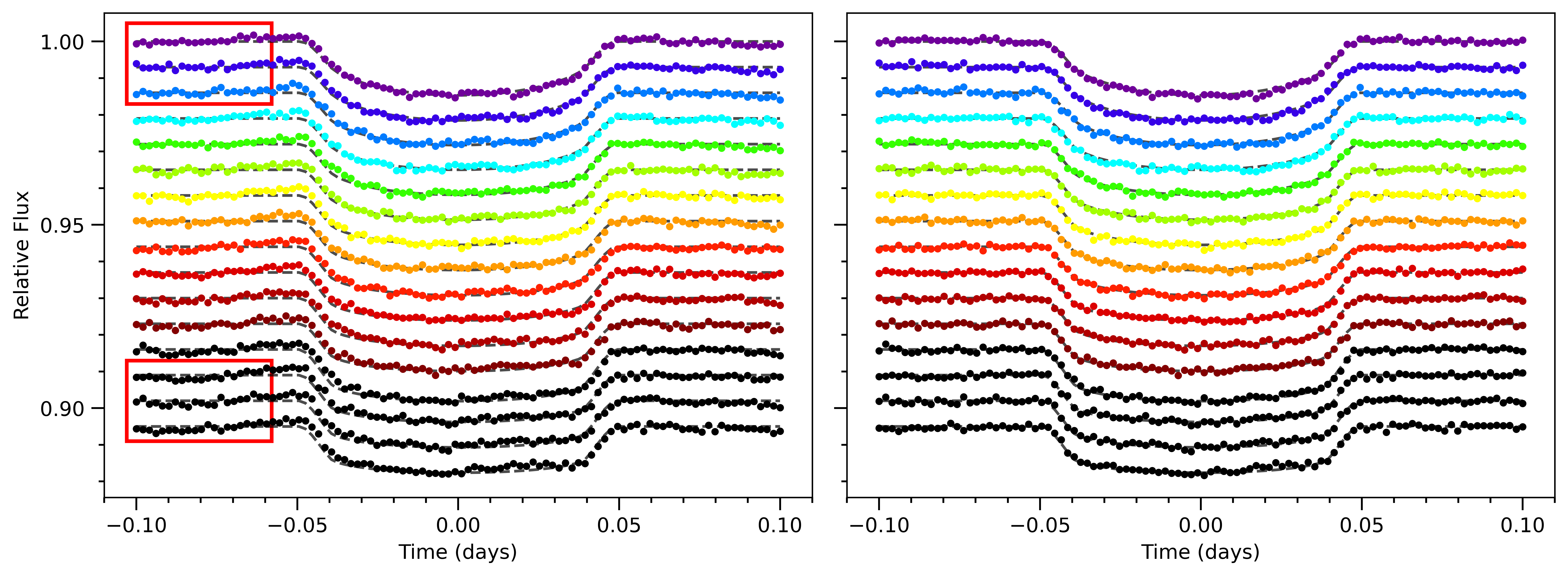}
        \caption{Raw light curves (left) and common-mode corrected light curves (right) for one simulated dataset in Simulation 3 (containing wavelength-correlated systematics). The injected transit signal is plotted as a black dotted line. The variation of the systematics in wavelength is not obvious by eye but can be noticed by comparing the regions highlighted in red boxes. Despite the systematics varying in wavelength, the common-mode correction appears to remove visual signs of systematics.}
        \label{fig:sim III light curves}
    \end{figure*}

    \begin{figure*}
    	\includegraphics[width=\textwidth]{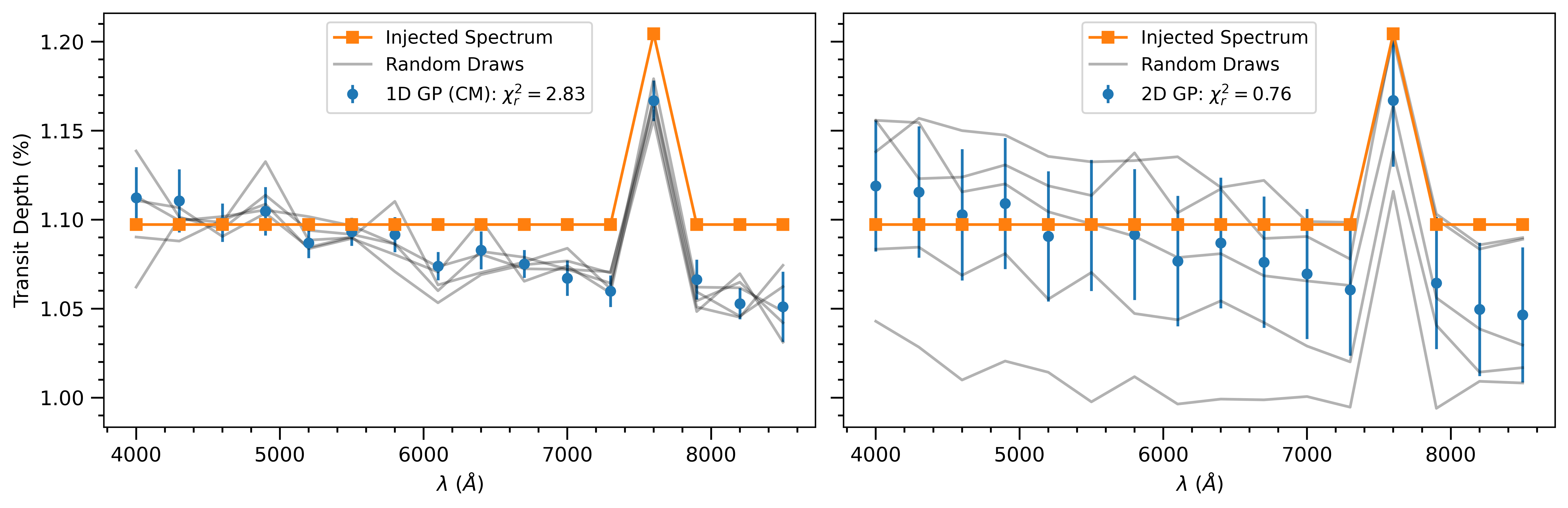}
        \caption{Example of retrieved transmission spectra with 1D GP method (left) analysing common-mode corrected light curves (right plot in Fig.~\ref{fig:sim III light curves}) compared with 2D GP method (right) analysing uncorrected light curves (left plot in Fig.~\ref{fig:sim III light curves}). The error bars in blue only convey the mean and standard deviation of each transit depth measurement, random draws are taken from the covariance matrix to convey potential correlations between transit depths, helping to visualise the increased uncertainty in the offset and scattering slope of the spectrum from the 2D GP method. Only the 1D GP method erroneously detects a negative scattering slope. While it is not visually apparent, the 2D GP method gives a stronger constraint on the detection of potassium ($13.9\sigma$ compared to $8.7\sigma$).}
        \label{fig:sim III transmission spectrum}
    \end{figure*}
    
    Interestingly, the 2D GP fitting for the wavelength length scale overestimated the uncertainty in the slope of the transmission spectrum (Z-score variance of $0.67 \pm 0.04$), while both methods which assumed constant wavelength systematics retrieved the slope with ideal Z-score statistics and with $\approx 15\%$ smaller average standard deviations. Fitting for the wavelength length scale appears to have reduced precision in the constraint of the scattering slope. Since systematics varying in wavelength could produce an apparent slope in the transmission spectrum, it makes sense that removing the assumption that the systematics are constant in wavelength has increased the uncertainty in our constraints.

    While this loss in precision may not be desirable, we must also consider the risk of incorrectly assuming systematics are constant in wavelength. For Simulation 3, we generated systematics that gradually varied in wavelength using long wavelength length scales (>1/2 the wavelength range of the dataset but <8 times the wavelength range of the dataset). We then analysed these datasets with the same three methods as for the common-mode systematics. In this case, only the 2D GP method that fit for the wavelength length scale produced robust retrievals of the injected signals across all accuracy metrics, although in this case the constraints on the slope may have been slightly too small with a Z-score sample variance of $s^2(Z_m) = 1.18 \pm 0.07$ (although the other two methods both had $s^2(Z_m) > 7$). We note that when fitting for $l_\lambda$, the slope constraints for Simulations 3 and 4 may be affected by the choice of prior bounds as $l_\lambda$ was often consistent with its maximum prior bound for both sets of simulations.
    
    An example of one of the synthetic datasets from Simulation 3 is shown in the left plot of Fig.~\ref{fig:sim III light curves}, along with the corresponding common-mode corrected light curves (right plot). We note that the systematics may appear to be constant in wavelength, despite being generated with a wavelength length scale of $4170\angstrom$. The resulting transmission spectra for the 1D GPs with a common-mode correction compared to the 2D GP method fitting for $l_\lambda$ are included in Fig.~\ref{fig:sim III transmission spectrum}. This was chosen as a particularly extreme example where the common-mode 1D GP method incorrectly detected a non-zero slope at $5.8\sigma$, compared to $2.1\sigma$ for a 2D GP fitting for $l_\lambda$. Out of the datasets that were simulated to have a flat spectrum, the 1D GP method with a common-mode correction retrieved a slope $>3\sigma$ away from zero in 26.7\% of them. This is compared to 0.4\% of these $>3\sigma$ outliers with the 2D GP method (consistent with the expected 0.3\% for a normally-distributed variable). This clearly demonstrates the potential for false detections of scattering slopes if systematics are incorrectly assumed to be constant in wavelength.
    
    Overall, the 2D GP fitting for the wavelength length scale performed well on Simulations 3 and 4 compared to methods which assumed that the systematics were common-mode. While there was some loss in precision on the scattering slope for common-mode systematics, this was likely due to a more conservative approach which performed significantly better on long wavelength length scale systematics.

    \subsection{Conclusions from simulations}
    \label{sec:sim_summary}

    2D GPs allow for hyperparameters to be shared between light curves, which we showed in Simulation 1 has the potential to improve the reliability of retrievals. 2D GPs with a broad wavelength length scale prior are able to accurately account for systematics across a wide range of wavelength length scales - including when the systematics are wavelength-independent (Simulation 1), varying in wavelength (Simulations 2 and 3), or constant in wavelength (Simulation 4).

    When systematics are correlated in wavelength with a short wavelength length (Simulation 2), analysing these data with 1D GPs or the Hybrid method resulted in constraints on broad features (i.e. the radius ratio and scattering slope) that were too small and constraints on sharp features (e.g. a K feature) that were too large.
    
    The effect of incorrectly assuming wavelength-correlated systematics are common-mode was examined in Simulation 3. This was also found to significantly underestimate the uncertainty in the scattering slope, which may help explain previous conflicting scattering slope measurements in the literature. We note that stellar activity (e.g. from unocculted starspots) has previously been used to explain conflicting scattering slopes between different transits (e.g. \citealt{Rackham2017, May2020}). Since unaccounted for wavelength-correlated systematics also has the potential to affect the measured transmission spectrum slope, it is possible that stellar properties such as star-spot covering fractions deduced from the slope of the transmission spectrum may be inaccurate.
    
    We conclude our 2D GP method is a much safer approach to fitting noise which may have wavelength-correlated systematics present and was the only model tested that could accurately account for systematics across all four sets of simulations, although common-mode systematics may cause the uncertainty in the scattering slope to be slightly overestimated. As our method does not require a separate common-mode correction, the uncertainty in the radius ratio can be fully accounted for in the transmission spectrum.


\section{Re-analysis of VLT/FORS2 observations}
\label{sec:4}


    To test our method on real data, we performed a re-analysis of the VLT/FORS2 transit observations of WASP-31b first analysed in \citet{Gibson2017}. We aimed to identify whether wavelength-correlated systematics are present in these observations and how the constraints on atmospheric parameters changed after accounting for these systematics. Based on the results in Sect.~\ref{sec:3}, if there is strong evidence for wavelength-varying systematics from the 2D GP analyses (based on the recovered wavelength length scale) then we should be skeptical of the original analysis as it did not account for wavelength-varying systematics - making the constraints unreliable.
    
    WASP-31b is a low-density hot Jupiter. It was initially reported in \cite{Anderson2011} with a mass of $M_\mathrm{J} = 0.478 \pm 0.029$ and radius of $R_\mathrm{J} = 1.549 \pm 0.050$. It has a 3.4 day orbit around a late F-type dwarf (V = 11.7).
    
    The data consist of two transits of WASP-31b, the first using the GRIS600B grism taken on the night of 2016 February 15 and the second using the GRIS600RI grism with the GG435 order blocking filter taken on 2016 March 3. These are referred to as the 600B and 600RI datasets, with the extracted light curves covering the wavelength ranges $3868-6238\angstrom$ and $5206-8476\angstrom$ respectively. See \citet{Gibson2017} for more details about the observations and subsequent data reduction. The same wavelength calibration and spectral extraction were used as in the previous analysis and the same $150 \angstrom$ wide wavelength bins were used for extracting the spectroscopic light curves.
    
    As noted in Sect.~\ref{sec:1}, these data conflict with previous observations from HST/STIS which identified a strong K feature at 4.3$\sigma$ significance \citep{Sing2015} not replicated in the original VLT analysis in \citet{Gibson2017}. This conflict was the initial reason for choosing these observations to re-analyse, in order to study if accounting for wavelength-correlated noise may resolve this discrepancy. We leave the re-analysis of the HST/STIS light curves for future work.
    
    \subsection{Previous analysis with 1D GPs}
    
    The approach to fit the light curves used in \citet{Gibson2017} was as follows: After basic data reduction steps were performed (such as background subtraction, aperture extraction and wavelength calibration), the flux from the target star was divided by flux from a comparison star to account for changes in flux due to the Earth's atmosphere.
    
    A common-mode correction was then performed - as described in Sect.~\ref{sec:common_mode}. However, in addition to dividing the spectroscopic light curves through by the common-mode systematics, the residuals of the white light curve (found by subtracting the GP predictive mean from the white light curve) were subtracted from each spectroscopic light curve. This was to account for high-frequency systematics - which are similar to common-mode systematics but occur on a faster timescale, making them indistinguishable from white noise in the white light curve fit. The presence of high-frequency systematics can make the residuals of each spectroscopic light curve correlated - which this process aims to remove. Because the white light curve residuals are subtracted from all spectroscopic light curves, this procedure assumes that these systematics are constant in wavelength. After these corrections were performed, each spectroscopic light curve was fit separately using a transit model combined with a 1D GP - accounting for remaining systematics that are correlated in time but assumed to be wavelength-independent. The amplitude of white noise was also fit separately for each wavelength.

    The exact cause of each of these systematics is unclear. In \citet{Gibson2017}, the authors speculate that the common-mode systematics are most likely due to instrumentation (e.g. inhomogeneities in components such as the grism or derotator) while the high-frequency systematics could be due to varying atmospheric throughput potentially from uneven cloud cover. Stellar activity can in principle also cause variations in flux during transit (i.e. from starspot crossings) or at any time (i.e. due to varying starspot coverage or asteroseismology). However, photometric monitoring described in \citet{Sing2015} suggests that WASP-31 is a quiet star with little photometric variability - so we consider it unlikely that stellar activity is a significant contributor to systematics in these data. Remaining wavelength-independent systematics are often found to be particularly strong near telluric features such as at the $\sim$7550$\angstrom - 7750\angstrom$ $\text{O}_{\text{2}}$ A band (see Fig. 7 in \citealt{Sedaghati2016} for an example), which suggests these systematics could be due to atmospheric variability or perhaps the sharp changes in flux increase instrumental systematics.
    
    \subsection{Choosing the kernel function}
    \label{sec:kernel choice}

    Examining the systematics accounted for in the previous analysis suggests that the kernel of our 2D GP should be able to account for (a) `common-mode' (CM) systematics, (b) high-frequency systematics (HFS), (c) any remaining time-correlated systematics in the spectroscopic light curves (which we call wavelength-specific systematics or WSS) and (d) the changing amplitude of white noise across different wavelength channels. It should also match the form of Eq.~(\ref{eq:kronecker kernel rakitsch}) to make this analysis computationally tractable. The kernel chosen for the 2D GP analysis is given in Eq.~(\ref{eq:VLT kernel fn}) and our choice of priors for all parameters is outlined in Appendix~\ref{app:priors}.

    \begin{subequations}\label{eq:VLT kernel fn}
    \begin{align}
        \mathbf{K}_{ij} &= \left[h_\mathrm{CM}^2 \exp\left(-\frac{|\lambda_i - \lambda_j|^2}{2 l_\mathrm{\lambda_\mathrm{CM}}^2}\right) + h_\mathrm{WSS; \lambda_i}^2 \delta_{\lambda_i \lambda_j}\right] \exp\left(-\frac{|t_i - t_j|^2}{2 l_{t}^2}\right) \nonumber \\
        &+ \left[h_\mathrm{HFS}^2 \exp\left(-\frac{|\lambda_i - \lambda_j|^2}{2 l_\mathrm{\lambda_\mathrm{HFS}}^2}\right) + \sigma_{\lambda_i}^2\delta_{\lambda_i \lambda_j}\right] \delta_{t_i t_j}.
        \tag{\ref{eq:VLT kernel fn}}
    \end{align}

    This equation can be multiplied out to reveal the four individual terms used to account for (a)-(d):
    
    \begin{align}
        &h_\mathrm{CM}^2 \exp\left(-\frac{|\lambda_i - \lambda_j|^2}{2 l_\mathrm{\lambda_\mathrm{CM}}^2}\right) \exp\left(-\frac{|t_i - t_j|^2}{2 l_{t}^2}\right), \label{eq:CM_term}\\
        &h_\mathrm{HFS}^2 \exp\left(-\frac{|\lambda_i - \lambda_j|^2}{2 l_\mathrm{\lambda_\mathrm{HFS}}^2}\right) \delta_{t_i t_j}, \label{eq:HFS_term} \\
        &h_\mathrm{WSS; \lambda_i}^2 \delta_{\lambda_i \lambda_j} \exp\left(-\frac{|t_i - t_j|^2}{2 l_{t}^2}\right), \label{eq:WSS_term} \\
        &\sigma_\mathrm{\lambda_i}^2\delta_{\lambda_i \lambda_j} \delta_{t_i t_j}. \label{eq:wn_term}
    \end{align}
    \end{subequations}

\begin{figure*}
    	\includegraphics[width=\textwidth]{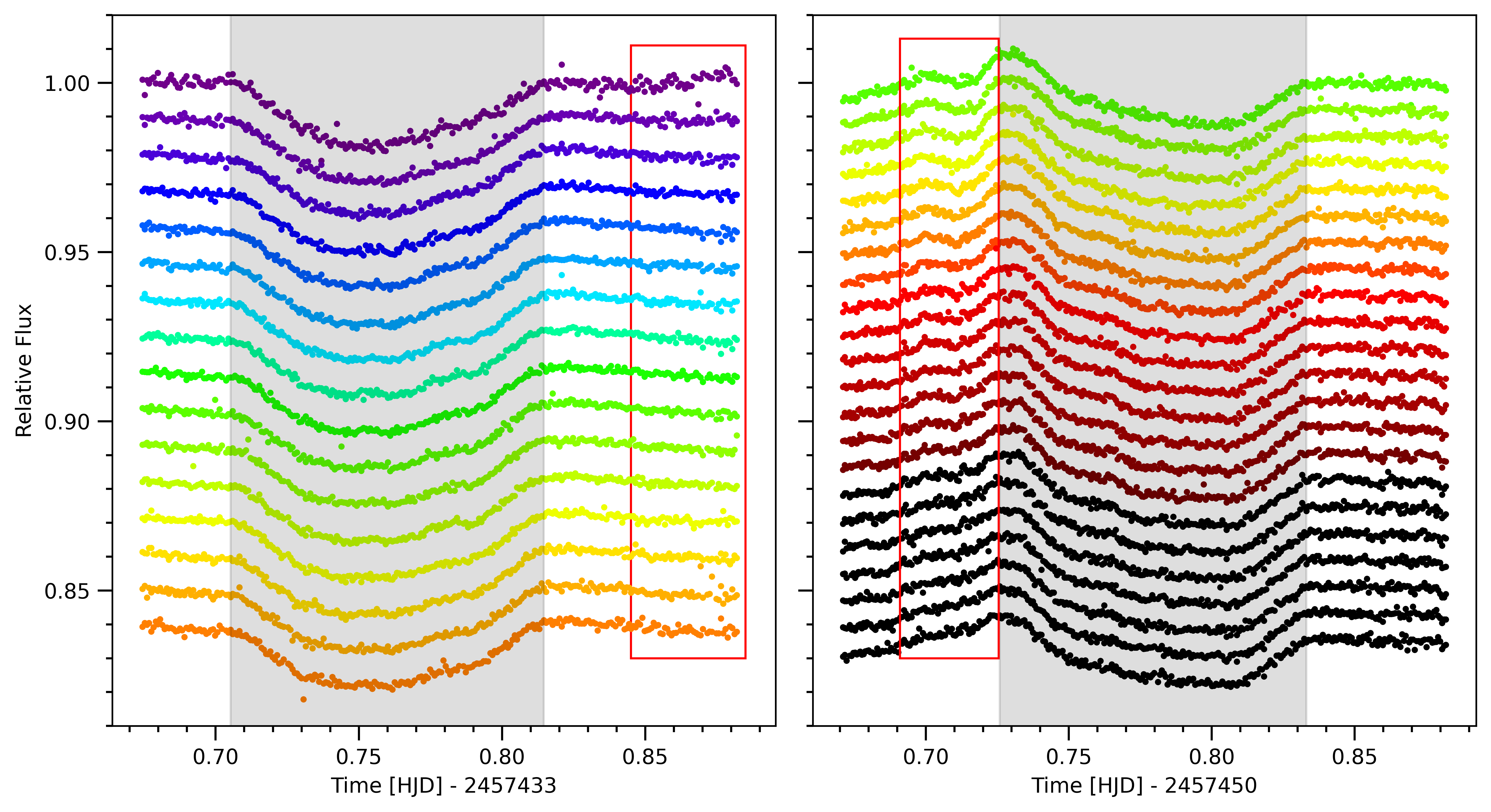}
        \caption{600B grism (left) and 600RI grism (right) spectroscopic light curves with best-fit transit region shaded in grey. Left: The light curves appear to have very similar systematics with the exception of the top one or two light curves in which the region in the red box shows a small increase in flux at the end of the observation. This could suggest the presence of wavelength-specific systematics (WSS) in these data. Right: The slowly varying shape of the systematics - as highlighted in the red box - suggest that the `common-mode' systematics are not actually common-mode but gradually vary in wavelength.}
        \label{fig:VLT noise example}
    \end{figure*}

    Equation~(\ref{eq:CM_term}) was included to account for (a), but instead of assuming {\it a priori} that the systematics are common-mode as in the previous analysis, we instead fit for a wavelength length scale $l_\mathrm{\lambda_\mathrm{CM}}$ with a broad prior range that can account for wavelength-varying or common-mode systematics. The right plot in Fig.~\ref{fig:VLT noise example} appears to show the presence of systematics that vary quite gradually in wavelength across the 600RI dataset, similar to the simulated light curves in Fig.~\ref{fig:sim III light curves} containing wavelength-varying systematics. Correlation in time was fit with the time length scale $l_{t}$ and we assumed the height scale of these systematics did not vary over the dataset and therefore fit for a single parameter $h_{\mathrm{CM}}$.

    Equation~(\ref{eq:HFS_term}) accounts for (b) high-frequency systematics. We fit for a wavelength length scale $l_\mathrm{\lambda_\mathrm{HFS}}$ and a single height scale $h_{HFS}$ for these systematics. The difference in our kernel choice between (a) and (b) is that we are assuming the high-frequency systematics are independent in time (also assumed in the original analysis). The Kronecker delta term $\delta_{t_i t_j}$ represents this independence in time in our kernel function.

    Equation~(\ref{eq:WSS_term}) accounts for remaining systematics that are correlated in time (with time length scale $l_{t}$) but wavelength-independent (represented by the Kronecker delta term $\delta_{\lambda_i \lambda_j}$). We tested replacing this Kronecker delta term with a squared-exponential kernel in wavelength to avoid assuming these systematics are wavelength-independent, but it had little effect on the results (similar to the analyses in Sect.~\ref{sec:indep_systematics}). Wavelength-independent systematics appear as light curves having systematics of a different shape to neighbouring light curves - such as the region highlighted in the left plot of Fig.~\ref{fig:VLT noise example}. The height scales of these systematics are fit independently for each wavelength (given by $h_\mathrm{WSS;\lambda_i}$) because some wavelength channels may cover telluric features that can have particularly large amplitude systematics. The time length scale $l_t$ is shared with Eq.~(\ref{eq:CM_term}) in order to fit the form of Eq.~(\ref{eq:kronecker kernel rakitsch}). This assumes that any common-mode or wavelength-varying systematics accounted for by Eq.~(\ref{eq:CM_term}) have the same timescale as any wavelength-independent systematics, which is not necessarily true but is an unavoidable assumption for this method. However as shown in Sect.~\ref{sec:shared_params}, sharing hyperparameters between light curves can provide more robust constraints when it is a safe assumption that the parameters are constant.
    
    The $\sigma_\mathrm{\lambda_i}$ terms in Eq.~(\ref{eq:wn_term}) fit for the amplitude of white noise. This is done separately for each wavelength channel as the white noise amplitude can be strongly wavelength-dependent for real observations. There is a Kronecker delta term $\delta_{\lambda_i \lambda_j}$ that accounts for the noise being independent in wavelength and a second Kronecker delta term $\delta_{t_i t_j}$ to account for the noise being independent in time.

\begin{figure*}
    	\includegraphics[width=\textwidth]{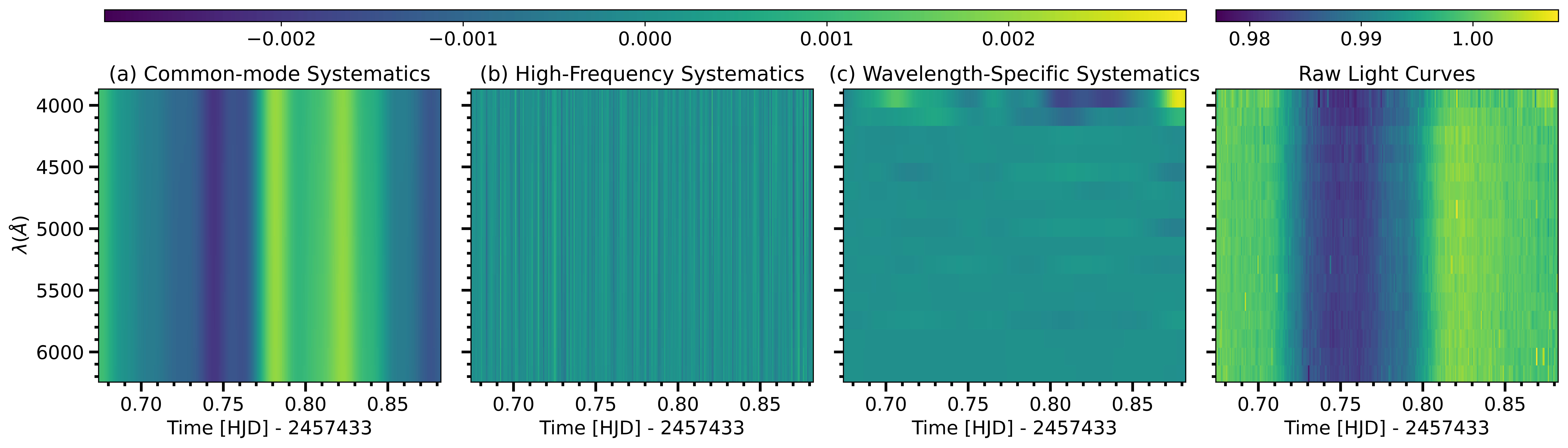}
        \caption{Plots showing GP predictive means fit by 2D GP in 600B dataset to account for (in order from left to right): (a) common-mode systematics, (b) high-frequency systematics and (c) wavelength-specific systematics. The rightmost plot is the raw light curves included for reference. Specifically, the middle plots show the change in the GP mean caused by including these systematics in the fit, as discussed in Sect.~\ref{sec:kernel choice}.}
        \label{fig:VLT gp mean}
    \end{figure*}

    By calculating the GP mean at the location of best-fit, but with different height scales $h_\mathrm{CM}$, $h_\mathrm{HFS}$ or $h_\mathrm{WSS}$ set to zero, it is possible to roughly visualise each type of systematics in these datasets. The common-mode systematics were visualised by setting both $h_\mathrm{HFS}$ and $h_\mathrm{WSS}$ to zero when calculating the GP mean, while the high-frequency and wavelength-specific systematics were visualised by examining the change in the GP mean when just $h_\mathrm{HFS}$ is set to zero or just $h_\mathrm{WSS}$ is set to zero respectively\footnote{The common-mode systematics have a much larger amplitude than the other systematics, requiring these different approaches.}. This is shown for the 600B data in Fig~\ref{fig:VLT gp mean}. The autocorrelation of the residuals given each of these terms are set to zero is examined in Appendix~\ref{app:autocorrelation}.
    
    Equation~(\ref{eq:VLT kernel fn}) can also be written in the Kronecker product form of Eq.~(\ref{eq:kronsum}) by choosing:
    \begin{align}
        (\mathbf{K}_\lambda)_{ij} &= h_\mathrm{CM}^2 \exp\left(-\frac{|\lambda_i - \lambda_j|^2}{2 l_\mathrm{\lambda_\mathrm{CM}}^2}\right) + h_\mathrm{WSS; \lambda_i}^2 \delta_{\lambda_i \lambda_j}, \\
        (\mathbf{K}_t)_{ij} &= \exp\left(-\frac{|t_i - t_j|^2}{2 l_{t}^2}\right),
        \label{eq:VLT covariance matrices}
    \end{align}
    and similarly for $\mathbf{\Sigma}_\lambda$ and $\mathbf{\Sigma}_t$ (in this case $\mathbf{\Sigma}_t$ is the identity matrix).

    \subsection{Re-analysis procedure}
    \label{sec:VLT procedure}
    
    We compare the original 1D GP analysis from \citet{Gibson2017} to multiple analyses with 2D GPs to examine how different assumptions affect the results. In addition, a re-analysis using 1D GPs was performed to enable a closer comparison of the differences between 1D and 2D GPs. This 1D GP re-analysis was similar to the original but more closely matched the 2D GP method implemented in this work, such as fitting each light curve for the transit depth $\rho^2$ instead of the radius ratio $\rho$, using the same priors on corresponding parameters as the 2D GP fit and also using a squared-exponential kernel instead of the Matérn 3/2 kernel chosen in the original analysis. Log-uniform priors on all of the hyperparameters were used which is also a slight change from the original analysis.
    
    The spectroscopic light curves were binned using the same wavelength bins as the original analysis, although there may potentially have been more outliers clipped at the edges of the datasets. Each light curve had separate parameters for transit depth $\rho^2$, limb-darkening parameters $c_1$ and $c_2$, out-of-transit flux $F_\mathrm{oot}$ and linear trend in baseline flux $T_\mathrm{grad}$. The light curves shared the central transit time parameter $T_0$, system scale $a/R_\mathrm{*}$, period $P$ and impact parameter $b$. The period was fixed to the mean literature value given in \citet{Patel2022} of $P = 3.4059095\pm0.0000047$ days.
    
    For the 1D GP re-analysis, a common-mode correction was performed (see Sect.~\ref{sec:common_mode}) with the white light curve for each dataset extracted across the same broad wavelength ranges as in \cite{Gibson2017}. It was fit for $T_0$, $a/R_\mathrm{*}$, $\rho^2$, $b$, $c_1$, $c_2$, $F_\mathrm{oot}$, $T_\mathrm{grad}$, $h$, $l_{t}$ and $\sigma$ using DE-MC. $T_0$, $a/R_\mathrm{*}$ and $b$ were fixed to their best-fit values when fitting the spectroscopic light curves. The same Gaussian priors were placed on $a/R_\mathrm{*}$, $b$ and on the radius ratio of the white light curve as in \citet{Gibson2017}. These priors were taken from previously reported values in \citet{Sing2015}. The priors were included because the large amplitude common-mode systematics made constraining parameters that affect the overall transmission spectrum difficult. To replicate this constraint for the 2D GP analyses - which do not perform a common-mode correction - these Gaussian priors were kept on $a/R_\mathrm{*}$ and $b$ and also on the mean radius ratio\footnote{specifically the square root of the mean transit depth as we were fitting for transit depths} when fitting the spectroscopic light curves.
    
    As the data alone was not sufficient to accurately constrain the limb-darkening parameters, Gaussian priors were placed on $c_1$ and $c_2$ based on predicted values from theoretical models using \textsc{PyLDTk}. The mean values given by \textsc{PyLDTk} were set as the prior means but the prior uncertainties were increased to a standard deviation of 0.1 to account for uncertainty in the limb-darkening models (matching the procedure in \citealt{Gibson2017}).
    
    Outliers in the spectroscopic light curves were clipped by performing a best-fit to the data with a GP - optimising both the transit parameters and hyperparameters - and clipping 4$\sigma$ outliers (evaluated using the GP predictive mean and variance). The outliers were replaced with the GP predictive mean evaluated at the location of the points. This was performed using the 2D GP predictive mean for the 2D analysis (see Appendix~\ref{app:gp_mean}) and the 1D GP predictive means for each light curve for the 1D GP re-analysis. Replacing these outliers as opposed to removing them ensures the flux observations still lie on a complete 2D grid. This is required for our optimised 2D GP method and also matches the procedure in the original analysis.
    
    The 1D GP re-analysis was similar to the procedure described in Sect.~\ref{sec:MCMC} but with each (corrected) spectroscopic light curve being fit for $\rho^2$, $c_1$, $c_2$, $F_\mathrm{oot}$, $T_\mathrm{grad}$, $h$, $l_{t}$ and $\sigma$. As more parameters were being fit, the chains were run for longer with 100 independent chains having 500 burn-in steps and another 500 steps run for the chain. If necessary, chains were extended until convergence occurred (measured using the Gelmin-Rubin statistic as in Sect.~\ref{sec:MCMC}). The 2D GP analyses used blocked Gibbs and slice sampling in combination with NUTS sampling for inference (see Appendix~\ref{app:VLT MCMC} for more details). Sampling was performed with a burn-in length of 1000 and a chain length of 1000 with four independent chains. Most of the 2D GP analyses were fitting seven parameters for each light curve ($\rho^2$, $c_1$, $c_2$, $F_\mathrm{oot}$, $T_\mathrm{grad}$, $h_\mathrm{WSS; \lambda_i}$, $\sigma$) with five wavelength-independent parameters ($h_\mathrm{CM}$, $l_{t}$, $l_\mathrm{\lambda_\mathrm{CM}}$, $h_\mathrm{HFS}$, $l_\mathrm{\lambda_\mathrm{HFS}}$). Exceptions to this are explained in Sect.~\ref{sec:VLT results}. For the 600B data with 16 light curves, this results in 117 parameters. The 600RI data had 22 light curves and therefore had 159 parameters to fit. The use of NUTS permitted convergence to occur for all parameters within this relatively small number of sampling steps.
    
    Basic atmospheric retrievals similar to those described in Sect.~\ref{sec:metrics} were performed. We fit a radius and slope to each transmission spectrum in addition to measuring the change in the radius ratio at the bin centred on the Na feature $\Delta \rho_\mathrm{Na}$ for both the 600B and 600RI data as well as the change in radius ratio at the K feature $\Delta \rho_\mathrm{K}$ for the 600RI data. The 600RI data have a $150\angstrom$ bin centred on $7681\angstrom$ which is wide enough to cover the full K I doublet. The 600B and 600RI datasets have $150\angstrom$ bins centred at $5893\angstrom$ and $5881\angstrom$ respectively which both cover the Na I doublet. This same atmospheric retrieval was also performed on the transmission spectra taken from the original analysis.
    
    \subsection{Re-analysis results}
    \label{sec:VLT results}

    \begin{figure*}
    	\includegraphics[width=\textwidth]{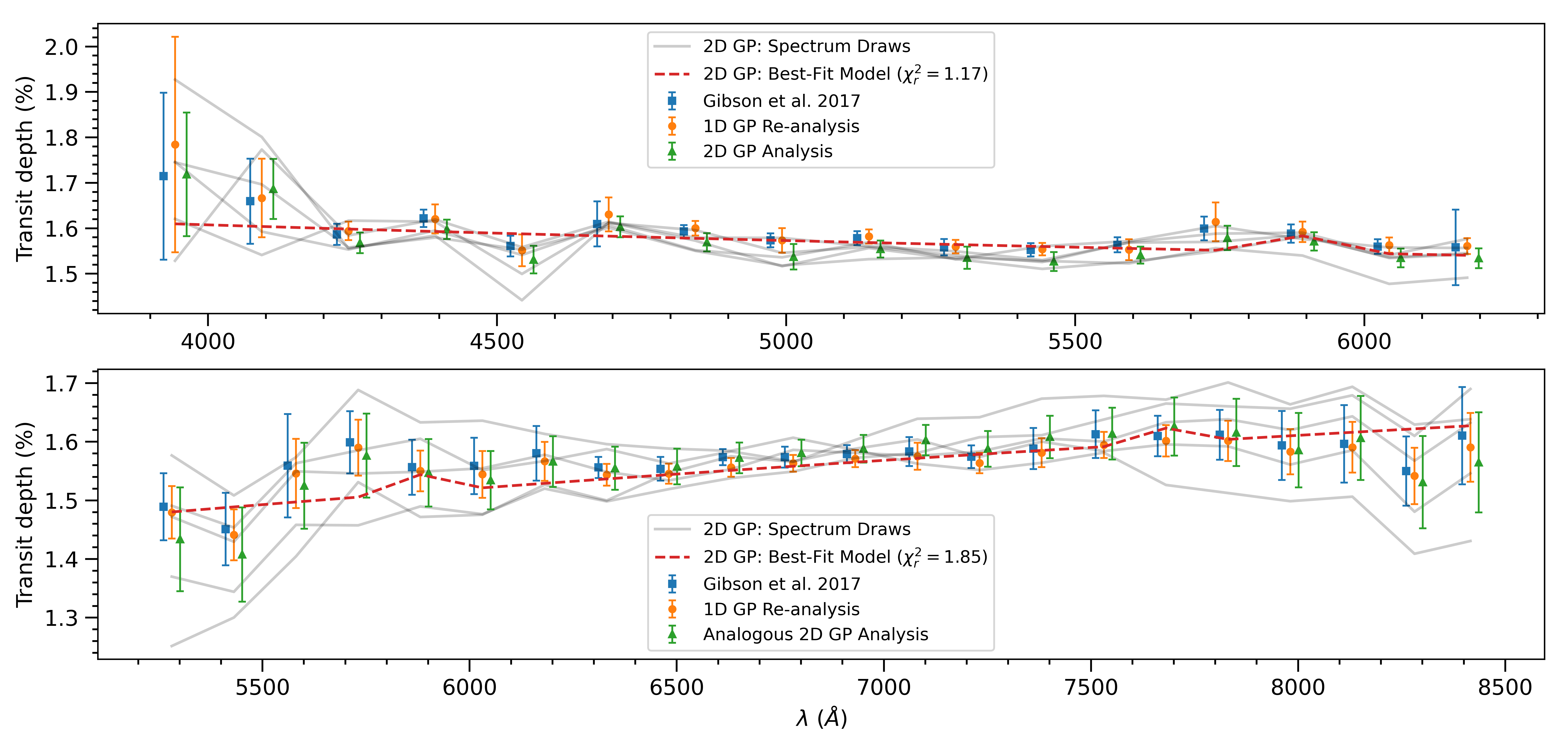}
        \caption{600B (top) and 600RI (bottom) retrieved transmission spectra comparing original analysis, 1D GP re-analysis and analogous 2D GP analysis. The 2D GP best-fit atmospheric retrieval is shown as a red dotted line and random draws taken from the 2D GP spectrum are shown in grey. Top: All three analyses show largely consistent results. Bottom: The error bars of each method may appear consistent but the random draws from the 2D GP spectrum demonstrates a large uncertainty in the slope of this spectrum after accounting for the covariance between transit depths.}
        \label{fig:Trans Spec}
    \end{figure*}
    
    \begin{table*}
    	\centering
    	\caption{Results of atmospheric retrievals on the 600B data analysed using different methods.}
    	\label{tab:VLT_blue}
    	\begin{tabular}{lcccc}
    		\hline
    		Method & Mean radius ($\bar{\rho}$) & Slope ($\alpha$) & $\Delta \rho_\mathrm{Na}$ & $\chi_\mathrm{r}^2$\\
    		\hline
            Gibson et al. 2017 Spectrum & 0.12565 $\pm$ 0.00021 & -3.38 $\pm$ 1.34 & 0.00125 $\pm$ 0.00086 & 0.73\\
    1D GP Re-analysis & 0.12572 $\pm$ 0.00024 & -3.31 $\pm$ 1.29 & 0.00120 $\pm$ 0.00106 & 0.70\\
    Analogous 2D GP Analysis & 0.12539 $\pm$ 0.00026 & -3.92 $\pm$ 1.46 & 0.00142 $\pm$ 0.00069 & 1.17\\
    2D GP: Lower maximum prior on $l_{\lambda}$ & 0.12537 $\pm$ 0.00026 & -3.89 $\pm$ 1.62 & 0.00138 $\pm$ 0.00068 & 1.17\\
    2D GP: Without Mean Radius prior & 0.12348 $\pm$ 0.00362 & -3.92 $\pm$ 1.41 & 0.00154 $\pm$ 0.00069 & 1.20\\
    2D GP: Varying $T_0$, $a/R_\mathrm{*}$, $b$ & 0.12538 $\pm$ 0.00026 & -4.05 $\pm$ 1.53 & 0.00138 $\pm$ 0.00070 & 1.20\\
    2D GP: Direct Atmospheric Retrieval & 0.12536 $\pm$ 0.00026 & -4.17 $\pm$ 1.45 & 0.00144 $\pm$ 0.00069 & N/A\\
    		\hline
    	\end{tabular}
    \end{table*}
    
    \begin{table*}
    	\centering
    	\caption{Results of atmospheric retrievals on the 600RI data analysed by different methods.}
    	\label{tab:VLT_red}
    	\begin{tabular}{lccccc}
    		\hline
    		Method & Mean radius ($\bar{\rho}$) & Slope ($\alpha$) & $\Delta \rho_\mathrm{Na}$ & $\Delta \rho_\mathrm{K}$ & $\chi_\mathrm{r}^2$\\
    		\hline
            Gibson et al. 2017 Spectrum & 0.12539 $\pm$ 0.00025 & 4.68 $\pm$ 2.04 & 0.00039 $\pm$ 0.00195 & 0.00059 $\pm$ 0.00143 & 0.31\\
    1D Re-analysis & 0.12493 $\pm$ 0.00021 & 5.27 $\pm$ 1.64 & 0.00070 $\pm$ 0.00147 & 0.00061 $\pm$ 0.00110 & 0.47\\
    Analogous 2D GP Analysis & 0.12484 $\pm$ 0.00036 & 8.09 $\pm$ 8.02 & 0.00123 $\pm$ 0.00074 & 0.00100 $\pm$ 0.00062 & 1.85\\
    2D GP: Without Mean Radius Prior & 0.08687 $\pm$ 0.00896 & 11.2 $\pm$ 12.1 & 0.00156 $\pm$ 0.00090 & 0.00139 $\pm$ 0.00073 & 1.58\\
    2D GP: Varying $T_0$, $a/R_\mathrm{*}$, $b$ & 0.12488 $\pm$ 0.00034 & 8.42 $\pm$ 8.44 & 0.00097 $\pm$ 0.00072 & 0.00111 $\pm$ 0.00061 & 1.68\\
    2D GP: Direct Atmospheric Retrieval & 0.12485 $\pm$ 0.00036 & -1.87 $\pm$ 10.9 & 0.00118 $\pm$ 0.00088 & 0.00094 $\pm$ 0.00074 & N/A\\
    		\hline
    	\end{tabular}
    \end{table*}

    A comparison of the transmission spectra from the 1D GP analyses to our initial 2D GP analysis is plotted in Fig.~\ref{fig:Trans Spec}. The results of all atmospheric retrievals performed are included in Tables~\ref{tab:VLT_blue} and \ref{tab:VLT_red}. The scattering slope reported in these tables was given in terms of the $\alpha$ parameter from Eq.~(\ref{eq:scattering}) to make it easier to compare to Rayleigh scattering ($\alpha = -4$). The $\alpha$ values were calculated by taking the literature values of $H = 1220$ km for the scale height and $R_* = 1.12$\(M_\odot\) for the stellar radius as reported in \citet{Sing2015} and \citet{Anderson2011} respectively. The $\chi_r^2$ of the best-fitting atmospheric model for each recovered transmission spectrum (accounting for covariance for the 2D GP analyses) is also included. Unlike in Sect.~\ref{sec:3} where the true transmission spectra were known, these $\chi_r^2$ values can only be used to help quantify how consistent each analysis is with the specific choice of atmospheric model.

    Appendix~\ref{app:VLT_visualise} includes visualisations of the correlation matrix for both the 600B and 600RI transmission spectra, 2D GP best-fits to the spectroscopic light curves and corner plots for two of the 2D GP analyses.
    
    \subsubsection{1D GP re-analysis and the analogous 2D GP analysis}
    \label{sec:analogous_2D_GP}
    
    Our initial 2D GP analysis attempted to closely resemble the 1D GP analysis and so we refer to this as the `analogous 2D GP analysis'. However, the 2D GP has a different kernel with wavelength length scale parameters $l_\mathrm{\lambda_\mathrm{CM}}$ and $l_\mathrm{\lambda_\mathrm{HFS}}$ that can account for how the `common-mode' and high-frequency systematics may actually vary in wavelength. It shares the same time length scale parameter $l_{t}$ across all light curves that could help provide more robust constraints (see Sect.~\ref{sec:shared_params}). It also retrieves the full covariance matrix of the transmission spectrum instead of assuming all transit depths are independent. Figure~\ref{fig:Trans Spec} compares the resulting transmission spectra from the original \citet{Gibson2017} analysis, the 1D GP re-analysis and the analogous 2D GP re-analysis. Random draws taken from the 2D GP covariance matrix are included to visualise the correlated uncertainties in the 2D GP transmission spectrum (as the error bars do not convey this information).

    \begin{figure*}
    	\includegraphics[width=\textwidth]{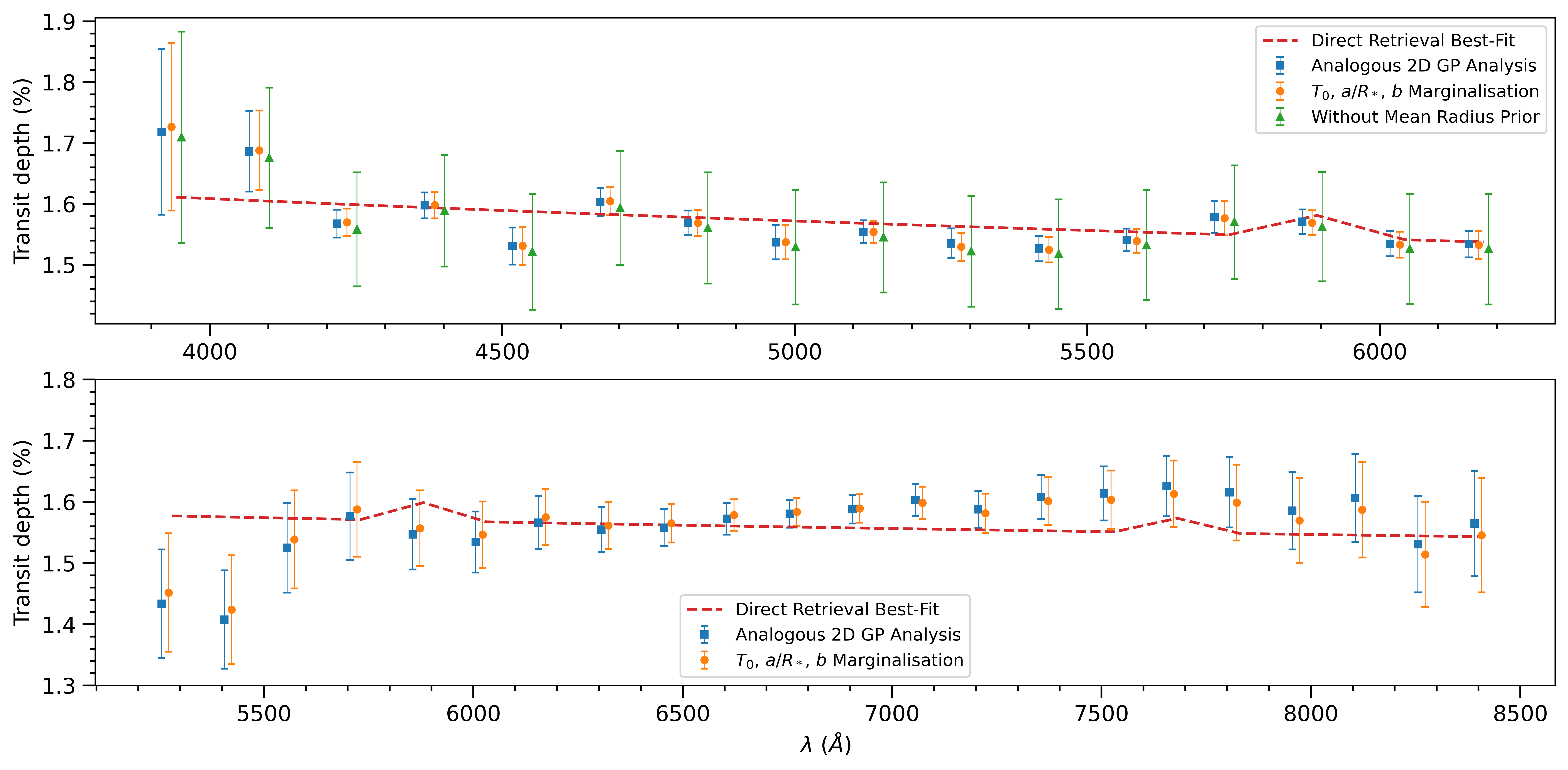}
        \caption{Similar to Fig.~\ref{fig:Trans Spec} but showing different 2D GP analyses tested. As error bars do not convey the correlations between the data points, it is difficult to see that the 600B fit with no mean radius prior - which appears to have much larger uncertainties - actually has a similar constraint on the slope and Na feature compared to the other methods. For the direct retrieval of atmospheric parameters from the data, a dotted line is included that shows the best-fit atmospheric model (as the transmission spectrum is not directly retrieved with this method).}
        \label{fig:Trans Spec 2D GPs}
    \end{figure*}

    For the 600B dataset, it was found the data were consistent with containing common-mode systematics as the posterior of the wavelength length scale was concentrated towards the maximum prior limit of 50 times the wavelength range of the data. This does not rule out that the systematics could be varying very gradually in wavelength, specifically the wavelength length scale was constrained to be $l_\mathrm{\lambda_\mathrm{CM}}>6772\angstrom$ ($\approx$3 times the wavelength range of the dataset) to 99\% confidence. This constraint implies that systematics that vary over length scales shorter than 3 times the wavelength range of the dataset must be distinguishable from common-mode systematics and so if we constrain other wavelength length scales to be below this value we can safely assume those systematics are varying in wavelength. For example, the high-frequency systematics in this dataset were strongly constrained to be varying in wavelength with $l_\mathrm{\lambda_\mathrm{HFS}}$ = 1030{\raisebox{0.5ex}{\tiny $\substack{+120 \\ -100}$}}$\angstrom$. The 600RI data were significantly affected by wavelength-varying systematics as $l_\mathrm{\lambda_\mathrm{CM}}$ = 2980{\raisebox{0.5ex}{\tiny $\substack{+370 \\ -330}$}}$\angstrom$ for data with a wavelength range of $3135\angstrom$. The high-frequency systematics in these data were also found to vary in wavelength with $l_\mathrm{\lambda_\mathrm{HFS}}$ = 2170{\raisebox{0.5ex}{\tiny $\substack{+350 \\ -250}$}}$\angstrom$.

    Since the wavelength length scale $l_\mathrm{\lambda_\mathrm{CM}}$ for the 600B dataset was concentrated towards an arbitrary choice of maximum prior limit, this analysis was repeated with the prior limit reduced to 15 times the wavelength range of the data. There was little difference in the results other than a slight increase ($\sim$11\%) in the uncertainty of the slope of the transmission spectrum. This increase in uncertainty makes sense as the shorter prior limit forces the wavelength length scale to take smaller values, which are likely to increase uncertainty in the slope.

    Note there is a risk that the wavelength-correlated systematics identified are actually from wavelength-varying inaccuracies in our transit model (e.g. due to unaccounted for limb-asymmetries; see \citealt{Powell2019}). We performed another test where we masked out the transit region of each dataset and performed an identical analysis on the masked datasets except without fitting for a transit model. The linear baseline parameters $F_\mathrm{oot}$ and $T_\mathrm{grad}$ were still fit for to account for changes in baseline flux. The resulting constraints on the hyperparameters that fit for wavelength correlation ($h_\mathrm{CM}$, $l_\mathrm{\lambda_\mathrm{CM}}$, $h_\mathrm{HFS}$ and $l_\mathrm{\lambda_\mathrm{HFS}}$) were all consistent with the values from the previous analyses to $2.2\sigma$. The systematics were similarly constrained to be varying in wavelength except for the 600B wavelength length scale $l_\mathrm{\lambda_\mathrm{CM}}$ which was again consistent with common-mode systematics.
    
    Our results strongly contradict the assumptions of the original analysis that only common-mode systematics and wavelength-independent systematics are present in the 600B and 600RI datasets. Based on the results of Sects.~\ref{sec:corr_systematics} and \ref{sec:common_mode}, these incorrect assumptions could lead uncertainties on broad features such as the scattering slope to be underestimated and could cause uncertainties on sharp features such as the strength of Na and K features to be overestimated. Tables~\ref{tab:VLT_blue} and \ref{tab:VLT_red} appear consistent with this interpretation as the 2D GP analyses all show increased uncertainty in the scattering slope and decreased uncertainty on the constraints of Na and K. The 600RI data were particularly affected, for example the uncertainty in the scattering slope was $\sim$3.9 times larger compared to the original analysis. This makes sense as both the `common-mode' systematics and high-frequency systematics were found to vary in wavelength for this dataset. The constraint of $\alpha = 8.09 \pm 8.02$ from the 2D GP analysis is weak enough to be consistent with a positive slope ($\alpha > 0$), a flat spectrum ($\alpha = 0$), Rayleigh-scattering ($\alpha = -4$) or super-Rayleigh scattering ($\alpha < -4$). This is in contrast to the 1D GP analyses that are both consistent with either a positive slope or a flat spectrum. The tighter constraints on Na and K in the 2D GP analyses ($\sim$50\% smaller) still did not lead to a detection of either species in any analysis ($<2.3\sigma$ for all analyses). We note that the shared time length scale $l_t$ in the 2D GP analysis could have also tightened the constraints on Na and K - similar to results in \citet{Ahrer2022a}.
    
    We found that the $\chi_\mathrm{r}^2$ of the best-fitting atmospheric models were increased significantly for the 2D GP analyses, particularly on the 600RI data. We should expect that $\chi_\mathrm{r}^2 \in $ [0.46, 1.75] to 95\% confidence for the 600RI data. The 1D GP analyses are on the lower end of this confidence interval while the analogous 2D GP analysis is on the higher end\footnote{If this is surprising note that the correlations in the uncertainties restricts the space of models consistent with the recovered spectra.}. Since the 1D GP analyses do not account for wavelength-correlated systematics, they could certainly be overestimating uncertainties. However, it is not clear if the high $\chi_r^2$ values for the 2D GP analyses indicate that the 2D GP analyses are underestimating uncertainties (potentially due to a sub-optimal choice of kernel) or if the chosen atmospheric model was not flexible enough. The $\chi_\mathrm{r}^2$ value is still close to the 95\% confidence interval however so this may also be due to random chance.
    
    The runtime of the 2D GP analysis was longer than the 1D GP analysis by about one order of magnitude. The combined runtime of the MCMCs for the 1D GP re-analysis took 28 minutes for the 600B data and 54 minutes for the 600RI data. The MCMCs for each of the analogous 2D GP analyses on the 600B and 600RI datasets took 6.5 - 7 hours each\footnote{running all analyses on a 2020 M1 MacBook Pro using the CPU}.
    
    \subsubsection{Fitting without a mean radius prior}
    \label{sec:mean prior}
    
    The original analysis placed a Gaussian prior on the radius ratio of the white light curve which the 2D GP analyses match by placing the same prior on the square root of the mean transit depth (which should approximately match the radius ratio of the white light curve). This prior is in addition to the uniform priors already used when fitting for each transit depth. Fitting the transmission spectrum without this prior on the mean radius ratio combined with the lack of a common-mode correction permits the retrieval of the overall offset of the transmission spectrum.
    
    We repeated the analogous 2D GP analyses without this prior on the mean radius ratio. By examining Table~\ref{tab:VLT_blue}, we can see for the 600B data that the uncertainty of the radius ratio increases significantly without this prior and is $\sim$13.9 times larger than the analogous 2D GP analysis. Other than this however, the constraints on the other parameters are almost identical with no significant difference in the scattering-slope or the retrieval of the Na feature.
    
    The corresponding analysis of the 600RI data resulted in a much weaker constraint on the offset of the transmission spectrum. The mean radius ratio was retrieved to be $\bar{\rho} = 0.08687 \pm 0.00896$, which is highly inconsistent with the value of $\bar{\rho} = 0.12348 \pm 0.00362$ from the 600B data (that shares some overlapping wavelength bins). Fitting the white light curve with a 1D GP without including the Gaussian radius prior also results in a low radius ratio of $\rho_\mathrm{white} = 0.10277 \pm 0.00891$. As these results are inconsistent with previous observations, the simplest explanation may be that the systematics present in this dataset happened to be particularly similar to a transit signal, as suggested in Sect.~\ref{sec:indep_systematics}. The other atmospheric parameters were not significantly changed from the analogous 2D GP analysis - as can be seen in Table~\ref{tab:VLT_red} - so the choice of including this prior mainly affects the constraint on the radius ratio.

    \subsubsection{Marginalising over uncertainty in $T_0$, $a/R_\mathrm{*}$ and $b$}
    \label{sec:Tab marginalisation}
    
    Joint-fitting the spectroscopic light curves allows us to vary the central transit time $T_0$, system scale $a/R_\mathrm{*}$ and impact parameter $b$ as parameters within the MCMC. We recommend this approach as it accounts for how the uncertainty in these parameters affects the transmission spectrum. We can examine if this has a significant effect by comparing our results to the analogous 2D GP analysis that matched the 1D GP procedure of fixing these parameters.
    
    The central transit time was allowed to freely vary with a uniform prior in the 600B data but the 600RI data struggled to tightly constrain $T_0$ (as was also seen in the 600RI white light curve analysis). It was decided to constrain $T_0$ using the retrieved value from the 600B data as a Gaussian prior. This assumes that WASP-31b does not have significant transit timing variations (TTVs) and that the literature value used for the period is accurate. While there is some evidence to question these assumptions (i.e. \citealt{Patel2022, Bonomo2017, Exoclock2022}), we ignore these here to demonstrate the method on both datasets. The Gaussian priors on $a/R_\mathrm{*}$ and $b$ from \citealt{Sing2015} that were used in the white light curve fitting in the original analysis were still kept on these parameters for the 600B analysis. The retrieved constraints on $a/R_\mathrm{*}$ and $b$ from the 600B data were then used as Gaussian priors on the 600RI data. A more rigorous but computationally expensive approach could have been to perform a joint fit on both light curves with common $a/R_\mathrm{*}$ and $b$ parameters and $T_0$ values offset by 5 periods (the observations are 5 orbits apart).
    
    The effect of marginalising over $T_0$, $a/R_\mathrm{*}$ and $b$ appears to have had little effect on the retrieval of the 600B data. The mean radius $\bar{\rho}$ would likely be the parameter most affected by varying these parameters (as these parameters are shared by all light curves) but the mean radius prior could be restricting this effect. The retrieved central transit time was $T_0$ (HJD - 2457433) $= 0.753597 \pm 0.000578$ which is very close to the value retrieved from the white light curve of $T_0$ (HJD - 2457433) $= 0.753609 \pm 0.000662$. The marginalisation over these parameters also had little effect on the 600RI data as the atmospheric retrieval produced similar results to the analogous 2D GP analysis.
    
    \subsubsection{Directly fitting atmospheric forward models to the data}
    \label{sec:atmo_fit}
    
    Typically a transmission spectrum is first obtained by individually fitting each spectroscopic light curve and then an atmospheric retrieval is performed on the resulting transmission spectrum. With all spectroscopic light curves being fit simultaneously, we can perform an atmospheric retrieval while fitting the light curves. The transit depths can be generated from atmospheric forward models with the atmospheric parameters directly retrieved in the MCMC fitting the light curves, similar to how high-resolution transmission spectroscopy atmospheric retrievals are performed (e.g. \citealt{Maguire2023}).
    
    This would reduce the number of parameters to fit, which may be useful if scaling this method to fit more light curves simultaneously. This also avoids approximating the posterior of the transmission spectrum. Typically the transmission spectrum is extracted by taking the mean and standard deviation (or the covariance matrix in our method) of each transit depth from the MCMC chains, which approximates the posterior as a Gaussian distribution. While different distributions could potentially model the posterior better, there would still likely be some modelling error which could be avoided by directly fitting atmospheric models to the light curves.
    
    The results from using this method closely match the analogous 2D GP analysis for the 600B and 600RI data, with the exception of the retrieved slope of the 600RI data which has a significantly reduced mean and larger uncertainty. Explaining why this discrepancy occurred is difficult as we do not recover individual transit depths with this method and no $\chi_\mathrm{r}^2$ can be retrieved to determine the goodness-of-fit. The increased complication in interpreting the results is a major downside of this approach. 
    
    With this method, the systematics may be fit differently depending on the choice of atmospheric model. This is quite concerning as imperfections in our models could result in the GP fitting larger systematics to the data. While it is beneficial to avoid approximating the posterior of the transmission spectrum, this method appears to complicate the choice of atmospheric model and makes it harder to visualise if the data are consistent with a model. This method is therefore not recommended for testing out different atmospheric retrievals but could be an interesting test if a specific atmospheric model has already been chosen.

    \subsection{Benchmarking}
    \label{sec:benchmarking}
    
    \begin{figure*}
    	\includegraphics[width=\textwidth]{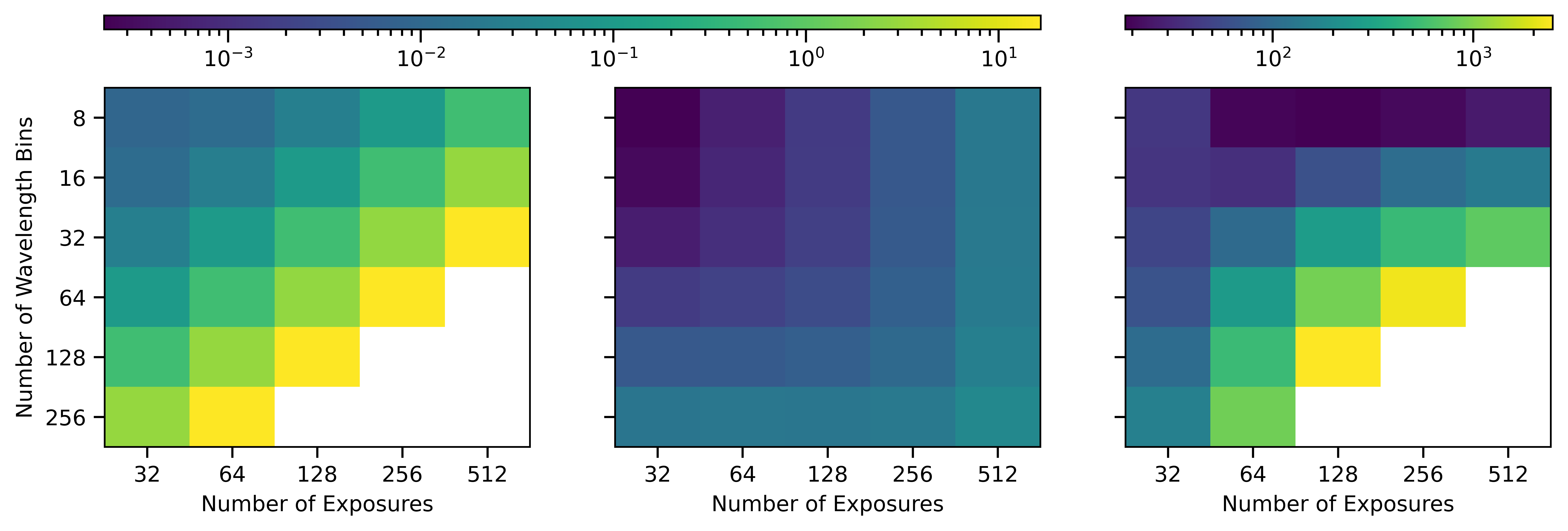}
        \caption{Comparison of log-likelihood calculation runtime (in seconds on logarithmic scale) between general approach of Cholesky factorisation (left) compared to our method (centre) as well as relative performance improvement (right). Cholesky factorisation was not calculated when $MN \ge 32768$ (shown in white) due to memory limitations.}
        \label{fig:benchmarking}
    \end{figure*}

    We tested how the runtime for this method would be affected as a function of the number of time exposures and wavelength bins. The kernel function used to fit the VLT datasets was used (Eq.~\ref{eq:VLT kernel fn}) with the kernel hyperparameters set as the best-fit values of the 600RI analogous 2D GP analysis. As some hyperparameters ($h_\mathrm{WSS; \lambda_i}$ and $\sigma_{\lambda_i}$) were fit separately for each wavelength, changing the number of wavelength bins required us to interpolate these best-fit values. One of the two time covariance matrices produced by this kernel function is the identity matrix. This makes the eigendecomposition of this matrix trivial which was exploited in these calculations. For a large number of time points relative to wavelength bins, the eigendecomposition of these two time covariance matrices is typically the bottleneck in the log-likelihood calculation and so avoiding one eigendecomposition can almost half the runtime.
    
    For each number of wavelengths and time points tested, a random draw of noise described by this kernel function was first taken. Runtime of the mean function was not included so the log-likelihood calculation was performed using this random draw of noise as the residuals. The mean function used in this paper would have had little effect as even for $N = 512$ and $M = 256$ the runtime for the mean function was only $\sim$3.5ms.
    
    The gradient calculation of the log-likelihood was not performed for this test. However, the runtime would likely not be significantly affected as eigendecomposition is generally the computational bottleneck regardless of whether the gradients are calculated and the same eigendecomposition can be stored to calculate both the log-likelihood and gradients of the log-likelihood (see Appendix~\ref{app:gradients}).
    
    Log-likelihood calculations were performed using both our method as well as using Cholesky factorisation on the full $MN \times MN$ covariance matrix. The benefit of the Cholesky method is that it would place no restrictions on the kernel function or on the data needing to lie on a complete grid. However, its poor runtime scaling of $\mathcal{O}(M^3 N^3)$ would make it unfeasible for most real datasets.

    Figure~\ref{fig:benchmarking} shows the results of this runtime comparison performed on a quad-core Intel Core i5-6500 CPU at 3.20GHz. Calculations where $MN \ge 32768$ were avoided for the Cholesky factorisation as they required $\ge8$GB of memory to store the $MN \times MN$ covariance matrix. This was not an issue for our new method as it only needs to store separate time and wavelength covariance matrices.
    
    The Kronecker product sum method has a runtime scaling of $\mathcal{O}(M^3 + N^3)$ due to the expensive eigendecomposition step which has the effect that when one dimension is much longer than the other it tends to act as the computational bottleneck. The 600B and 600RI datasets analysed had dimensions of ($M = 16$, $N = 262$) and ($M = 22$, $N = 319$) respectively, so the number of time points far exceeded the number of wavelength channels chosen. As can be seen in the central plot of Fig.~\ref{fig:benchmarking}, increasing the number of wavelength channels would have little effect on the log-likelihood calculation runtime due to the eigendecomposition of the time covariance matrix dominating the runtime when $N \approx 256$. There were however seven wavelength-dependent parameters that each light curve was individually fit for, so convergence time of the MCMC would increase as the number of parameters would increase. Convergence time with No U-Turn Sampling has been shown to scale as $\mathcal{O}(d^{\frac{5}{4}})$ for $d$ parameters under certain assumptions \citep{Neal2011}.
    
    As each MCMC chain is run independently with NUTS, parallelisation is trivial to implement. The same \textsc{JAX} code can also be run on a GPU, which could significantly improve performance. Utilising parallelisation and GPU acceleration could permit this method to be scaled to larger datasets from JWST.


\section{Discussion}
\label{sec:5}

    
    The re-analysis of the VLT/FORS2 data confirms that systematics that vary in wavelength can be present in real observations and can be mistaken for common-mode systematics. As demonstrated in Sect.~\ref{sec:common_mode}, this can significantly impact atmospheric retrievals and so we recommend a principled method to account for them using 2D GPs. Both the 600B and 600RI datasets contained high-frequency systematics that were found to gradually vary in wavelength. The re-analysis of the 600RI data also revealed that time-correlated systematics previously assumed to be common-mode were strongly constrained to vary in wavelength. Accounting for this led to a weaker constraint on the scattering slope but produced tighter constraints on the Na and K features, consistent with the results of Sects.~\ref{sec:corr_systematics} and \ref{sec:common_mode}.

    Multiple 2D GP analyses were performed either with a different choice of priors or different parameters being fit. For the 600B data, each of these analyses were consistent with Rayleigh scattering and no strong evidence for Na was found (<$2.3\sigma$ for all analyses). The results of the 600RI analyses were less consistent and show that different assumptions made when fitting these data can significantly affect the retrieved radius and slope of the transmission spectrum, although evidence for sodium or potassium always remained below $2\sigma$. This means that there is still a conflict on the presence of potassium between the VLT/FORS2 data and the original analysis of the HST observations from \citet{Sing2015}. We leave the re-analysis of the HST data using our new method for future work.

    We have demonstrated that our method can avoid the need for a common-mode correction, which avoids artificially reducing uncertainty in the offset of the transmission spectrum. This is even more important when considering eclipse spectra, where the overall offset is critical to the interpretation of the spectrum (e.g. \citealt{Evans2013, Bell2017}). Joint-fitting spectroscopic light curves also permitted us to marginalise over uncertainty in the central transit time, system-scale and impact parameter - helping to improve the rigour of transmission spectroscopy. Approximating the posterior of the transmission spectrum for atmospheric retrievals can also be avoided if atmospheric forward models are fit directly to the data.
    
    One disadvantage of assuming noise follows a Gaussian process is that it makes the analysis very sensitive to outliers. Student-t processes have been applied to transmission spectroscopy as they less sensitive to outliers \citep{Wilson2021}. Optimisations presented in this work could also permit the application of 2D Student-t processes to transmission spectroscopy as both methods have similar likelihood functions.

    Our method could be used to account for wavelength-correlated systematics in JWST observations. For example, NIRCam observations are contaminated by 1/f correlated read-out noise that is read parallel to the spectral trace (in the dispersion direction) on the detector, introducing wavelength-correlated systematics that are visible in the residuals of the light curves \citep{Ahrer2022}. In addition, wavelength-correlated systematics have been identified in NIRISS/SOSS observations \citep{Holmberg2023}. Scaling our method to larger datasets typical of JWST observations could be enabled by utilising parallelisation and using GPUs.
    
    In addition, there has been recent work attempting to fit low-resolution ground-based transmission spectroscopy datasets without the use of a comparison star \citep{Panwar2022, Spyratos2023}. Combining 2D GPs with this approach could be beneficial due to the wavelength-dependent effects of atmospheric extinction no longer being accounted for with a comparison star.
    
    GPs have a broad variety of uses within astronomy \citep{Aigrain2022} and astronomical data often include noise that is correlated in two-dimensions (i.e. x and y pixel positions on a detector), so there could be many other useful applications for this method. Our optimisation should have significant performance advantages over other 2D GP optimisations (e.g. \citealt{Gordon2020, sleaf}) when there are non-trivial correlations across both input dimensions and neither dimension is very large. This could benefit fields that already use these other GP optimisations (e.g. exoplanet detection). GPs have been proposed to be used in high-resolution time series spectroscopy (e.g. \citealt{Meech2022}), our method may enable much faster application to large datasets. Interpolation of 2D datasets can also be performed efficiently with this method (see Appendix~\ref{app:gp_mean}) which could have very broad applications. For example, 2D GPs have already been used to interpolate supernova light curves (e.g. \citealt{Fakhouri2015, Boone2019}) which this method might help optimise. 


\section{Conclusions}
\label{sec:6}


    We have developed a new method that can reliably recover transmission spectra in the presence of both time and wavelength correlated systematics, have tested it on a range of simulated data to show its advantages over current approaches and demonstrated it on real data of WASP-31b observed using VLT/FORS2, despite the presence of large amplitude systematics. To summarise our conclusions:

    \begin{itemize}
    \item We developed a 2D GP framework exploiting the Kronecker product structure of the kernel for data described on a (potential non-uniform) 2D grid.

    \item Our method is capable of handling correlations in time and wavelength when extracting the transmission spectra, enabling us to recover the covariance matrix of the spectra, which in turn can be used for atmospheric retrievals.
    
    \item Simulations show that our method provides robust retrievals in the presence of wavelength-correlated systematics, whereas these same systematics can cause 1D GPs to overestimate uncertainties on sharp spectral features and recover erroneous scattering slopes.

    \item Previous detections of extreme scattering slopes that are difficult to give physical explanations for could be explained by unaccounted for wavelength-correlated systematics.
    
    \item For the two VLT/FORS2 datasets analysed, both datasets were found to contain wavelength-correlated systematics - contradicting the assumptions of their original analysis. Our method recovered significantly weaker constraints on the scattering slope from the 600RI data but tighter constraints on sodium for both datasets as well as on potassium for the 600RI data.
    
    \item Our method removes the need for the common-mode correction - which normally introduces arbitrary offsets to the transmission spectra - making the comparison between independent transit observations simpler.
    
    \item Other benefits of joint-fitting spectroscopic light curves include the ability to marginalise over the uncertainty in the central transit time $T_0$, system scale $a/R_\mathrm{*}$ and impact parameter $b$. The simulations demonstrated that sharing hyperparameters between light curves may also result in more robust retrievals.
    \end{itemize}
    
    Overall, our method presents several clear advantages over the use of 1D GPs with the potential to be applied to a wide variety of datasets in transmission and emission spectroscopy.

\begin{acknowledgements}

    We thank Sochiro Hattori for careful reading of the manuscript and providing helpful comments. We thank the team who led the observations analysed in this work (PI: Nikolov). The observations are publicly available in the ESO Science Archive Facility (http://archive.eso.org) under ESO programme 096.C-0765. Many of the calculations in this work were performed on the astro01 system maintained by the Trinity Centre for High Performance Computing (Research IT). This system is co-funded by the School of Physics and by the Irish Research Council grant award IRCLA/2022/3788. This research has made use of the NASA Exoplanet Archive, which is operated by the California Institute of Technology, under contract with the National Aeronautics and Space Administration under the Exoplanet Exploration Program.

    We are grateful to the developers of \textsc{NumPy}, \textsc{SciPy}, \textsc{Matplotlib}, \textsc{Astropy}, \textsc{pandas}, \textsc{iPython}, \textsc{batman}, \textsc{ArviZ}, \textsc{corner}, \textsc{PyLDTk}, \textsc{JAX}, \textsc{PyMC}, \textsc{jaxoplanet} and \textsc{exoplanet} as these packages were used extensively in this work \citep{numpy, scipy, matplotlib, astropy, pandas, ipython, batman, arviz_2019, corner, LDTK, JAX, Salvatier2015, jaxoplanet, exoplanet}.
    
\end{acknowledgements}

\bibliographystyle{aa}
\bibliography{main}

\begin{thebibliography}{81}
\expandafter\ifx\csname natexlab\endcsname\relax\def\natexlab#1{#1}\fi

\bibitem[{{Ahrer} {et~al.}(2022){Ahrer}, {Wheatley}, {Kirk}, {Gandhi}, {King},
  \& {Louden}}]{Ahrer2022a}
{Ahrer}, E., {Wheatley}, P.~J., {Kirk}, J., {et~al.} 2022, \mnras, 510, 4857

\bibitem[{{Ahrer} {et~al.}(2023){Ahrer}, {Stevenson}, {Mansfield}, {Moran},
  {Brande}, {Morello}, {Murray}, {Nikolov}, {Petit dit de la Roche},
  {Schlawin}, {Wheatley}, {Zieba}, {Batalha}, {Damiano}, {Goyal}, {Lendl},
  {Lothringer}, {Mukherjee}, {Ohno}, {Batalha}, {Battley}, {Bean}, {Beatty},
  {Benneke}, {Berta-Thompson}, {Carter}, {Cubillos}, {Daylan}, {Espinoza},
  {Gao}, {Gibson}, {Gill}, {Harrington}, {Hu}, {Kreidberg}, {Lewis}, {Line},
  {L{\'o}pez-Morales}, {Parmentier}, {Powell}, {Sing}, {Tsai}, {Wakeford},
  {Welbanks}, {Alam}, {Alderson}, {Allen}, {Anderson}, {Barstow}, {Bayliss},
  {Bell}, {Blecic}, {Bryant}, {Burleigh}, {Carone}, {Casewell}, {Changeat},
  {Chubb}, {Crossfield}, {Crouzet}, {Decin}, {D{\'e}sert}, {Feinstein},
  {Flagg}, {Fortney}, {Gizis}, {Heng}, {Iro}, {Kempton}, {Kendrew}, {Kirk},
  {Knutson}, {Komacek}, {Lagage}, {Leconte}, {Lustig-Yaeger}, {MacDonald},
  {Mancini}, {May}, {Mayne}, {Miguel}, {Mikal-Evans}, {Molaverdikhani},
  {Palle}, {Piaulet}, {Rackham}, {Redfield}, {Rogers}, {Roy}, {Rustamkulov},
  {Shkolnik}, {Sotzen}, {Taylor}, {Tremblin}, {Tucker}, {Turner}, {de
  Val-Borro}, {Venot}, \& {Zhang}}]{Ahrer2022}
{Ahrer}, E.-M., {Stevenson}, K.~B., {Mansfield}, M., {et~al.} 2023, \nat, 614,
  653

\bibitem[{{Aigrain} \& {Foreman-Mackey}(2023)}]{Aigrain2022}
{Aigrain}, S. \& {Foreman-Mackey}, D. 2023, \araa, 61, 329

\bibitem[{{Alderson} {et~al.}(2023){Alderson}, {Wakeford}, {Alam}, {Batalha},
  {Lothringer}, {Adams Redai}, {Barat}, {Brande}, {Damiano}, {Daylan},
  {Espinoza}, {Flagg}, {Goyal}, {Grant}, {Hu}, {Inglis}, {Lee}, {Mikal-Evans},
  {Ramos-Rosado}, {Roy}, {Wallack}, {Batalha}, {Bean}, {Benneke},
  {Berta-Thompson}, {Carter}, {Changeat}, {Col{\'o}n}, {Crossfield},
  {D{\'e}sert}, {Foreman-Mackey}, {Gibson}, {Kreidberg}, {Line},
  {L{\'o}pez-Morales}, {Molaverdikhani}, {Moran}, {Morello}, {Moses},
  {Mukherjee}, {Schlawin}, {Sing}, {Stevenson}, {Taylor}, {Aggarwal}, {Ahrer},
  {Allen}, {Barstow}, {Bell}, {Blecic}, {Casewell}, {Chubb}, {Crouzet},
  {Cubillos}, {Decin}, {Feinstein}, {Fortney}, {Harrington}, {Heng}, {Iro},
  {Kempton}, {Kirk}, {Knutson}, {Krick}, {Leconte}, {Lendl}, {MacDonald},
  {Mancini}, {Mansfield}, {May}, {Mayne}, {Miguel}, {Nikolov}, {Ohno}, {Palle},
  {Parmentier}, {Petit dit de la Roche}, {Piaulet}, {Powell}, {Rackham},
  {Redfield}, {Rogers}, {Rustamkulov}, {Tan}, {Tremblin}, {Tsai}, {Turner}, {de
  Val-Borro}, {Venot}, {Welbanks}, {Wheatley}, \& {Zhang}}]{G3952022}
{Alderson}, L., {Wakeford}, H.~R., {Alam}, M.~K., {et~al.} 2023, \nat, 614, 664

\bibitem[{{Anderson} {et~al.}(2011){Anderson}, {Collier Cameron}, {Hellier},
  {Lendl}, {Lister}, {Maxted}, {Queloz}, {Smalley}, {Smith}, {Triaud}, {West},
  {Brown}, {Gillon}, {Pepe}, {Pollacco}, {S{\'e}gransan}, {Street}, \&
  {Udry}}]{Anderson2011}
{Anderson}, D.~R., {Collier Cameron}, A., {Hellier}, C., {et~al.} 2011, \aap,
  531, A60

\bibitem[{{Appenzeller} {et~al.}(1998){Appenzeller}, {Fricke}, {F{\"u}rtig},
  {G{\"a}ssler}, {H{\"a}fner}, {Harke}, {Hess}, {Hummel}, {J{\"u}rgens},
  {Kudritzki}, {Mantel}, {Meisl}, {Muschielok}, {Nicklas}, {Rupprecht},
  {Seifert}, {Stahl}, {Szeifert}, \& {Tarantik}}]{Appenzeller1998}
{Appenzeller}, I., {Fricke}, K., {F{\"u}rtig}, W., {et~al.} 1998, The
  Messenger, 94, 1

\bibitem[{{Astropy Collaboration} {et~al.}(2022){Astropy Collaboration},
  {Price-Whelan}, {Lim}, {Earl}, {Starkman}, {Bradley}, {Shupe}, {Patil},
  {Corrales}, {Brasseur}, {N{\"o}the}, {Donath}, {Tollerud}, {Morris},
  {Ginsburg}, {Vaher}, {Weaver}, {Tocknell}, {Jamieson}, {van Kerkwijk},
  {Robitaille}, {Merry}, {Bachetti}, {G{\"u}nther}, {Aldcroft},
  {Alvarado-Montes}, {Archibald}, {B{\'o}di}, {Bapat}, {Barentsen},
  {Baz{\'a}n}, {Biswas}, {Boquien}, {Burke}, {Cara}, {Cara}, {Conroy},
  {Conseil}, {Craig}, {Cross}, {Cruz}, {D'Eugenio}, {Dencheva}, {Devillepoix},
  {Dietrich}, {Eigenbrot}, {Erben}, {Ferreira}, {Foreman-Mackey}, {Fox},
  {Freij}, {Garg}, {Geda}, {Glattly}, {Gondhalekar}, {Gordon}, {Grant},
  {Greenfield}, {Groener}, {Guest}, {Gurovich}, {Handberg}, {Hart},
  {Hatfield-Dodds}, {Homeier}, {Hosseinzadeh}, {Jenness}, {Jones}, {Joseph},
  {Kalmbach}, {Karamehmetoglu}, {Ka{\l}uszy{\'n}ski}, {Kelley}, {Kern},
  {Kerzendorf}, {Koch}, {Kulumani}, {Lee}, {Ly}, {Ma}, {MacBride}, {Maljaars},
  {Muna}, {Murphy}, {Norman}, {O'Steen}, {Oman}, {Pacifici}, {Pascual},
  {Pascual-Granado}, {Patil}, {Perren}, {Pickering}, {Rastogi}, {Roulston},
  {Ryan}, {Rykoff}, {Sabater}, {Sakurikar}, {Salgado}, {Sanghi}, {Saunders},
  {Savchenko}, {Schwardt}, {Seifert-Eckert}, {Shih}, {Jain}, {Shukla}, {Sick},
  {Simpson}, {Singanamalla}, {Singer}, {Singhal}, {Sinha}, {Sip{\H{o}}cz},
  {Spitler}, {Stansby}, {Streicher}, {{\v{S}}umak}, {Swinbank}, {Taranu},
  {Tewary}, {Tremblay}, {de Val-Borro}, {Van Kooten}, {Vasovi{\'c}}, {Verma},
  {de Miranda Cardoso}, {Williams}, {Wilson}, {Winkel}, {Wood-Vasey}, {Xue},
  {Yoachim}, {Zhang}, {Zonca}, \& {Astropy Project Contributors}}]{astropy}
{Astropy Collaboration}, {Price-Whelan}, A.~M., {Lim}, P.~L., {et~al.} 2022,
  \apj, 935, 167

\bibitem[{{Barrag{\'a}n} {et~al.}(2022){Barrag{\'a}n}, {Aigrain}, {Rajpaul}, \&
  {Zicher}}]{PyanetiII}
{Barrag{\'a}n}, O., {Aigrain}, S., {Rajpaul}, V.~M., \& {Zicher}, N. 2022,
  Monthly Notices of the Royal Astronomical Society, 509, 866

\bibitem[{{Bell} {et~al.}(2017){Bell}, {Nikolov}, {Cowan}, {Barstow}, {Barman},
  {Crossfield}, {Gibson}, {Evans}, {Sing}, {Knutson}, {Kataria}, {Lothringer},
  {Benneke}, \& {Schwartz}}]{Bell2017}
{Bell}, T.~J., {Nikolov}, N., {Cowan}, N.~B., {et~al.} 2017, \apjl, 847, L2

\bibitem[{{Bishop} \& {Nasrabadi}(2007)}]{Bishop}
{Bishop}, C.~M. \& {Nasrabadi}, N.~M. 2007, Journal of Electronic Imaging, 16,
  049901

\bibitem[{{Bonomo} {et~al.}(2017){Bonomo}, {Desidera}, {Benatti}, {Borsa},
  {Crespi}, {Damasso}, {Lanza}, {Sozzetti}, {Lodato}, {Marzari}, {Boccato},
  {Claudi}, {Cosentino}, {Covino}, {Gratton}, {Maggio}, {Micela}, {Molinari},
  {Pagano}, {Piotto}, {Poretti}, {Smareglia}, {Affer}, {Biazzo}, {Bignamini},
  {Esposito}, {Giacobbe}, {H{\'e}brard}, {Malavolta}, {Maldonado}, {Mancini},
  {Martinez Fiorenzano}, {Masiero}, {Nascimbeni}, {Pedani}, {Rainer}, \&
  {Scandariato}}]{Bonomo2017}
{Bonomo}, A.~S., {Desidera}, S., {Benatti}, S., {et~al.} 2017, \aap, 602, A107

\bibitem[{{Boone}(2019)}]{Boone2019}
{Boone}, K. 2019, \aj, 158, 257

\bibitem[{Bradbury {et~al.}(2018)Bradbury, Frostig, Hawkins, Johnson, Leary,
  Maclaurin, Necula, Paszke, Vander{P}las, Wanderman-{M}ilne, \& Zhang}]{JAX}
Bradbury, J., Frostig, R., Hawkins, P., {et~al.} 2018, {JAX}: composable
  transformations of {P}ython+{N}um{P}y programs

\bibitem[{Brandt(2014)}]{Brandt_2014}
Brandt, S. 2014, Data Analysis (Springer International Publishing)

\bibitem[{{Brown}(2001)}]{Brown2001}
{Brown}, T.~M. 2001, \apj, 553, 1006

\bibitem[{{Carter} \& {Winn}(2009)}]{Carter2009}
{Carter}, J.~A. \& {Winn}, J.~N. 2009, \apj, 704, 51

\bibitem[{{Charbonneau} {et~al.}(2002){Charbonneau}, {Brown}, {Noyes}, \&
  {Gilliland}}]{Charbonneau2002}
{Charbonneau}, D., {Brown}, T.~M., {Noyes}, R.~W., \& {Gilliland}, R.~L. 2002,
  The Astrophysical Journal, 568, 377

\bibitem[{{Delisle} {et~al.}(2022){Delisle}, {Unger}, {Hara}, \&
  {S{\'e}gransan}}]{sleaf}
{Delisle}, J.~B., {Unger}, N., {Hara}, N.~C., \& {S{\'e}gransan}, D. 2022,
  \aap, 659, A182

\bibitem[{{Diamond-Lowe} {et~al.}(2020){Diamond-Lowe}, {Berta-Thompson},
  {Charbonneau}, {Dittmann}, \& {Kempton}}]{Lowe2020}
{Diamond-Lowe}, H., {Berta-Thompson}, Z., {Charbonneau}, D., {Dittmann}, J., \&
  {Kempton}, E. M.~R. 2020, \aj, 160, 27

\bibitem[{{Espinoza} {et~al.}(2019){Espinoza}, {Rackham}, {Jord{\'a}n}, {Apai},
  {L{\'o}pez-Morales}, {Osip}, {Grimm}, {Hoeijmakers}, {Wilson}, {Bixel},
  {McGruder}, {Rodler}, {Weaver}, {Lewis}, {Fortney}, \&
  {Fraine}}]{Espinoza2019}
{Espinoza}, N., {Rackham}, B.~V., {Jord{\'a}n}, A., {et~al.} 2019, Monthly
  Notices of the Royal Astronomical Society, 482, 2065

\bibitem[{{Evans} {et~al.}(2013){Evans}, {Pont}, {Sing}, {Aigrain}, {Barstow},
  {D{\'e}sert}, {Gibson}, {Heng}, {Knutson}, \& {Lecavelier des
  Etangs}}]{Evans2013}
{Evans}, T.~M., {Pont}, F., {Sing}, D.~K., {et~al.} 2013, \apjl, 772, L16

\bibitem[{{Fakhouri} {et~al.}(2015){Fakhouri}, {Boone}, {Aldering},
  {Antilogus}, {Aragon}, {Bailey}, {Baltay}, {Barbary}, {Baugh}, {Bongard},
  {Buton}, {Chen}, {Childress}, {Chotard}, {Copin}, {Fagrelius}, {Feindt},
  {Fleury}, {Fouchez}, {Gangler}, {Hayden}, {Kim}, {Kowalski}, {Leget},
  {Lombardo}, {Nordin}, {Pain}, {Pecontal}, {Pereira}, {Perlmutter},
  {Rabinowitz}, {Ren}, {Rigault}, {Rubin}, {Runge}, {Saunders}, {Scalzo},
  {Smadja}, {Sofiatti}, {Strovink}, {Suzuki}, {Tao}, {Thomas}, {Weaver}, \&
  {Nearby Supernova Factory}}]{Fakhouri2015}
{Fakhouri}, H.~K., {Boone}, K., {Aldering}, G., {et~al.} 2015, \apj, 815, 58

\bibitem[{Foreman-Mackey(2016)}]{corner}
Foreman-Mackey, D. 2016, The Journal of Open Source Software, 1, 24

\bibitem[{Foreman-Mackey {et~al.}(2017)Foreman-Mackey, Agol, Ambikasaran, \&
  Angus}]{Foreman_Mackey_2017}
Foreman-Mackey, D., Agol, E., Ambikasaran, S., \& Angus, R. 2017, The
  Astronomical Journal, 154, 220

\bibitem[{Foreman-Mackey \& Garcia(2023)}]{jaxoplanet}
Foreman-Mackey, D. \& Garcia, L.~J. 2023, {jaxoplanet}: Astronomical time
  series analysis with {JAX}

\bibitem[{{Foreman-Mackey} {et~al.}(2021){Foreman-Mackey}, {Luger}, {Agol},
  {Barclay}, {Bouma}, {Brandt}, {Czekala}, {David}, {Dong}, {Gilbert},
  {Gordon}, {Hedges}, {Hey}, {Morris}, {Price-Whelan}, \& {Savel}}]{exoplanet}
{Foreman-Mackey}, D., {Luger}, R., {Agol}, E., {et~al.} 2021, {exoplanet:
  Gradient-based probabilistic inference for exoplanet data \& other
  astronomical time series}

\bibitem[{{Gelman} \& {Rubin}(1992)}]{GR}
{Gelman}, A. \& {Rubin}, D.~B. 1992, Statistical Science, 7, 457

\bibitem[{{Gibson}(2014)}]{Gibson2014}
{Gibson}, N.~P. 2014, Monthly Notices of the Royal Astronomical Society, 445,
  3401

\bibitem[{{Gibson} {et~al.}(2012{\natexlab{a}}){Gibson}, {Aigrain}, {Pont},
  {Sing}, {D{\'e}sert}, {Evans}, {Henry}, {Husnoo}, \& {Knutson}}]{Gibson2012b}
{Gibson}, N.~P., {Aigrain}, S., {Pont}, F., {et~al.} 2012{\natexlab{a}},
  \mnras, 422, 753

\bibitem[{{Gibson} {et~al.}(2012{\natexlab{b}}){Gibson}, {Aigrain}, {Roberts},
  {Evans}, {Osborne}, \& {Pont}}]{Gibson2012}
{Gibson}, N.~P., {Aigrain}, S., {Roberts}, S., {et~al.} 2012{\natexlab{b}},
  \mnras, 419, 2683

\bibitem[{{Gibson} {et~al.}(2019){Gibson}, {de Mooij}, {Evans}, {Merritt},
  {Nikolov}, {Sing}, \& {Watson}}]{Gibson2019}
{Gibson}, N.~P., {de Mooij}, E. J.~W., {Evans}, T.~M., {et~al.} 2019, Monthly
  Notices of the Royal Astronomical Society, 482, 606

\bibitem[{{Gibson} {et~al.}(2017){Gibson}, {Nikolov}, {Sing}, {Barstow},
  {Evans}, {Kataria}, \& {Wilson}}]{Gibson2017}
{Gibson}, N.~P., {Nikolov}, N., {Sing}, D.~K., {et~al.} 2017, Monthly Notices
  of the Royal Astronomical Society, 467, 4591

\bibitem[{{Gordon} {et~al.}(2020){Gordon}, {Agol}, \&
  {Foreman-Mackey}}]{Gordon2020}
{Gordon}, T.~A., {Agol}, E., \& {Foreman-Mackey}, D. 2020, The Astronomical
  Journal, 160, 240

\bibitem[{{Greene} {et~al.}(2023){Greene}, {Bell}, {Ducrot}, {Dyrek}, {Lagage},
  \& {Fortney}}]{Greene2023}
{Greene}, T.~P., {Bell}, T.~J., {Ducrot}, E., {et~al.} 2023, \nat, 618, 39

\bibitem[{Harris {et~al.}(2020)Harris, Millman, van~der Walt, Gommers,
  Virtanen, Cournapeau, Wieser, Taylor, Berg, Smith, Kern, Picus, Hoyer, van
  Kerkwijk, Brett, Haldane, del R{\'{i}}o, Wiebe, Peterson,
  G{\'{e}}rard-Marchant, Sheppard, Reddy, Weckesser, Abbasi, Gohlke, \&
  Oliphant}]{numpy}
Harris, C.~R., Millman, K.~J., van~der Walt, S.~J., {et~al.} 2020, Nature, 585,
  357

\bibitem[{Hoerl \& Kennard(1970)}]{Ridge}
Hoerl, A.~E. \& Kennard, R.~W. 1970, Technometrics, 12, 55

\bibitem[{Hoffman \& Gelman(2014)}]{Hoffman2011}
Hoffman, M.~D. \& Gelman, A. 2014, Journal of Machine Learning Research, 15,
  1593–1623

\bibitem[{{Holmberg} \& {Madhusudhan}(2023)}]{Holmberg2023}
{Holmberg}, M. \& {Madhusudhan}, N. 2023, \mnras, 524, 377

\bibitem[{Hunter(2007)}]{matplotlib}
Hunter, J.~D. 2007, Computing in Science \& Engineering, 9, 90

\bibitem[{{Ih} \& {Kempton}(2021)}]{Ih2021}
{Ih}, J. \& {Kempton}, E. M.~R. 2021, The Astronomical Journal, 162, 237

\bibitem[{Jensen {et~al.}(1995)Jensen, Kjærulff, \& Kong}]{blocked_Gibbs}
Jensen, C.~S., Kjærulff, U., \& Kong, A. 1995, International Journal of
  Human-Computer Studies, 42, 647

\bibitem[{{Kokori} {et~al.}(2022){Kokori}, {Tsiaras}, {Edwards}, {Rocchetto},
  {Tinetti}, {Bewersdorff}, {Jongen}, {Lekkas}, {Pantelidou}, {Poultourtzidis},
  {W{\"u}nsche}, {Aggelis}, {Agnihotri}, {Arena}, {Bachschmidt}, {Bennett},
  {Benni}, {Bernacki}, {Besson}, {Betti}, {Biagini}, {Brandebourg}, {Bretton},
  {Brincat}, {Cal{\'o}}, {Campos}, {Casali}, {Ciantini}, {Crow}, {Dauchet},
  {Dawes}, {Deldem}, {Deligeorgopoulos}, {Dymock}, {Eenm{\"a}e}, {Evans},
  {Esseiva}, {Falco}, {Ferratfiat}, {Fowler}, {Futcher}, {Gaitan}, {Horta},
  {Guerra}, {Hurter}, {Jones}, {Kang}, {Kiiskinen}, {Kim}, {Laloum}, {Lee},
  {Lomoz}, {Lopresti}, {Mallonn}, {Mannucci}, {Marino}, {Mario}, {Marquette},
  {Michelet}, {Miller}, {Mollier}, {Molina}, {Montigiani}, {Mortari}, {Morvan},
  {Mugnai}, {Naponiello}, {Nastasi}, {Neito}, {Pace}, {Papadeas}, {Paschalis},
  {Pereira}, {Perroud}, {Phillips}, {Pintr}, {Pioppa}, {Popowicz}, {Raetz},
  {Regembal}, {Rickard}, {Roberts}, {Rousselot}, {Rubia}, {Savage}, {Sedita},
  {Shave-Wall}, {Sioulas}, {{\v{S}}koln{\'\i}k}, {Smith}, {St-Gelais},
  {Stouraitis}, {Strikis}, {Thurston}, {Tomacelli}, {Tomatis}, {Trevan},
  {Valeau}, {Vignes}, {Vora}, {Vra{\v{s}}{\v{t}}{\'a}k}, {Walter}, {Wenzel},
  {Wright}, \& {Z{\'\i}bar}}]{Exoclock2022}
{Kokori}, A., {Tsiaras}, A., {Edwards}, B., {et~al.} 2022, \apjs, 258, 40

\bibitem[{{Kreidberg}(2015)}]{batman}
{Kreidberg}, L. 2015, \pasp, 127, 1161

\bibitem[{Kumar {et~al.}(2019)Kumar, Carroll, Hartikainen, \&
  Martin}]{arviz_2019}
Kumar, R., Carroll, C., Hartikainen, A., \& Martin, O. 2019, Journal of Open
  Source Software, 4, 1143

\bibitem[{{Lecavelier Des Etangs} {et~al.}(2008){Lecavelier Des Etangs},
  {Pont}, {Vidal-Madjar}, \& {Sing}}]{Etangs2008}
{Lecavelier Des Etangs}, A., {Pont}, F., {Vidal-Madjar}, A., \& {Sing}, D.
  2008, \aap, 481, L83

\bibitem[{Liu \& Nocedal(1989)}]{LBFGS}
Liu, D.~C. \& Nocedal, J. 1989, Mathematical Programming, 45, 503

\bibitem[{Llorente {et~al.}(2023)Llorente, Martino, Curbelo, López-Santiago,
  \& Delgado}]{model_selection}
Llorente, F., Martino, L., Curbelo, E., López-Santiago, J., \& Delgado, D.
  2023, WIREs Computational Statistics, 15, e1595

\bibitem[{{Maguire} {et~al.}(2023){Maguire}, {Gibson}, {Nugroho}, {Ramkumar},
  {Fortune}, {Merritt}, \& {de Mooij}}]{Maguire2023}
{Maguire}, C., {Gibson}, N.~P., {Nugroho}, S.~K., {et~al.} 2023, \mnras, 519,
  1030

\bibitem[{{Mandel} \& {Agol}(2002)}]{mandelagol}
{Mandel}, K. \& {Agol}, E. 2002, The Astrophysical Journal, 580, L171

\bibitem[{{May} {et~al.}(2020){May}, {Gardner}, {Rauscher}, \&
  {Monnier}}]{May2020}
{May}, E.~M., {Gardner}, T., {Rauscher}, E., \& {Monnier}, J.~D. 2020, The
  Astronomical Journal, 159, 7

\bibitem[{{McCullough} {et~al.}(2014){McCullough}, {Crouzet}, {Deming}, \&
  {Madhusudhan}}]{McCullough2014}
{McCullough}, P.~R., {Crouzet}, N., {Deming}, D., \& {Madhusudhan}, N. 2014,
  \apj, 791, 55

\bibitem[{{McGruder} {et~al.}(2020){McGruder}, {L{\'o}pez-Morales}, {Espinoza},
  {Rackham}, {Apai}, {Jord{\'a}n}, {Osip}, {Alam}, {Bixel}, {Fortney}, {Henry},
  {Kirk}, {Lewis}, {Rodler}, \& {Weaver}}]{McGruder2020}
{McGruder}, C.~D., {L{\'o}pez-Morales}, M., {Espinoza}, N., {et~al.} 2020, \aj,
  160, 230

\bibitem[{{Meech} {et~al.}(2022){Meech}, {Aigrain}, {Brogi}, \&
  {Birkby}}]{Meech2022}
{Meech}, A., {Aigrain}, S., {Brogi}, M., \& {Birkby}, J.~L. 2022, \mnras, 512,
  2604

\bibitem[{{Nasedkin} {et~al.}(2023){Nasedkin}, {Molli{\`e}re}, {Wang},
  {Cantalloube}, {Kreidberg}, {Pueyo}, {Stolker}, \& {Vigan}}]{Evert2023}
{Nasedkin}, E., {Molli{\`e}re}, P., {Wang}, J., {et~al.} 2023, \aap, 678, A41

\bibitem[{{Neal}(2011)}]{Neal2011}
{Neal}, R. 2011, in Handbook of Markov Chain Monte Carlo, 113--162

\bibitem[{Neal(2003)}]{Neal2003}
Neal, R.~M. 2003, The Annals of Statistics, 31, 705

\bibitem[{{Nikolov} {et~al.}(2016){Nikolov}, {Sing}, {Gibson}, {Fortney},
  {Evans}, {Barstow}, {Kataria}, \& {Wilson}}]{Nikolov2016}
{Nikolov}, N., {Sing}, D.~K., {Gibson}, N.~P., {et~al.} 2016, The Astrophysical
  Journal, 832, 191

\bibitem[{pandas~development team(2020)}]{pandas}
pandas~development team, T. 2020, pandas-dev/pandas: Pandas

\bibitem[{{Panwar} {et~al.}(2022){Panwar}, {D{\'e}sert}, {Todorov}, {Bean},
  {Stevenson}, {Huitson}, {Fortney}, \& {Bergmann}}]{Panwar2022}
{Panwar}, V., {D{\'e}sert}, J.-M., {Todorov}, K.~O., {et~al.} 2022, \mnras,
  510, 3236

\bibitem[{{Parviainen} \& {Aigrain}(2015)}]{LDTK}
{Parviainen}, H. \& {Aigrain}, S. 2015, Monthly Notices of the Royal
  Astronomical Society, 453, 3821

\bibitem[{{Patel} \& {Espinoza}(2022)}]{Patel2022}
{Patel}, J.~A. \& {Espinoza}, N. 2022, \aj, 163, 228

\bibitem[{P\'erez \& Granger(2007)}]{ipython}
P\'erez, F. \& Granger, B.~E. 2007, Computing in Science and Engineering, 9, 21

\bibitem[{{Powell} {et~al.}(2019){Powell}, {Louden}, {Kreidberg}, {Zhang},
  {Gao}, \& {Parmentier}}]{Powell2019}
{Powell}, D., {Louden}, T., {Kreidberg}, L., {et~al.} 2019, \apj, 887, 170

\bibitem[{{Rackham} {et~al.}(2017){Rackham}, {Espinoza}, {Apai},
  {L{\'o}pez-Morales}, {Jord{\'a}n}, {Osip}, {Lewis}, {Rodler}, {Fraine},
  {Morley}, \& {Fortney}}]{Rackham2017}
{Rackham}, B., {Espinoza}, N., {Apai}, D., {et~al.} 2017, \apj, 834, 151

\bibitem[{{Rackham} {et~al.}(2018){Rackham}, {Apai}, \&
  {Giampapa}}]{Rackham2018}
{Rackham}, B.~V., {Apai}, D., \& {Giampapa}, M.~S. 2018, \apj, 853, 122

\bibitem[{{Rajpaul} {et~al.}(2015){Rajpaul}, {Aigrain}, {Osborne}, {Reece}, \&
  {Roberts}}]{Rajpaul2015}
{Rajpaul}, V., {Aigrain}, S., {Osborne}, M.~A., {Reece}, S., \& {Roberts}, S.
  2015, \mnras, 452, 2269

\bibitem[{Rakitsch {et~al.}(2013)Rakitsch, Lippert, Borgwardt, \&
  Stegle}]{Rakitsch2013}
Rakitsch, B., Lippert, C., Borgwardt, K., \& Stegle, O. 2013, in Advances in
  Neural Information Processing Systems, Vol.~26

\bibitem[{Rasmussen \& Williams(2006)}]{Rasmussen}
Rasmussen, C.~E. \& Williams, C. K.~I. 2006, Gaussian processes for machine
  learning., Adaptive computation and machine learning (MIT Press), I--XVIII,
  1--248

\bibitem[{{Rustamkulov} {et~al.}(2023){Rustamkulov}, {Sing}, {Mukherjee},
  {May}, {Kirk}, {Schlawin}, {Line}, {Piaulet}, {Carter}, {Batalha}, {Goyal},
  {L{\'o}pez-Morales}, {Lothringer}, {MacDonald}, {Moran}, {Stevenson},
  {Wakeford}, {Espinoza}, {Bean}, {Batalha}, {Benneke}, {Berta-Thompson},
  {Crossfield}, {Gao}, {Kreidberg}, {Powell}, {Cubillos}, {Gibson}, {Leconte},
  {Molaverdikhani}, {Nikolov}, {Parmentier}, {Roy}, {Taylor}, {Turner},
  {Wheatley}, {Aggarwal}, {Ahrer}, {Alam}, {Alderson}, {Allen}, {Banerjee},
  {Barat}, {Barrado}, {Barstow}, {Bell}, {Blecic}, {Brande}, {Casewell},
  {Changeat}, {Chubb}, {Crouzet}, {Daylan}, {Decin}, {D{\'e}sert},
  {Mikal-Evans}, {Feinstein}, {Flagg}, {Fortney}, {Harrington}, {Heng}, {Hong},
  {Hu}, {Iro}, {Kataria}, {Kempton}, {Krick}, {Lendl}, {Lillo-Box}, {Louca},
  {Lustig-Yaeger}, {Mancini}, {Mansfield}, {Mayne}, {Miguel}, {Morello},
  {Ohno}, {Palle}, {Petit dit de la Roche}, {Rackham}, {Radica},
  {Ramos-Rosado}, {Redfield}, {Rogers}, {Shkolnik}, {Southworth}, {Teske},
  {Tremblin}, {Tucker}, {Venot}, {Waalkes}, {Welbanks}, {Zhang}, \&
  {Zieba}}]{PRISM2022}
{Rustamkulov}, Z., {Sing}, D.~K., {Mukherjee}, S., {et~al.} 2023, \nat, 614,
  659

\bibitem[{{Saatchi}(2011)}]{Saatchi2011}
{Saatchi}, Y. 2011, PhD thesis, University of Cambridge

\bibitem[{Salvatier {et~al.}(2016)Salvatier, Wiecki, \&
  Fonnesbeck}]{Salvatier2015}
Salvatier, J., Wiecki, T.~V., \& Fonnesbeck, C. 2016, PeerJ Computer Science,
  2, e55

\bibitem[{{Seager} \& {Sasselov}(2000)}]{Seager2000}
{Seager}, S. \& {Sasselov}, D.~D. 2000, \apj, 537, 916

\bibitem[{{Sedaghati} {et~al.}(2016){Sedaghati}, {Boffin},
  {Je{\v{r}}abkov{\'a}}, {Garc{\'\i}a Mu{\~n}oz}, {Grenfell}, {Smette},
  {Ivanov}, {Csizmadia}, {Cabrera}, {Kabath}, {Rocchetto}, \&
  {Rauer}}]{Sedaghati2016}
{Sedaghati}, E., {Boffin}, H.~M.~J., {Je{\v{r}}abkov{\'a}}, T., {et~al.} 2016,
  \aap, 596, A47

\bibitem[{{Sedaghati} {et~al.}(2017){Sedaghati}, {Boffin}, {MacDonald},
  {Gandhi}, {Madhusudhan}, {Gibson}, {Oshagh}, {Claret}, \&
  {Rauer}}]{Sedaghati2017}
{Sedaghati}, E., {Boffin}, H. M.~J., {MacDonald}, R.~J., {et~al.} 2017, Nature,
  549, 238

\bibitem[{{Sedaghati} {et~al.}(2021){Sedaghati}, {MacDonald},
  {Casasayas-Barris}, {Hoeijmakers}, {Boffin}, {Rodler}, {Brahm}, {Jones},
  {S{\'a}nchez-L{\'o}pez}, {Carleo}, {Figueira}, {Mehner}, \&
  {L{\'o}pez-Puertas}}]{Sedaghati2021}
{Sedaghati}, E., {MacDonald}, R.~J., {Casasayas-Barris}, N., {et~al.} 2021,
  \mnras, 505, 435

\bibitem[{{Sing} {et~al.}(2016){Sing}, {Fortney}, {Nikolov}, {Wakeford},
  {Kataria}, {Evans}, {Aigrain}, {Ballester}, {Burrows}, {Deming},
  {D{\'e}sert}, {Gibson}, {Henry}, {Huitson}, {Knutson}, {Lecavelier Des
  Etangs}, {Pont}, {Showman}, {Vidal-Madjar}, {Williamson}, \&
  {Wilson}}]{Sing2016}
{Sing}, D.~K., {Fortney}, J.~J., {Nikolov}, N., {et~al.} 2016, \nat, 529, 59

\bibitem[{{Sing} {et~al.}(2015){Sing}, {Wakeford}, {Showman}, {Nikolov},
  {Fortney}, {Burrows}, {Ballester}, {Deming}, {Aigrain}, {D{\'e}sert},
  {Gibson}, {Henry}, {Knutson}, {Lecavelier des Etangs}, {Pont},
  {Vidal-Madjar}, {Williamson}, \& {Wilson}}]{Sing2015}
{Sing}, D.~K., {Wakeford}, H.~R., {Showman}, A.~P., {et~al.} 2015, Monthly
  Notices of the Royal Astronomical Society, 446, 2428

\bibitem[{Spyratos {et~al.}(2023)Spyratos, Nikolov, Constantinou, Southworth,
  Madhusudhan, Sedaghati, Ehrenreich, \& Mancini}]{Spyratos2023}
Spyratos, P., Nikolov, N.~K., Constantinou, S., {et~al.} 2023, Monthly Notices
  of the Royal Astronomical Society, 521, 2163

\bibitem[{Virtanen {et~al.}(2020)Virtanen, Gommers, Oliphant, Haberland, Reddy,
  Cournapeau, Burovski, Peterson, Weckesser, Bright, {van der Walt}, Brett,
  Wilson, Millman, Mayorov, Nelson, Jones, Kern, Larson, Carey, Polat, Feng,
  Moore, {VanderPlas}, Laxalde, Perktold, Cimrman, Henriksen, Quintero, Harris,
  Archibald, Ribeiro, Pedregosa, {van Mulbregt}, \& {SciPy 1.0
  Contributors}}]{scipy}
Virtanen, P., Gommers, R., Oliphant, T.~E., {et~al.} 2020, Nature Methods, 17,
  261

\bibitem[{{Wilson} {et~al.}(2021){Wilson}, {Gibson}, {Lothringer}, {Sing},
  {Mikal-Evans}, {de Mooij}, {Nikolov}, \& {Watson}}]{Wilson2021}
{Wilson}, J., {Gibson}, N.~P., {Lothringer}, J.~D., {et~al.} 2021, \mnras, 503,
  4787

\bibitem[{{Zieba} {et~al.}(2023){Zieba}, {Kreidberg}, {Ducrot}, {Gillon},
  {Morley}, {Schaefer}, {Tamburo}, {Koll}, {Lyu}, {Acu{\~n}a}, {Agol}, {Iyer},
  {Hu}, {Lincowski}, {Meadows}, {Selsis}, {Bolmont}, {Mandell}, \&
  {Suissa}}]{Zieba2023}
{Zieba}, S., {Kreidberg}, L., {Ducrot}, E., {et~al.} 2023, \nat, 620, 746

\end{thebibliography}


\begin{appendix}


    \section{Gaussian process regression}
    \label{app:gp_mean}

    This method can be used to efficiently perform Gaussian process regression (as demonstrated in \citealt{Rakitsch2013}), although the full predictive covariance matrix can be expensive to compute. The predicted mean and covariance of the GP conditioned on the data were used in this work to help visualise the fit of the systematics (Fig.~\ref{fig:VLT gp mean}) and also to clip outliers and replace with interpolated values (see Sect.~\ref{sec:VLT procedure}).
    
    Suppose we have a set of observations $\vec{y}\sim\mathcal{N}(\vec{\mu}(\vec{t}, \vec{\lambda}, \vec{\theta}),\mathbf{K}(\vec{t}, \vec{\lambda}, \vec{\theta}))$, where we are using the best-fit values for all parameters $\vec{\theta}$. We can compute the predictive distribution for points $\vec{y}_*$ - which are located on a grid of $\vec{\lambda}_*$ wavelengths and $\vec{t}_*$ times - using:
    \begin{eqnarray}\label{eq:gp_mean}
        \mathbb{E}[\vec{y}_*] & = & \vec{\mu}_* + \mathbf{K}_*^T \mathbf{K}^{-1} (\vec{y} - \vec{\mu}), \\
        \text{Var}[\vec{y}_*] & = & \mathbf{K}_{**} - \mathbf{K}_*^T \mathbf{K}^{-1} \mathbf{K}_*.
    \end{eqnarray}
    Where $\vec{\mu}_*$ is the mean function computed at the locations of $\vec{y}_*$, $\mathbf{K}_*$ is the covariance between $\vec{y}$ and $\vec{y}_*$ and $\mathbf{K}_{**}$ is the covariance between values of $\vec{y}_*$ with itself.

    Calculating $\mathbb{E}[\vec{y}_*]$ is straight-forward as we can break down the calculation of $\mathbf{K}^{-1} \vec{r}$ using Eq.~(\ref{eq:K_inv_R_rakitsch}) and $\mathbf{K}_*$ is in general a sum of two Kronecker products similar to $\mathbf{K}$ so we can use Eq.~(\ref{eq:Algo14}) to perform this matrix-vector multiplication.

    Calculating all of the terms in the covariance matrix $\mathrm{Var}[\vec{y}_*]$ would be computationally expensive and memory-intensive as it would require the calculation and storage of $M_* N_* \times M_* N_*$ terms. If it is desired to take random draws from the predictive distribution then no method was found to efficiently do this, as it would likely require the calculation of the full covariance matrix $\mathrm{Var}[\vec{y}_*]$, followed by Cholesky factorisation or eigendecomposition of that matrix. However, if we are just interested in calculating the diagonal terms of the covariance matrix (such as for identifying outliers) then an optimisation was found which does not require the storage of any $M_* N_* \times M_* N_*$ matrices in memory. This method has no calculations that would scale worse in runtime than the cube of any of the dimensions $M$, $N$, $M_*$ or $N_*$. The details of this algorithm have been omitted for brevity but the implementation is available on the \textsc{luas} GitHub repository.

    \section{Error propagation with a covariance matrix}
    \label{app:error_prop}
    While it is well-known how to propagate uncertainties when they are independent, propagating uncertainties given any general covariance matrix is less well-known. It is included here for reference.
    
    Given a vector $\vec{x}$ that follows a multivariate-normal distribution $\vec{x} \sim \mathcal{N}(\vec{\mu}_{\vec{x}},\,\mathbf{\Sigma}_{\vec{x}})$ and values $\vec{y}$ we wish to calculate which are a linear transformation of $\vec{x}$ (i.e. $\vec{y} = \mathbf{T} \vec{x}$ for some matrix $\mathbf{T}$) then $\vec{y}$ follows a normal distribution $\vec{y} \sim \mathcal{N}(\vec{\mu}_{\vec{y}},\,\mathbf{\Sigma}_{\vec{y}})$ and we can calculate $\vec{\mu}_{\vec{y}}$ and $\mathbf{\Sigma}_{\vec{y}}$ using the following equations (see \citealt{Brandt_2014} for derivations):
    \begin{align}
        \vec{\mu}_{\vec{y}} &= \mathbf{T} \vec{\mu}_{\vec{x}}, \label{eq:error_prop1} \\
        \mathbf{\Sigma}_{\vec{y}} &= \mathbf{T} \mathbf{\Sigma}_{\vec{x}} \mathbf{T}^T.
        \label{eq:error_prop2}
    \end{align}
    For example, to compute the average transit depth given a transmission spectrum of $M$ transit depths described by $\vec{\rho}^2 \sim \mathcal{N}(\vec{\mu}_\mathrm{\vec{\rho}^2},\,\mathbf{\Sigma}_\mathrm{\vec{\rho}^2})$, we must take the mean of $\vec{\rho}^2$. We can do this by taking $\mathbf{T} = \frac{1}{M}\vec{1}^T$ where $\vec{1}$ is an $M$-long vector where every element is equal to one. If the covariance matrix $\mathbf{\Sigma_{\vec{x}}}$ is diagonal with constant variance $\sigma^2$ then this reduces to the familiar form $\sigma_{\bar{x}}^2 = \sigma^2/N$ or $\sigma_{\bar{x}} = \sigma/\sqrt{N}$ for the standard error of the mean.
    
    This can also be used to bin a transmission spectrum to larger bin sizes. For example, if binning each pair of light curves together then the vector $\vec{a} = \text{vec}(\frac{1}{2}, \frac{1}{2}, 0, 0, ...)$ could be used to propagate the uncertainty for binning the first pair of bins. The covariance matrix of the full (binned) transmission spectrum could be obtained by using the matrix $\mathbf{T} = \text{vec}(\frac{1}{2}, \frac{1}{2}) \otimes \mathbb{I}_{\frac{N}{2}}$ (assuming an even number of bins) for Equations~\ref{eq:error_prop1} and ~\ref{eq:error_prop2}.

    \section{Priors for the simulations and the VLT/FORS2 analyses}
    \label{app:priors}

    Priors used for the simulations are included in Table~\ref{tab:sim_priors}. These same priors were used for all methods where the parameter is relevant, that is the 1D GP method did not use the parameter $l_{\lambda}$ but used the same priors on the other parameters.

    \begin{table*}
    	\centering
    	\caption{Priors used when fitting parameters in Simulations 1-4.}
    	\label{tab:sim_priors}
    	\begin{tabular}{llll} 
    		\hline
    		Parameter & Prior Distribution & Prior Bounds & Simulated Bounds\\
    		\hline
            $\rho^2$: Transit depth & Uniform & $(0, 1)$ & See Table~\ref{tab:mf} \\
            $h$: Height scale & Log-uniform & $(0.000001, 0.01)$ & $(0.0005,0.001)$\\
            $l_{t}$: Time length scale (days) & Log-uniform & $(0.002,0.4)$ & $(0.004,0.1)$\\
            $l_{\lambda}$: Wavelength length scale (\angstrom) & Log-uniform & $(75,225000)$ & See Table~\ref{tab:sim_1-4_hp}\\
            $\sigma$: White noise amplitude & Log-uniform & $(0.00001,0.01)$ & $(0.0001,0.001)$\\
    		\hline
    	\end{tabular}
    \end{table*}

    \begin{table*}
    	\centering
    	\caption{Priors used for fitting the 600B and 600RI VLT/FORS2 datasets.}
    	\label{tab:VLT_priors}
    	\begin{tabular}{lll} 
    		\hline
    		Parameter & Prior Distribution & Prior Range\\
    		\hline
            $T_0$: Central transit time (days - predicted time) & Uniform & $(-0.01, 0.01)$\\
            $P$: Period (days) & Fixed & $3.4059095$\\
            $a/R_*$: System scale & Gaussian & ($\mu = 8.19$, $\sigma = 0.1$)\\
            $\rho^2$: Transit depth & Uniform & $(0, 1)$\\
            $\bar{\rho}_\mathrm{600B}$: Mean radius ratio (600B) & Gaussian & $(\mu = 0.12546, \sigma = 0.00026)$\\
            $\bar{\rho}_\mathrm{600RI}$: Mean radius ratio (600RI) & Gaussian & $(\mu = 0.125054, \sigma = 0.00035)$\\
            $b$: Impact parameter & Gaussian & ($\mu = 0.761$, $\sigma = 0.018$)\\
            $F_\mathrm{oot}$: Baseline flux & Uniform & $(0.99, 1.01)$\\
            $T_\mathrm{grad}$: Linear Trend in Baseline flux & Uniform & $(-0.1, 0.1)$\\
            $h_\mathrm{CM}$: Common-mode systematics height scale & Log-uniform & $(0.000001, 0.01)$\\
            $h_\mathrm{HFS}$: High-frequency systematics height scale & Log-uniform & $(0.000001, 0.01)$\\
            $h_\mathrm{WSS}$: Wavelength-specific systematics height scale & Log-uniform & $(0.000001, 0.01)$\\
            $l_{t}$: Time length scale (days) & Log-uniform & $(1x$ observation cadence, 2x total time range$)$\\
            $l_\mathrm{\lambda_\mathrm{CM}}$: Wavelength length scale & Log-uniform & $(1$ bin width, 50x total wavelength range$)$\\
            $l_\mathrm{\lambda_\mathrm{HFS}}$: Wavelength length scale & Log-uniform & $(1$ bin width, 50x total wavelength range$)$\\
            $\sigma$: White noise amplitude & Log-uniform & $(0.00001,0.01)$\\
    		\hline
    	\end{tabular}
    \end{table*}

    The default priors used for both VLT/FORS2 datasets are given in Table~\ref{tab:VLT_priors}. The central transit time priors are given relative to the predicted central transit times from \citet{Patel2022}. When analyses in Sect.~\ref{sec:VLT results} diverged from these default values it was explicitly stated. Where parameters are shared between the 1D GP re-analysis and the 2D GP analyses, the same priors were used for both. The differences for the 1D GP re-analysis are that: it used the mean radius ratio prior during the white light curve fit instead of during the spectroscopic fit, the high-frequency systematics were recovered from the residuals of the white light curve fit and therefore $h_\mathrm{HFS}$ was never fit for and finally the wavelength length scales $l_\mathrm{\lambda_\mathrm{CM}}$ and $l_\mathrm{\lambda_\mathrm{HFS}}$ were not fit for.
    
    We note that while prior bounds were included for all uniform and log-uniform priors, the only parameters that had MCMC values close to these bounds were $l_\mathrm{\lambda_\mathrm{CM}}$ for the 600B data and the wavelength-specific systematics height scale $h_\mathrm{WSS}$ for many of the wavelength channels (it was fit independently for each wavelength). The implications for $l_\mathrm{\lambda_\mathrm{CM}}$ were discussed in Sect.~\ref{sec:analogous_2D_GP}, while for $h_\mathrm{WSS}$ (which could sometimes be consistent with the minimum prior bound) this suggests that some wavelength channels were consistent with having no wavelength-independent systematics present in them at all. The choice of minimum prior bound therefore could have affected the results, with lower minimum prior bounds more strongly weighting the probability of negligible systematics being present and higher minimum prior bounds forcing the probability of non-negligible systematics to be higher. The rest of the parameters were constrained within the prior bounds (see the corner plots for some of these parameters in Figs.~\ref{fig:corner_blue} and \ref{fig:corner_red}) and therefore were likely unaffected by them, other than by the choice of a uniform prior (for most mean function parameters) or a log-uniform prior (for hyperparameters).

    Gaussian priors were also taken from \citet{Sing2015} for the mean radius ratio, $a/R_*$ and $b$ - as were also used in \citet{Gibson2017}. Gaussian priors on the limb-darkening parameters $c_1$ and $c_2$ were calculated using \textsc{PyLDTk}, as discussed in Sect.~\ref{sec:VLT procedure}.

    For the direct atmospheric retrievals in Sect.~\ref{sec:atmo_fit}, the Gaussian prior on the mean radius ratio $\bar{\rho}$ was still used and the remaining parameters ($m$, $\Delta \rho_\mathrm{Na}$, $\Delta \rho_\mathrm{K}$) were varied uniformly and with the constraint that the radius ratio was positive for all wavelengths.

    \section{MCMC sampling for the simulations}
    \label{app:sim MCMC}

    We initially found that some simulations would not be efficiently sampled by varying all parameters together in the same No U-Turn Sampling step. As a result, blocked Gibbs sampling could be used to speed up convergence. The approach used for all the simulations was to vary all transit depths simultaneously with NUTS in one blocked Gibbs sampling step followed by varying the hyperparameters together in one or more blocked Gibbs sampling steps. Separately varying the transit depths and the hyperparameters would not be efficient if significant correlations existed between them but for the simulations it appears this was not a significant issue. For real datasets such as the VLT/FORS2 data, the approach used was different because the correlations between the parameters and hyperparameters were found to be more significant.
    
    Work performed after the simulations has revealed that the improved efficiency when varying parameters separately was likely caused by the mass matrix we generated from a Laplace approximation not accounting for the transformations performed by \textsc{PyMC} to keep parameters within the desired prior bounds. As different parameters can have significantly different prior bounds, this likely had a significant impact on sampling efficiency and future updates to the \textsc{luas} package may be able to account for this without using blocked Gibbs sampling. However, blocked Gibbs sampling is a valid MCMC method \citep{blocked_Gibbs} and the choice of mass matrix should only affect the sampling efficiency and not whether the sampling is valid. We present in this section the procedure used for the simulations in this work, noting that it may be improved upon in future work.
    
    A Laplace approximation was used to generate the mass matrix for each blocked Gibbs step of NUTS. However, it was found that parameters being fit close to a prior limit would often converge slowly with NUTS, likely because of the unaccounted for transformations performed by \textsc{PyMC} and also because they were poorly approximated by a Gaussian (since a Laplace approximation works by approximating the posterior as Gaussian). For the simulations performed in this work, only the wavelength and time length scales $l_{\lambda}$ and $l_{t}$ were occasionally being fit close to their prior limits so these were the parameters most affected. To deal with this issue, if the best-fit values of $l_{t}$ and $l_{\lambda}$ were identified to be outside of a certain range (given in Table~\ref{tab:slice}), they were instead sampled in a separate Gibbs step using slice sampling (see \cite{Neal2003} for an explanation of slice sampling). Slice sampling is a method of sampling that samples under the area of the probability density by alternating between vertical and horizontal `slices'. Univariate slice sampling - as implemented in \textsc{PyMC} - was often found to converge faster than NUTS for these length scale parameters when they were close to prior limits. Future work may be done to try to avoid the need for slice sampling and improve sampling efficiency. However, blocked Gibbs sampling between NUTS and slice sampling is valid\footnote{Alternating between NUTS and other MCMC methods is especially common when sampling both continuous and discrete variables \citep{Salvatier2015}.} and the results of the simulations demonstrate that the retrievals using this inference approach were generally consistent with the true simulated values. 
    
    Table~\ref{tab:slice} summarises the 2D GP sampling steps used depending on the best-fit values of the hyperparameters. All hyperparameters were drawn from log-uniform distributions within the ranges listed. NUTS 1 and NUTS 2 refer to two different blocked Gibbs steps that use NUTS, that is all parameters being varied in step `NUTS 2' are being varied together. The transit depth parameters were always varied together in a single step of NUTS. If the best-fit values of $l_{t}$ and/or $l_{\lambda}$ prior to running the MCMC were not within the ranges listed then they were independently varied with univariate slice sampling in additional Gibbs sampling steps (resulting in a maximum of four steps of Gibbs sampling). Both the Hybrid and 2D GP method followed this description with the exception that the Hybrid method did not have a wavelength length scale.
    
    \begin{table}
    	\centering
    	\caption{Overview of MCMC blocked Gibbs sampling steps for Simulations 1-4.}
    	\label{tab:slice}
    	\begin{tabular}{lll} 
    		\hline
    		Parameter & Best-fit Value & Step\\
    		\hline
            $\rho^2$: Transit depth & Any & NUTS 1\\
            $h$: Height scale &  Any & NUTS 2\\
            $l_{t}$: Time length scale (days) & $\in (0.004,0.03)$ & NUTS 2\\
            $l_{t}$ Time length scale (days) & $\not\in (0.004,0.03)$ & Slice\\
            $l_{\lambda}$: Wavelength length scale ($\angstrom$) & $\in (400,4500)$ & NUTS 2\\
            $l_{\lambda}$: Wavelength length scale ($\angstrom$) & $\not\in (400,4500)$ & Slice\\
            $\sigma$: White noise amplitude & Any & NUTS 2\\
    		\hline
    	\end{tabular}
    \end{table}

    \section{MCMC sampling for the VLT/FORS2 analyses}
    \label{app:VLT MCMC}
    
    Fitting both VLT/FORS2 datasets with 2D GPs required fitting over 100 parameters with an MCMC. The approach to do this was the same for both datasets. It was similar to the method followed for the simulations, although many more parameters were being fit in this case with more parameters combined together into the same blocked Gibbs steps. It was found that adjusting the mass matrix from the Laplace approximation by scaling the rows and columns relating to certain parameters could improve sampling efficiency. It was not known at the time why this was the case but the transformations by PyMC to keep parameters within certain prior bounds are likely the cause. We could still account for this by running NUTS on each group of parameters (i.e. $\rho^2$, $F_\mathrm{oot}$, $T_\mathrm{grad}$, etc.) individually and using dual-averaging (as described in \citealt{Hoffman2011} as implemented in \textsc{PyMC}) to optimise the step size $\epsilon$ for each group of parameters. This was performed for each group of parameters for 200 steps with a single chain of NUTS, which was long enough for the step sizes to have converged near to `optimal' values. The approximate covariance matrix of the posterior obtained using the Laplace approximation was scaled using these values which significantly improved sampling efficiency, likely because it accounted for the different transformations applied to each group of parameters. This procedure allowed for all the mean function parameters, the common-mode height scale $h_\mathrm{CM}$ and length scale in time $l_{t}$ to be varied within a single blocked Gibbs step of No U-Turn Sampling (112 parameters for the 600RI data). A simpler procedure later identified would have been to take the Laplace approximation with respect to the log-posterior calculated from the transformed parameters which \textsc{PyMC} uses. However, as this only affects the choice of mass matrix to tune NUTS, it should not impact the validity of our results but has allowed us to improve sampling efficiency and reduce runtimes.
    
    Similar to what was discussed in Appendix~\ref{app:sim MCMC}, for parameters close to the prior limit it was often more efficient to sample them independently using slice sampling rather than use NUTS. It was found to be more efficient for the wavelength-length scale $l_{\lambda}$ and for some of the wavelength-specific systematic height scales $h_\mathrm{WSS; \lambda_i}$. In particular, the $h_\mathrm{WSS; \lambda_i}$ values which were largely consistent with zero and had probabilities that dropped off sharply for larger values could be efficiently sampled with NUTS and $h_\mathrm{WSS; \lambda_i}$ values that were constrained to be larger than zero and had approximately Gaussian distributions were also typically sampled well with NUTS. However, parameters that were both consistent with zero but also had significant probability mass larger than zero appeared to have long correlation lengths with NUTS and were sampled more efficiently with slice sampling (e.g. see the posterior of $h_\mathrm{WSS; 5743\text{\AA}}$ in Fig.~\ref{fig:corner_blue}). Since it was difficult before running an MCMC to distinguish between the parameters that were constrained to be greater than zero and the parameters that were better sampled with slice sampling, all of the larger parameters were fit with slice sampling. The threshold was chosen that if the best-fit value $h_\mathrm{WSS; \lambda_i} > 5 \times 10^{-5}$ then it would be fit with slice sampling. This is not a particularly efficient method but avoided handpicking particular terms and was still faster to converge than using NUTS for these parameters. For larger datasets - such as from JWST - a more efficient approach may need to be found if a similar kernel function is used.
    
    The remaining $h_\mathrm{WSS; \lambda_i}$ parameters that were below the $5 \times 10^{-5}$ threshold were all fit together with a single blocked Gibbs step of NUTS. However, using a mass matrix calculated from a Laplace approximation was found to be slow for these parameters and so the mass-matrix was automatically adapted by \textsc{PyMC} during the burn-in phase for these parameters.
    
    Finally, the white noise parameters and high-frequency systematics parameters were sampled together in a separate blocked Gibbs sampling step with a mass-matrix taken from the Laplace approximation. Table~\ref{tab:VLT MCMC} summarises which parameters were being sampled in which blocked Gibbs sampling steps. For the other 2D GP analyses performed, any of the mean function parameters were always varied in the first blocked Gibbs sampling step with NUTS (such as $T_0$, $a/R_\mathrm{*}$, $b$ or the atmospheric parameters in the direct retrieval).
    
    \begin{table}
    	\centering
    	\caption{Overview of MCMC blocked Gibbs sampling steps for all 2D GP analyses of both VLT/FORS2 datasets.}
    	\label{tab:VLT MCMC}
    	\begin{tabular}{ll} 
    		\hline
    		Parameter & Step\\
    		\hline
            All Mean Function Parameters & NUTS 1\\
            $h_\mathrm{CM}$: CM Height Scale & NUTS 1\\
            $l_{t}$: Time length scale (days) & NUTS 1\\
            $l_\mathrm{\lambda_\mathrm{CM}}$: CM Wavelength length scale ($\angstrom$) & Slice\\
            $h_\mathrm{WSS; \lambda_i}$: WSS Height Scales ($>5\times10^{-5}$) & Slice\\
            $h_\mathrm{WSS; \lambda_i}$: WSS Height Scales ($<5\times10^{-5}$) & NUTS 2\\
            $\sigma$: White noise amplitude & NUTS 3\\
            $h_\mathrm{HFS}$: HFS Height Scale & NUTS 3\\
            $l_\mathrm{\lambda_\mathrm{HFS}}$: HFS Wavelength Length Scale & NUTS 3\\
    		\hline
    	\end{tabular}
    \end{table}

    \section{Examining the choice of kernel for the VLT/FORS2 data using autocorrelation}
    \label{app:autocorrelation}

    \begin{figure*}
    	\includegraphics[width=\textwidth]{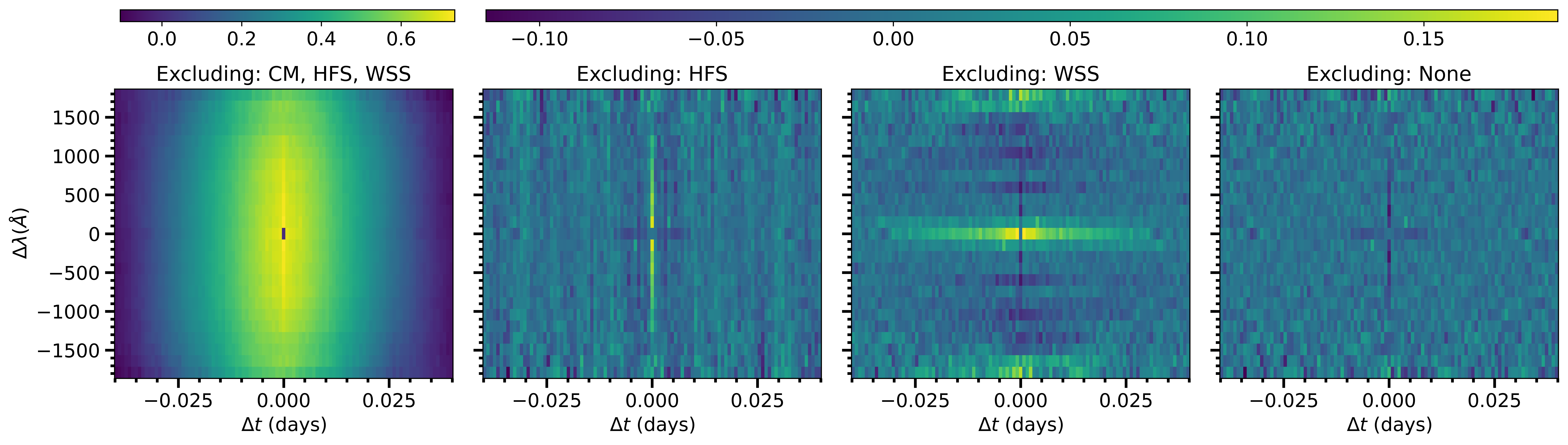}
        \caption{Autocorrelation plots of residuals minus GP mean if (from left to right): no correlated noise is fit, no high-frequency systematics are fit, no wavelength-specific systematics are fit, or if the full kernel in Eq.~(\ref{eq:VLT kernel fn}) is fit.}
        \label{fig:VLT autocorr}
    \end{figure*}

    A useful approach to try visualise correlated noise in the VLT/FORS2 data is to calculate the autocorrelation of the residuals. The autocorrelation of a signal is the correlation of a signal with an offset copy of itself and it can be used to identify correlations in a signal. In this case, we want to identify correlations in the noise of the data so our signal would be the flux observations minus our best fit mean function. We consider correlations in two dimensions - both in time and wavelength.
    
    The left-most plot in Fig.~\ref{fig:VLT autocorr} shows the autocorrelation of the data minus the best-fit transit mean function. This should approximately look like a heat map of the kernel function. Note, the autocorrelation at zero offset has been set to zero for all these plots to increase the contrast for the rest of the values.
    
    The second plot shows the autocorrelation of the residuals after subtraction of the best-fit GP mean if we set $h_\mathrm{HFS} = 0$. This visualises the approximate correlation in the high-frequency systematics - which appears as a vertical line. The correlation appears to reduce at increasing wavelength separation, suggesting a finite wavelength length scale. The third plot is similar to the second but instead if all the $h_\mathrm{WSS; \lambda_i}$ terms are set to zero. If our assumption that these systematics are fully wavelength-independent is true then we should only expect to find correlations at zero wavelength separation (i.e. a horizontal line in the middle of the plot) with other wavelength separations being independent from each other. 
    
    In the right-most plot, we see the autocorrelation of the residuals after subtraction of the best-fit GP mean including all the terms in our kernel function. The resulting plot appears to show almost no correlation with the exception of an apparent purple cross in the middle of the plot, implying some residual correlation from both high-frequency (wavelength-correlated but time-independent) systematics and wavelength-specific (wavelength-independent but time-correlated) systematics. This could suggest that the squared-exponential kernel function used for these systematics was not the ideal shape to account for them or perhaps that the best-fit did not converge to the optimal values. It could also be due to the restrictions on the kernel made such as the wavelength-specific systematics being forced to share the same length scale in time as the common-mode systematics. However, the remaining correlation is low in magnitude compared to the other autocorrelation plots so it is possible any remaining correlation has a negligible impact on the resulting transmission spectrum.

    \section{Tables of results for Simulations 1-4}
    \label{app:sim_results}

    The accuracy metrics from Sect.~\ref{sec:metrics} calculated for Simulations 1-4 are included in this section with Table~\ref{tab:example_table} explaining how to read and interpret each table. An interpretation of the results of each table is included in this section for convenience but repeats many of the points in Sect.~\ref{sec:3}.

    \begin{table*}
        \centering
        \caption{Explanation of the tables that report the results for each set of simulations.}
        \label{tab:example_table}
        \begin{tabular}{lp{0.82\textwidth}} 
            \hline
            Statistic & Explanation\\
            \hline
            $\chi_\mathrm{r}^2$ K-S Stat. & K-S statistic of the chi-squared values: Are the retrieved transmission spectra consistent within errors to the simulated values? Should be <0.072 to 99\% confidence.\\
            $\bar{\chi}^2_r$ & Mean chi-squared value: Are the transmission spectrum uncertainties accurately sized (=1), too large (<1) or too small (>1)?\\
            $s^2(\chi_\mathrm{r}^2)$ & Sample variance of chi-squared values: Are the transmission spectrum uncertainties distributed accurately ($\approx$0.125) or is it likely there are outliers present (>0.15)?\\
            $\bar{\sigma}_\rho$ & Mean standard deviation of the retrieved radius ratio.\\
            $s^2(Z_\rho)$ & Sample variance of radius ratio Z-scores: Are the radius ratio uncertainties accurate (=1), too large (<1) or too small (>1)?\\
            $\bar{\sigma}_m$ & Mean standard deviation of the slope.\\
            $s^2(Z_m$) & Sample variance of slope Z-scores: Are the slope uncertainties accurate (=0), too large (<1) or too small (>1)?\\
            $\bar{\sigma}_\mathrm{\Delta \rho_\mathrm{K}}$ & Mean standard deviation of the change in radius ratio at the K feature.\\
            $s^2(Z_\mathrm{\Delta \rho_\mathrm{K}})$ & Sample variance of K feature Z-scores: Are the uncertainties in the change of radius at the K feature accurate (=1), too large (<1) or too small (>1)?\\
            \hline
        \end{tabular}
    \end{table*}

    Table~\ref{tab:sim1} presents the results of Simulation 1 containing wavelength-independent systematics tested on three methods. Firstly, 1D GPs fitting each spectroscopic light curve separately were tested but did not produce reliable constraints as the $\chi_r^2$ mean and the variance of the Z-score for each atmospheric parameter are all larger than one. This suggests that the 1D GPs underestimate uncertainty on average despite having the correct kernel to fit these systematics. This can be explained by having a very flexible noise model that fits for a separate height scale, length scale and white noise amplitude for each spectroscopic light curve. The Hybrid method shares the same hyperparameters for all light curves and performs reliably across all accuracy metrics measured despite using the same kernel function as the 1D GPs. The 2D GP method is similar to the Hybrid method but fits for a wavelength length scale - which is a more conservative approach as it does not assume systematics are wavelength-independent {\it a priori}. This still results in reliable retrievals of all parameters measured and with no statistically significant increase in the mean uncertainties on the parameters, so the more conservative analysis does not result in a significant loss in precision (although there was a minor loss in precision on the transmission spectrum, as noted in Sect.~\ref{sec:indep_systematics}.
    
    \begin{table}
    	\centering
    	\caption{Results of Simulation 1 containing wavelength-independent systematics.}
    	\label{tab:sim1}
    	\begin{tabular}{llll} 
    		\hline
    		Method & 1D GP & Hybrid & 2D GP\\
    		\hline
            $\chi_\mathrm{r}^2$ K-S Stat. & 0.209 & 0.034 & 0.031\\
    $\bar{\chi}^2_r$ & 1.35 $\pm$ 0.04 & 1.00 $\pm$ 0.02 & 1.00 $\pm$ 0.02\\
    $s^2(\chi_\mathrm{r}^2)$ & 0.751 & 0.149 & 0.147\\
    $\bar{\sigma}_\rho$ ($\times10^{6}$) & 391 $\pm$ 8 & 405 $\pm$ 8 & 414 $\pm$ 8\\
    $s^2(Z_\rho$) & 1.85 $\pm$ 0.12 & 0.99 $\pm$ 0.06 & 0.94 $\pm$ 0.06\\
    $\bar{\sigma}_m$ ($\times10^{6}$) & 1720 $\pm$ 35 & 1770 $\pm$ 36 & 1800 $\pm$ 37\\
    $s^2(Z_m$) & 1.81 $\pm$ 0.11 & 0.96 $\pm$ 0.06 & 0.93 $\pm$ 0.06\\
    $\bar{\sigma}_\mathrm{\Delta \rho_\mathrm{K}}$ ($\times10^{6}$) & 1720 $\pm$ 40 & 1620 $\pm$ 32 & 1620 $\pm$ 33\\
    $s^2(Z_\mathrm{\Delta \rho_\mathrm{K}})$ & 1.30 $\pm$ 0.08 & 1.03 $\pm$ 0.07 & 1.04 $\pm$ 0.07\\
    		\hline
    	\end{tabular}
    \end{table}

    Table~\ref{tab:sim2} shows the results of Simulation 2 containing wavelength-correlated systematics with a relatively short wavelength length scale ($300\angstrom < l_{\lambda} < 2250\angstrom$). The same three methods are run for Simulation 2 as for Simulation 1, to demonstrate that the 1D GP and Hybrid methods that assume wavelength-independent systematics perform very unreliably when systematics are correlated in wavelength. They overestimate the uncertainty in the radius ratio and slope (Z-score variances > 1) and underestimate uncertainty in the strength of the K feature (Z-score variance < 1). The K-S statistics show that the $\chi_r^2$ distribution of the samples does not match the correct $\chi_r^2$ distribution, showing that the Hybrid method does not reliably retrieve the transmission spectrum even though the mean chi-squared value is consistent with one. The 2D GP method performs reliably across all accuracy metrics - showing that our method is robust at accounting for wavelength-correlated systematics within this range of wavelength length scales (although the sample variance of the $\chi_r^2$ values is slightly too high which could suggest a few outliers being present). The uncertainty in the strength of the K feature is both more reliable and also much smaller than the 1D GP and Hybrid methods, lending support to our result from the VLT/FORS2 analyses which produced tighter constraints on Na and K in all 2D GP analyses compared to the 1D GP analyses.

    \begin{table}
    	\centering
    	\caption{Results of Simulation 2 with short wavelength length scale ($300\angstrom < l_{\lambda} < 2250\angstrom$) systematics.}
    	\label{tab:sim2}
    	\begin{tabular}{llll} 
    		\hline
    		Method & 1D GP & Hybrid & 2D GP\\
    		\hline
            $\chi_\mathrm{r}^2$ K-S Stat. & 0.201 & 0.169 & 0.041\\
    $\bar{\chi}^2_r$ & 1.31 $\pm$ 0.05 & 1.04 $\pm$ 0.04 & 1.00 $\pm$ 0.02\\
    $s^2(\chi_\mathrm{r}^2)$ & 1.44 & 0.688 & 0.156\\
    $\bar{\sigma}_\rho$ ($\times10^{6}$) & 405 $\pm$ 8 & 406 $\pm$ 8 & 713 $\pm$ 18\\
    $s^2(Z_\rho$) & 5.77 $\pm$ 0.37 & 4.74 $\pm$ 0.30 & 1.01 $\pm$ 0.06\\
    $\bar{\sigma}_m$ ($\times10^{6}$) & 1780 $\pm$ 37 & 1760 $\pm$ 35 & 2400 $\pm$ 54\\
    $s^2(Z_m$) & 3.62 $\pm$ 0.23 & 2.62 $\pm$ 0.17 & 1.09 $\pm$ 0.07\\
    $\bar{\sigma}_\mathrm{\Delta \rho_\mathrm{K}}$ ($\times10^{6}$) & 1710 $\pm$ 37 & 1600 $\pm$ 31 & 465 $\pm$ 14\\
    $s^2(Z_\mathrm{\Delta \rho_\mathrm{K}})$ & 0.59 $\pm$ 0.04 & 0.55 $\pm$ 0.03 & 1.02 $\pm$ 0.06\\
    		\hline
    	\end{tabular}
    \end{table}

    Table~\ref{tab:sim3} shows the results of Simulation 3 containing longer wavelength length scale systematics ($2250\angstrom < l_{\lambda} < 36000\angstrom$). While the 2D GP method was identical to Simulations 1 and 2, in these simulations we were investigating how the assumption of systematics being constant in wavelength performs when the systematics are actually gradually varying in wavelength. We tested our 2D GP method against the 1D GP method with an initial common-mode (CM) correction as well as the same 2D GP method but with the wavelength length scale fixed to a large value which is close to assuming systematics are common-mode (labelled as 2D GP (CM) in the table although note this method does not perform a common-mode correction). Both the 1D GP (CM) and 2D GP (CM) methods were unreliable across almost all metrics whereas the 2D GP that fit for the wavelength length scale produced robust retrievals across all metrics (the variance in the slope Z-scores may be slightly too high although this is only a $2.6\sigma$ result). The asterisk for the 1D GP (CM) $\chi_r^2$ values is there to note that these values were calculated after subtracting the offset in the recovered transmission spectrum to the injected spectrum because this offset does not have a large effect on atmospheric retrievals but was significantly affecting the $\chi_r^2$ values. This offset is the result of the common-mode correction and so did not affect the other two methods which do not have this offset correction performed for the $\chi_r^2$ values.
    
    \begin{table}
    	\centering
    	\caption{Results of Simulation 3 containing long wavelength length scale ($2250\angstrom < l_{\lambda} < 36000\angstrom$) systematics.}
    	\label{tab:sim3}
    	\begin{tabular}{llll} 
    		\hline
    		Method & 1D GP (CM) & 2D GP (CM) & 2D GP\\
    		\hline
    $\chi^2_r$ K-S Stat. & 0.152* & 0.344 & 0.051\\
    $\bar{\chi}^2_r$ & 1.24 $\pm$ 0.06* & 2.07 $\pm$ 0.13 & 1.03 $\pm$ 0.02\\
    $s^2(\chi^2_r)$ & 1.72* & 8.69 & 0.160\\
    $\bar{\sigma}_\rho$ ($\times10^{6}$) & 122 $\pm$ 3 & 1317 $\pm$ 56 & 994 $\pm$ 31\\
    $s^2(Z_\rho$) & 163 $\pm$ 10 & 2.12 $\pm$ 0.13 & 1.02 $\pm$ 0.06\\
    $\bar{\sigma}_m$ ($\times10^{6}$) & 660 $\pm$ 18 & 639 $\pm$ 17 & 1241 $\pm$ 40\\
    $s^2(Z_m$) & 8.35 $\pm$ 0.53 & 7.14 $\pm$ 0.45 & 1.18 $\pm$ 0.07\\
    $\bar{\sigma}_{\Delta \rho_K}$ ($\times10^{6}$) & 577 $\pm$ 15 & 358 $\pm$ 11 & 348 $\pm$ 11\\
    $s^2(Z_{\Delta \rho_K})$ & 0.58 $\pm$ 0.04 & 1.11 $\pm$ 0.07 & 0.97 $\pm$ 0.06\\
    		\hline
    	\end{tabular}
    \end{table}
    
    \begin{table}
    	\centering
    	\caption{Results of Simulation 4 containing common-mode systematics.}
    	\label{tab:sim4}
    	\begin{tabular}{llll} 
    		\hline
    		Method & 1D GP (CM) & 2D GP (CM) & 2D GP\\
    		\hline
            $\chi_\mathrm{r}^2$ K-S Stat. & 0.236* & 0.045 & 0.032\\
    $\bar{\chi}^2_r$ & 0.81 $\pm$ 0.01* & 1.02 $\pm$ 0.02 & 1.00 $\pm$ 0.02\\
    $s^2(\chi_\mathrm{r}^2)$ & 0.101* & 0.141 & 0.132\\
    $\bar{\sigma}_\rho$ ($\times10^{6}$) & 93 $\pm$ 3 & 903 $\pm$ 28 & 909 $\pm$ 28\\
    $s^2(Z_\rho$) & 269 $\pm$ 17 & 1.02 $\pm$ 0.06 & 1.02 $\pm$ 0.06\\
    $\bar{\sigma}_m$ ($\times10^{6}$) & 406 $\pm$ 12 & 393 $\pm$ 11 & 473 $\pm$ 13\\
    $s^2(Z_m$) & 0.99 $\pm$ 0.06 & 0.89 $\pm$ 0.06 & 0.67 $\pm$ 0.04\\
    $\bar{\sigma}_\mathrm{\Delta \rho_\mathrm{K}}$ ($\times10^{6}$) & 387 $\pm$ 12 & 336 $\pm$ 10 & 337 $\pm$ 10\\
    $s^2(Z_\mathrm{\Delta \rho_\mathrm{K}})$ & 0.84 $\pm$ 0.05 & 1.08 $\pm$ 0.07 & 1.08 $\pm$ 0.07\\
    		\hline
    	\end{tabular}
    \end{table}

    Table~\ref{tab:sim4} shows the results of Simulation 4 that contains common-mode systematics. The same methods as for Simulation 3 were tested, although the assumption that the systematics are common-mode is correct for these simulations so both the 1D GP (CM) and 2D GP (CM) methods perform much more reliably. The 1D GP (CM) method is slightly more conservative compared to the 2D GP (CM) method on both the constraint of the slope and the strength of the K feature, this is may be because the 1D GPs are still accounting for wavelength-independent systematics in each spectroscopic light curve while the 2D GP (CM) method assumes that only common-systematics and white noise are present in the data. The 2D GP method that fits for a wavelength length scale performs reliably on all metrics except the constraint of the slope, where it overestimates the uncertainties on average. This is likely due to the challenge of constraining systematics to be fully common-mode as the 2D GP (CM) method is much better at constraining the slope. However, since we do not know for real data whether systematics are constant in wavelength or gradually varying in wavelength, the safest approach is to fit for the wavelength length scale and accept some loss in precision in the slope of the transmission spectrum if the systematics are common-mode.

    \section{Gradient calculations}
    \label{app:gradients}
    
    No U-Turn Sampling requires the computation of the gradient of the log-likelihood with respect to each parameter being varied. One of the benefits of \textsc{JAX} is that it can compute the gradient of functions with limited modification to \textsc{NumPy} code. Using \textsc{jaxoplanet} in combination with implementing the algorithms in Sect.~\ref{sec:rakitsch} in \textsc{JAX} was sufficient to efficiently and accurately compute the gradient of the log-likelihood with respect to any of the mean function parameters.
    
    However, these algorithms were not sufficient for computing the gradients with respect to many of the hyperparameters. It was noticed that the gradients \textsc{JAX} computed from these equations were inconsistent with finite difference methods. It is likely that the numerical stability of eigendecomposition was an issue when computing the gradients of hyperparameters in these formula.
    
    Fortunately, it is possible to rewrite the gradient of the log-likelihood in a way that is much less sensitive to numerical stability issues compared to the default method used by \textsc{JAX}. A method for analytically calculating the gradient of the log-likelihood with respect to hyperparameters was demonstrated in \cite{Rakitsch2013}. The method is included here, although the calculation of the derivative of the log-determinant of the covariance matrix has been changed. The method presented in \cite{Rakitsch2013} would require additional eigendecompositions of the matrices $\mathbf{K}_{{\lambda}}$, $\mathbf{K}_{{t}}$ and the transformed matrices $\mathbf{\tilde{\Sigma}}_{{\lambda}}$ and $\mathbf{\tilde{\Sigma}}_{{t}}$ to be calculated in order to calculate the gradient with respect to all hyperparameters. It also would have no way of getting the gradient of a hyperparameter which is included in both Kronecker product terms such as in $\mathbf{K}_{{\lambda}}$ and $\mathbf{\Sigma}_{{\lambda}}$. The method presented here is therefore faster, more general and is also more numerically stable.
    
    Suppose $\theta$ is a hyperparameter, with $\vec{r}$ independent of $\theta$ (i.e. $\frac{\partial\vec{r}}{\partial \theta} = 0$). We can find $\frac{\partial(\log L)}{\partial \theta}$ by calculating:
    \begin{align}
    \frac{\partial(\log L)}{\partial \theta} &= \frac{\partial}{\partial \theta} \left[-\frac{1}{2} \vec{r}^T \mathbf{K}^{-1} \vec{r} -\frac{1}{2} \log |\mathbf{K}| - \frac{N}{2} \log (2 \pi) \right] \\
    &= -\frac{1}{2} \vec{r}^T \frac{\partial \mathbf{K}^{-1}}{\partial \theta} \vec{r} -\frac{1}{2} \frac{\partial \log |\mathbf{K}|}{\partial \theta}.
    \label{eq:grad_calc}
    \end{align}
    where we can use the following two identities from matrix calculus:
    \begin{align}
    \frac{\partial \mathbf{K}^{-1}}{\partial \theta} = - \mathbf{K}^{-1} \frac{\partial \mathbf{K}}{\partial \theta} \mathbf{K}^{-1}, \label{eq:K_inv_derivative} \\ 
    \frac{\partial \log |\mathbf{K}|}{\partial \theta} = \text{Tr}\left(\mathbf{K}^{-1} \frac{\partial \mathbf{K}}{\partial \theta}\right). \label{eq:trace}
    \end{align}
    The advantage of these identities is that they allow us to rewrite the gradient of the log-likelihood in terms of $\frac{\partial \mathbf{K}}{\partial \theta}$ instead of $\frac{\partial \mathbf{K}^{-1}}{\partial \theta}$ or $\frac{\partial |\mathbf{K}|}{\partial \theta}$ (which by default \textsc{JAX} fails to calculate in a numerically stable way). It was found that this approach results in gradients of the log-likelihood that are consistent with finite-difference methods.
    
    Suppose $\theta$ is a parameter of $\mathbf{K}_{\lambda}$ and none of the other matrices in $\mathbf{K}$, then:
    \begin{align}
    \frac{\partial \mathbf{K}}{\partial \theta} &= \frac{\partial}{\partial \theta} \left[\mathbf{K}_{\lambda} \otimes \mathbf{K}_{t} + \mathbf{\Sigma}_{\lambda} \otimes \mathbf{\Sigma}_{t} \right] \\
    &= \frac{\partial \mathbf{K}_{\lambda}}{\partial \theta} \otimes \mathbf{K}_{t}.
    \label{eq:kronsum_derivative}
    \end{align}
    
    This allows us to efficiently calculate the first term in Eq.~(\ref{eq:grad_calc}) using the identity in Eq.~(\ref{eq:K_inv_derivative}):
    \begin{align}
    -\frac{1}{2} \vec{r}^T \frac{\partial \mathbf{K}^{-1}}{\partial \theta} \vec{r} &= \frac{1}{2} \vec{r}^T \mathbf{K}^{-1} \left[ \frac{\partial \mathbf{K}_{\lambda}}{\partial \theta} \otimes \mathbf{K}_{t} \right] \mathbf{K}^{-1} \vec{r} \\
    &= \frac{1}{2} \vec{\alpha}^T \left[ \frac{\partial \mathbf{K}_{\lambda}}{\partial \theta} \otimes \mathbf{K}_{t} \right] \vec{\alpha}.
    \label{eq:rTr_grad}
    \end{align}
    Where $\mathbf{K}^{-1} \vec{r} = \vec{\alpha}$ can be efficiently solved using Eq.~(\ref{eq:K_inv_R_rakitsch}) and $\frac{\partial \mathbf{K}}{\partial \theta} \vec{\alpha}$ can be solved using Eq.~(\ref{eq:Algo14}) as it is of the form $[\mathbf{A} \otimes \mathbf{B}] \vec{c}$
    
    The second term in Eq.~(\ref{eq:grad_calc}) may be computed as follows. We will use the identity in Eq.~(\ref{eq:trace}), but first it is helpful to expand out the $\mathbf{K}^{-1}$ term using the factorisation described in Sect.~\ref{sec:rakitsch}:
    \begin{align}
        \mathbf{K}^{-1} &= [\mathbf{Q}_{\mathbf{\Sigma}_{\lambda}} \mathbf{\Lambda}_{\mathbf{\Sigma}_{\lambda}}^{-\frac{1}{2}} \otimes \mathbf{Q}_{\mathbf{\Sigma}_{t}} \mathbf{\Lambda}_{\mathbf{\Sigma}_{t}}^{-\frac{1}{2}}] [\tilde{\mathbf{K}}_{\lambda} \otimes \tilde{\mathbf{K}}_{t} + \mathbb{I}]^{-1} \nonumber \\
        &\times [\mathbf{\Lambda}_{\mathbf{\Sigma}_{\lambda}}^{-\frac{1}{2}} \mathbf{Q}_{\mathbf{\Sigma}_{\lambda}}^T \otimes \mathbf{\Lambda}_{\mathbf{\Sigma}_{t}}^{-\frac{1}{2}} \mathbf{Q}_{\mathbf{\Sigma}_{t}}^T].
    \end{align}
    We then factorise the middle term similar to Eq.~(\ref{eq:K_sph_inv}) as:
    \begin{equation}
    [\tilde{\mathbf{K}}_{\lambda} \otimes \tilde{\mathbf{K}}_{t} + \mathbb{I}]^{-1} = [\mathbf{Q_\mathrm{\tilde{K}_{\lambda}}} \otimes \mathbf{Q_\mathrm{\tilde{K}_{t}}}] [\mathbf{\Lambda_\mathrm{\tilde{K}_{\lambda}}} \otimes \mathbf{\Lambda_\mathrm{\tilde{K}_{t}}} + \mathbb{I}]^{-1} [\mathbf{Q^T_\mathrm{\tilde{K}_{\lambda}}} \otimes \mathbf{Q^T_\mathrm{\tilde{K}_{t}}}].
    \end{equation}
    
    Plugging this in and using the properties of Kronecker product algebra:
    \begin{align}
        \mathbf{K}^{-1} &= [\mathbf{Q}_{\mathbf{\Sigma}_{\lambda}} \mathbf{\Lambda}_{\mathbf{\Sigma}_{\lambda}}^{-\frac{1}{2}} \mathbf{Q_\mathrm{\tilde{K}_{\lambda}}} \otimes \mathbf{Q}_{\mathbf{\Sigma}_{t}} \mathbf{\Lambda}_{\mathbf{\Sigma}_{t}}^{-\frac{1}{2}} \mathbf{Q_\mathrm{\tilde{K}_{t}}}] [\mathbf{\Lambda_\mathrm{\tilde{K}_{\lambda}}} \otimes \mathbf{\Lambda_\mathrm{\tilde{K}_{t}}} + \mathbb{I}]^{-1} \nonumber \\
        &\times [\mathbf{Q^T_\mathrm{\tilde{K}_{\lambda}}} \mathbf{\Lambda}_{\mathbf{\Sigma}_{\lambda}}^{-\frac{1}{2}} \mathbf{Q}_{\mathbf{\Sigma}_{\lambda}}^T \otimes \mathbf{Q^T_\mathrm{\tilde{K}_{t}}} \mathbf{\Lambda}_{\mathbf{\Sigma}_{t}}^{-\frac{1}{2}} \mathbf{Q}_{\mathbf{\Sigma}_{t}}^T] \nonumber \\
        &= [{\mathbf{W}}_{\lambda} \otimes \mathbf{W}_{t}] \mathbf{D}^{-1} [\mathbf{W^T_{\lambda}} \otimes \mathbf{W^T_{t}}],
    \end{align}
    where we have defined:
    \begin{align}
        &\mathbf{W}_{\lambda} = \mathbf{Q}_{\mathbf{\Sigma}_{\lambda}} \mathbf{\Lambda}_{\mathbf{\Sigma}_{\lambda}}^{-\frac{1}{2}} \mathbf{Q_\mathrm{\tilde{K}_{\lambda}}}, \\
        &\mathbf{W}_{t} = \mathbf{Q}_{\mathbf{\Sigma}_{t}} \mathbf{\Lambda}_{\mathbf{\Sigma}_{t}}^{-\frac{1}{2}} \mathbf{Q_\mathrm{\tilde{K}_{t}}},\\
        &\mathbf{D} = \mathbf{\Lambda_\mathrm{\tilde{K}_{\lambda}}} \otimes \mathbf{\Lambda_\mathrm{\tilde{K}_{t}}} + \mathbb{I}.
    \end{align}
    We note that while it may look as if we have found the eigendecomposition of the covariance matrix $\mathbf{K}$, the matrices $\mathbf{W}_{\lambda}$ and $\mathbf{W}_{t}$ can be clearly shown to not be eigenvector matrices as they are not orthogonal (i.e. the product $\mathbf{W}_{\lambda}^T \mathbf{W}_{\lambda}$ does not equal the identity matrix unlike for an actual eigendecomposition). Similarly, the diagonal matrix $\mathbf{D}$ does not contain the eigenvalues of $\mathbf{K}$ as can be seen by noting that the product of the diagonal entries would not in general produce the same determinant as given in Eq.~(\ref{eq:logdet2}). Nonetheless, it is a useful way of writing $\mathbf{K}^{-1}$ which we can now plug into Eq.~(\ref{eq:trace}):
    \begin{align}
    \frac{\partial \log |\mathbf{K}|}{\partial \theta} &= \text{Tr}\left([{\mathbf{W}}_{\lambda} \otimes \mathbf{W}_{t}] \mathbf{D}^{-1} [\mathbf{W^T_{\lambda}} \otimes \mathbf{W^T_{t}}] \frac{\partial \mathbf{K}}{\partial \theta}\right) \nonumber \\
    &= \text{Tr}\left([{\mathbf{W}}_{\lambda} \otimes \mathbf{W}_{t}] \mathbf{D}^{-1} [\mathbf{W^T_{\lambda}} \otimes \mathbf{W^T_{t}}] \left[\frac{\partial \mathbf{K}_{\lambda}}{\partial \theta} \otimes \mathbf{K}_{t}\right]\right).
    \end{align}
    
    We can use the cyclic property of the trace:
    \begin{equation}
        \text{Tr}(\mathbf{ABCD}) = \text{Tr}(\mathbf{BCDA}) = \text{Tr}(\mathbf{CDAB}) = \text{Tr}(\mathbf{DABC}).
    \end{equation}
    This allows us to simplify further:
    \begin{align}
    \frac{\partial \log |\mathbf{K}|}{\partial \theta} &= \text{Tr}\left(\mathbf{D}^{-1} [\mathbf{W^T_{\lambda}} \otimes \mathbf{W^T_{t}}] \left[\frac{\partial \mathbf{K}_{\lambda}}{\partial \theta} \otimes \mathbf{K}_{t}\right] [{\mathbf{W}}_{\lambda} \otimes \mathbf{W}_{t}] \right) \nonumber \\
    &= \text{Tr}\left(\mathbf{D}^{-1} \left[\mathbf{W^T_{\lambda}} \frac{\partial \mathbf{K}_{\lambda}}{\partial \theta} {\mathbf{W}}_{\lambda} \otimes \mathbf{W^T_{t}} \mathbf{K}_{t} \mathbf{W}_{t}\right]\right).
    \end{align}

    Finally, $\mathbf{D}^{-1}$ is a diagonal matrix and the trace only depends on the diagonal of the overall matrix so we can equivalently write this expression as a dot product of the diagonal of the two matrices and make use of the fact that the diagonal of a Kronecker product matrix is the Kronecker product of the diagonals:
    \begin{align}
    &\frac{\partial \log |\mathbf{K}|}{\partial \theta} = \text{diag}(\mathbf{D}^{-1})^T  \text{diag}\left(\mathbf{W^T_{\lambda}} \frac{\partial \mathbf{K}_{\lambda}}{\partial \theta} {\mathbf{W}}_{\lambda} \otimes \mathbf{W^T_{t}} \mathbf{K}_{t} \mathbf{W}_{t}\right) \nonumber \\
    &= \text{diag}(\mathbf{D}^{-1})^T \left[\text{diag}\left(\mathbf{W^T_{\lambda}} \frac{\partial \mathbf{K}_{\lambda}}{\partial \theta} {\mathbf{W}}_{\lambda}\right) \otimes \text{diag}\left(\mathbf{W^T_{t}} \mathbf{K}_{t} \mathbf{W}_{t}\right)\right],
    \label{eq:logdet_grad}
    \end{align}
    where diag$(\mathbf{A})$ denotes the vector formed from the diagonal of some matrix $\mathbf{A}$.
    
    Equations~\ref{eq:rTr_grad} and ~\ref{eq:logdet_grad} permit the efficient calculation of the gradient of the log-likelihood with respect to a hyperparameter of $\mathbf{K}_{\lambda}$. By taking advantage of the Kronecker product structure, the memory requirement is still $\mathcal{O}(M^2 + N^2)$ and we do not require the calculation of any additional eigendecompositions compared to the log-likelihood calculation. The mathematics follows similarly for calculating the gradients of parameters from $\mathbf{K}_{t}$, $\mathbf{\Sigma}_{\lambda}$ or $\mathbf{\Sigma}_{t}$.
    
    \section{Hessian calculations}
    \label{app:hessian}
    
    Similar calculations can be performed as in Appendix~\ref{app:gradients} in order to efficiently calculate the Hessian of the log-likelihood with respect to all of the mean function parameters and hyperparameters. This is implemented in the \textsc{luas} Github repository but is excluded here for brevity.
    
    The benefit of computing the Hessian of the log-likelihood is that we can use it to calculate a Laplace approximation of the posterior. This is an analytic way of estimating the posterior using the best-fit value of all the parameters and the Hessian at the location of best-fit. It works by taking a second-order Taylor expansion of the log-posterior around the mode of the posterior (see \citealt{Bishop} for more information about the Laplace approximation). It assumes that the posterior can be approximated as a Gaussian and works by determining what the posterior would need to be in order to have the same Hessian at the mean of the Gaussian. It was found that this approximation produces an accurate enough fit to the posterior that using it to calculate the mass-matrix of the No U-Turn Sampler significantly improved the efficiency of sampling.
    
    While the Laplace approximation is useful for tuning No U-Turn Sampling, it can also be seen as a very computationally inexpensive way of approximating the posterior (taking $<10$ seconds for the VLT/FORS2 datasets) which avoids using expensive inference techniques such as MCMC. While the results are not as robust as a full MCMC retrieval, by making this approximation it may permit the application of 2D GPs to very large datasets which could otherwise be unfeasible using a full exploration of the posterior. It is also a useful check before running an MCMC (which may take hours to run) to see if the Laplace approximation produces a reasonable fit to the data.
    
    \section{Numerical stability with eigendecomposition}
    \label{app:stability}
    
    The numerical stability of eigendecomposition is dependent on the condition number of the matrix in question. For any covariance matrix, the condition number is given by the ratio of the largest to smallest eigenvalue of that matrix. Matrix inversion is generally more prone to numerical errors for matrices with large condition numbers.
    
    This is often not a major issue for 1D GPs because it is common to have terms in the kernel that are added to the diagonal, such as white noise terms. Adding to the diagonal generally has the effect of reducing the condition number. For example, if a constant $c$ is added to all of the diagonal elements of a covariance matrix then it has the effect of increasing all the eigenvalues by $c$. This always decreases the ratio of the largest to smallest eigenvalues. However, due to the way this method splits up the covariance matrix into different sums of Kronecker products, the component wavelength and time covariance matrices may not all have terms added to their diagonals. For the kernel function used to fit the VLT/FORS2 data (shown in Eq.~\ref{eq:VLT kernel fn}), the matrix accounting for time-correlated systematics $\mathbf{K}_{t}$ is calculated from a squared-exponential kernel without any terms added to the diagonal, potentially resulting in a high condition number which may produce significant numerical errors.
    
    The approach used to deal with these numerical errors was to use a method that may be referred to as Ridge regression or Tikhonov regularisation \citep{Ridge}. A constant was added to the diagonal elements of this time covariance matrix $\mathbf{K}_{t}$ in order to increase all the eigenvalues, often referred to as regularising the matrix. The value of the constant used was likely larger than was necessary to guarantee stability as it was unclear how to determine an optimal value. The diagonal values - which were all equal to one - were increased by $10^{-5}$. To understand the effect this could have had, we can treat it similar to any of the other terms in the kernel and consider what type of noise it accounts for. The actual kernel function used for the VLT analysis was effectively:
    \begin{align}
        \mathbf{K}_{ij} = &\left[h_\mathrm{CM}^2 \exp\left(-\frac{|\lambda_i - \lambda_j|^2}{2 l_\mathrm{\lambda_\mathrm{CM}}^2}\right) + h_\mathrm{WSS; \lambda_i}^2 \delta_{\lambda_i \lambda_j}\right] \nonumber \\
        \times&\left[\exp\left(-\frac{|t_i - t_j|^2}{2 l_{t}^2}\right) + c\delta_{t_i t_j}\right] \nonumber \\
        +&\left[h_\mathrm{HFS}^2 \exp\left(-\frac{|\lambda_i - \lambda_j|^2}{2 l_\mathrm{\lambda_\mathrm{HFS}}^2}\right) + \sigma_\mathrm{\lambda_i}^2\delta_{\lambda_i \lambda_j}\right] \delta_{t_i t_j}. \label{eq:numerically stable kernel}
    \end{align}
    
    This new term $c\delta_{t_i t_j}$ multiplies both the $h_\mathrm{CM}$ and $h_\mathrm{WSS}$ terms in the wavelength kernel function it is multiplying. We can rearrange this kernel function as:
    
    \begin{align}
        \mathbf{K}_{ij} = &\left[h_\mathrm{CM}^2 \exp\left(-\frac{|\lambda_i - \lambda_j|^2}{2 l_\mathrm{\lambda_\mathrm{CM}}^2}\right) + h_\mathrm{WSS; \lambda_i}^2 \delta_{\lambda_i \lambda_j}\right] \nonumber \\
        \times&\left[\exp\left(-\frac{|t_i - t_j|^2}{2 l_{t}^2}\right)\right] \nonumber \\
        +&\left[h_\mathrm{HFS}^2 \exp\left(-\frac{|\lambda_i - \lambda_j|^2}{2 l_\mathrm{\lambda_\mathrm{HFS}}^2}\right) + c h_\mathrm{CM}^2 \exp\left(-\frac{|\lambda_i - \lambda_j|^2}{2 l_\mathrm{\lambda_\mathrm{CM}}^2}\right) \right. \nonumber \\
        &\left. + (\sigma_\mathrm{\lambda_i}^2 + c h_\mathrm{WSS; \lambda_i}^2)\delta_{\lambda_i \lambda_j}  \vphantom{\exp\left(-\frac{|\lambda_i - \lambda_j|^2}{2 l_\mathrm{\lambda_\mathrm{CM}}^2}\right)}\right] \delta_{t_i t_j}. 
        \label{eq:numerically stable kernel2}
    \end{align}
    
    This rearrangement of the equation shows that by adding the constant $c$ to the diagonal of the first covariance matrix for the time dimension, we can think of it as instead adding a second high-frequency systematics term with height scale $\sqrt{c} h_\mathrm{CM}$ and wavelength length scale $l_\mathrm{\lambda_\mathrm{CM}}$ as well as changing the amplitude of the white noise terms.
    
    Changing the amplitude of the white noise terms should not be an issue since we are freely fitting for each $\sigma_\mathrm{\lambda_i}$ term. The effect of this is simply that the retrieved $\sigma_\mathrm{\lambda_i}$ terms are slightly smaller than the actual level of white noise the kernel is fitting. For the best-fit values of the 600RI and 600B analogous 2D GP fits, the maximum decrease for any white noise value retrieved would be a decrease of $8.8 \times 10^{-9}$ with $c = 10^{-5}$.
    
    The only negative effect regularisation might have had was from this additional high-frequency systematics term with height scale $\sqrt{c} h_\mathrm{CM}$ and with a different wavelength length scale to the actual high-frequency systematics term in the kernel. Suppose the common-mode height scale $h_\mathrm{CM}$ is ten times greater than the high-frequency systematics height scale $h_\mathrm{HFS}$. The additional term in the kernel would have a height scale $\sqrt{c} h_\mathrm{CM}$ which would be 3.2\% of $h_\mathrm{HFS}$. If both the wavelength length scales $l_\mathrm{\lambda_\mathrm{CM}}$ and $l_\mathrm{\lambda_\mathrm{HFS}}$ were identical, then this would result in the MCMC retrieving $h_\mathrm{HFS}$ to be only $\approx0.05\%$ smaller but would have no effect on the retrieved transmission spectrum. Since the two wavelength length scales were different, this essentially resulted in the kernel function used for the high-frequency systematics very slightly deviating from a squared-exponential kernel as it would instead be the sum of two squared-exponential kernels with different length scales. Overall, this should have a very small effect.
    
    For future work, it may be possible to avoid the addition of a constant to the diagonals. This is because to calculate the log-likelihood, the eigenvalues of the $\mathbf{K}_{\lambda}$ and $\mathbf{K}_{t}$ matrices are not used but instead the eigenvalues of $[\tilde{\mathbf{K}}_{\lambda} \otimes \tilde{\mathbf{K}}_{t} + \mathbb{I}]$ are used. This addition of the identity matrix should regularise the matrix. The eigenvalues of the matrices $\mathbf{\Sigma}_{\lambda}$ and $\mathbf{\Sigma}_{t}$ will need to be calculated directly however, but as white noise terms are likely needed in any suitable kernel then these matrices can be chosen to include the white noise terms as this naturally regularises the matrices.

    \section{Extension of method to higher dimensions}
    \label{app:3D_GPs}

    The optimisations developed in this paper could be extended to higher dimensions with limited extra work. For the case of a 3D Gaussian process, given the data lie on a complete 3D grid and the covariance matrix is of the form:
    \begin{equation}\label{eq:kronsum3D}
    \mathbf{K} = \mathbf{K}_{\lambda} \otimes \mathbf{K}_{t} \otimes \mathbf{K}_\mathrm{n} + \mathbf{\Sigma}_{\lambda} \otimes \mathbf{\Sigma}_{t} \otimes \mathbf{\Sigma}_\mathrm{n},
    \end{equation}
    then $\mathbf{K}^{-1}$ could be factorised using a similar method by first taking the eigendecompositions of $\mathbf{\Sigma}_{\lambda}$, $\mathbf{\Sigma}_{t}$ and $\mathbf{\Sigma}_\mathrm{n}$ and using them to transform $\mathbf{K}_{\lambda}$, $\mathbf{K}_{t}$ and $\mathbf{K}_\mathrm{n}$. The eigendecompositions of these transformed matrices could then be calculated and $\mathbf{K}^{-1}$ could be factorised similarly to Eq.~(\ref{eq:K_inv_R_rakitsch}). \citet{Saatchi2011} contains algorithms to efficiently calculate matrix-vector products for any number of dimensions of Kronecker products including calculating $[\mathbf{A} \otimes \mathbf{B} \otimes \mathbf{C}] \vec{d}$ for three dimensions (analogous to Eq.~\ref{eq:Algo14}). Although it was not implemented, any of the algorithms described in this work could conceivably be generalised to higher dimensions using this.

    An example within transmission spectroscopy where this could be useful is to study if the systematics from different transit observations using the same instrument are correlated. In that case, the three dimensions could be wavelength, time, and transit number. Many hyperparameters would need to be shared between different transit observations however so this may not be a reasonable approach for most datasets.

    \section{Visualisations of 2D GP analyses of the VLT/FORS2 data}\label{app:VLT_visualise}
    
    \begin{figure*}
    	\includegraphics[width=\textwidth]{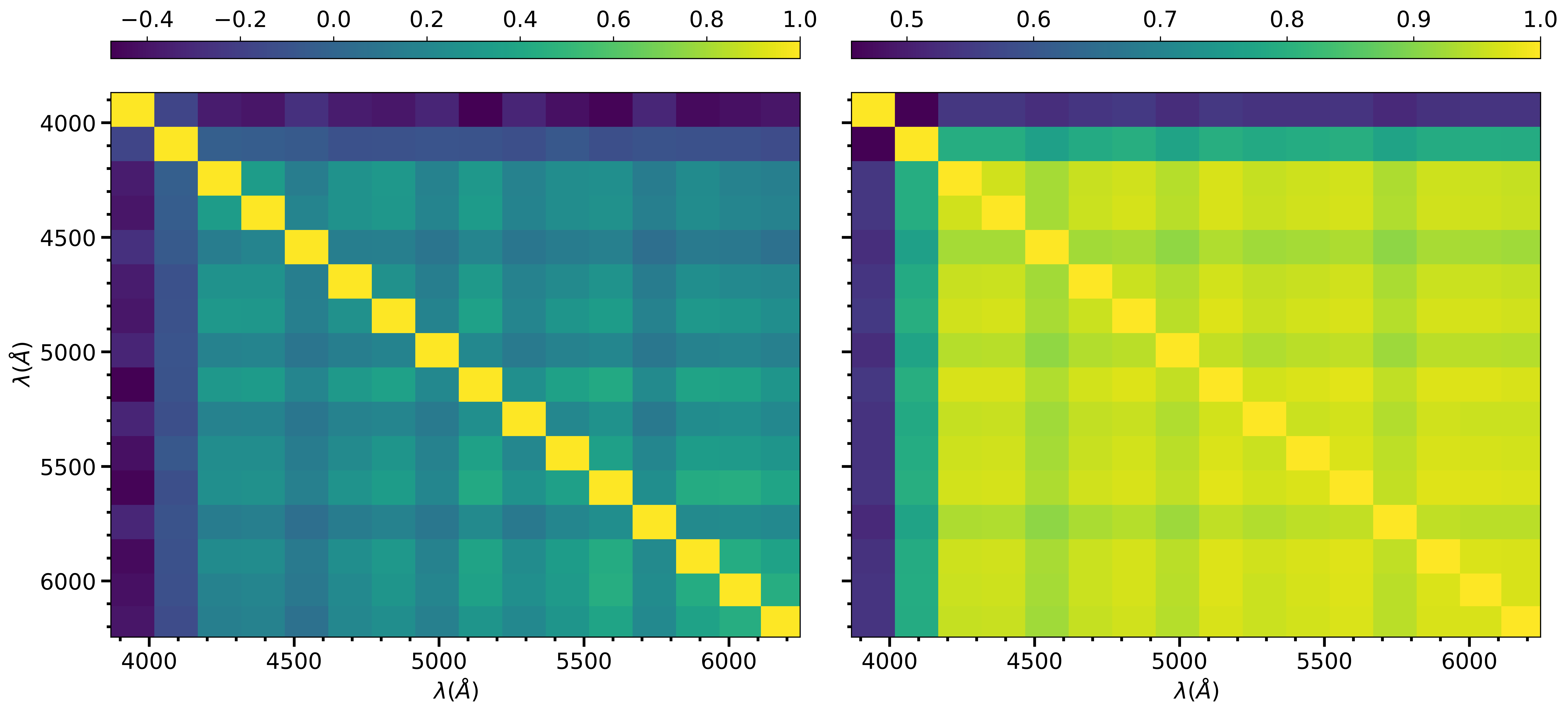}
        \caption{Heat map of correlation matrices of transmission spectra for 600B data. Left: From the analogous 2D GP analysis, visualising how the uncertainty in each transit depth is correlated across the spectrum. Right: For the 2D GP analysis with no added prior on the mean radius (from Sect.~\ref{sec:mean prior}). In both plots, the strength of correlation is not significantly affected by the wavelength separation between light curves for most wavelengths. The right plot in particular shows significant correlations that are mostly independent of wavelength-separation, consistent with significant uncertainty in the offset of the transmission spectrum as all wavelengths are approximately affected equally.}
        \label{fig:correlation_mat_blue}
    \end{figure*}
    
    \begin{figure*}
    	\includegraphics[width=\textwidth]{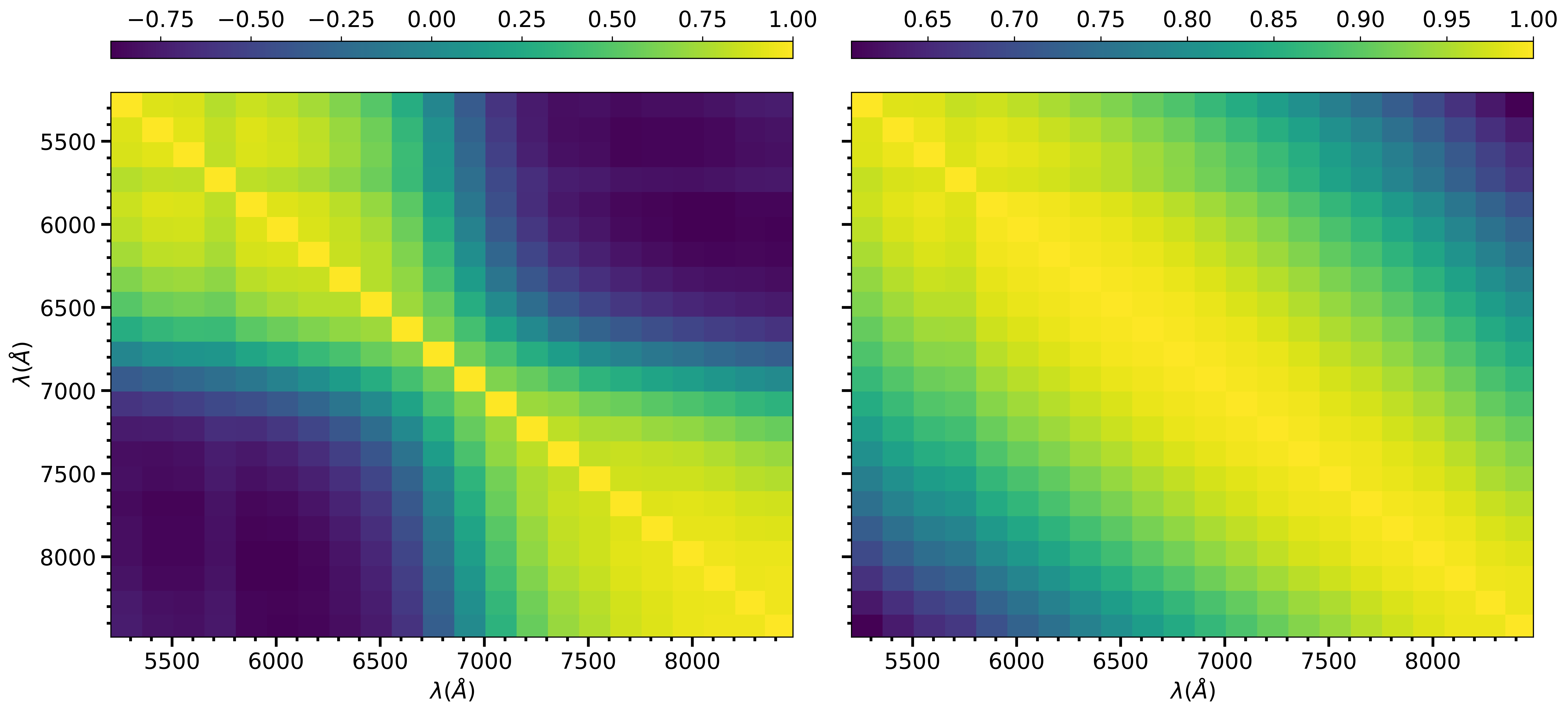}
        \caption{Same as Fig.~\ref{fig:correlation_mat_blue} but for 600RI data.  Both plots demonstrate that transit depths at large wavelength separations in the dataset are less correlated than for closer separations, which likely explains the significant uncertainty in the slope of the transmission spectrum for this analysis.}
        \label{fig:correlation_mat_red}
    \end{figure*}
    
    \begin{figure*}
    	\includegraphics[width=\textwidth]{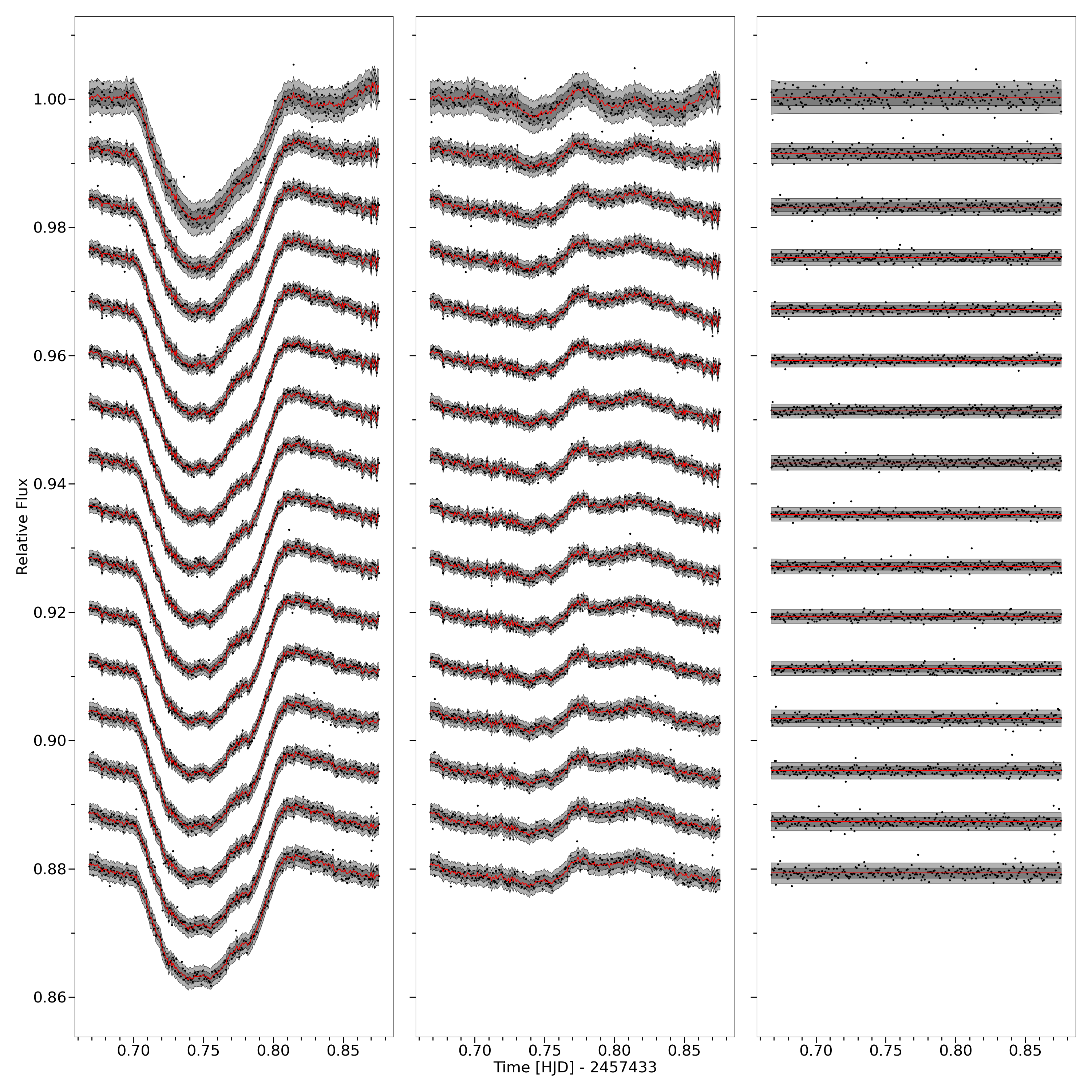}
        \caption{Fitted transit light curves for 600B data, showing GP predictive mean in red as well as $1\sigma$ and $2\sigma$ uncertainty in mean shaded in grey. Left: Fitting the raw light curves. Middle: The same fit minus the best-fit transit model, displaying just the GP fit to the systematics. Right: The same fit minus both the best-fit transit model and the GP mean (fitting the systematics) subtracted, showing the residual white noise.}
        \label{fig:blue residuals}
    \end{figure*}
    
    \begin{figure*}
    	\includegraphics[width=\textwidth]{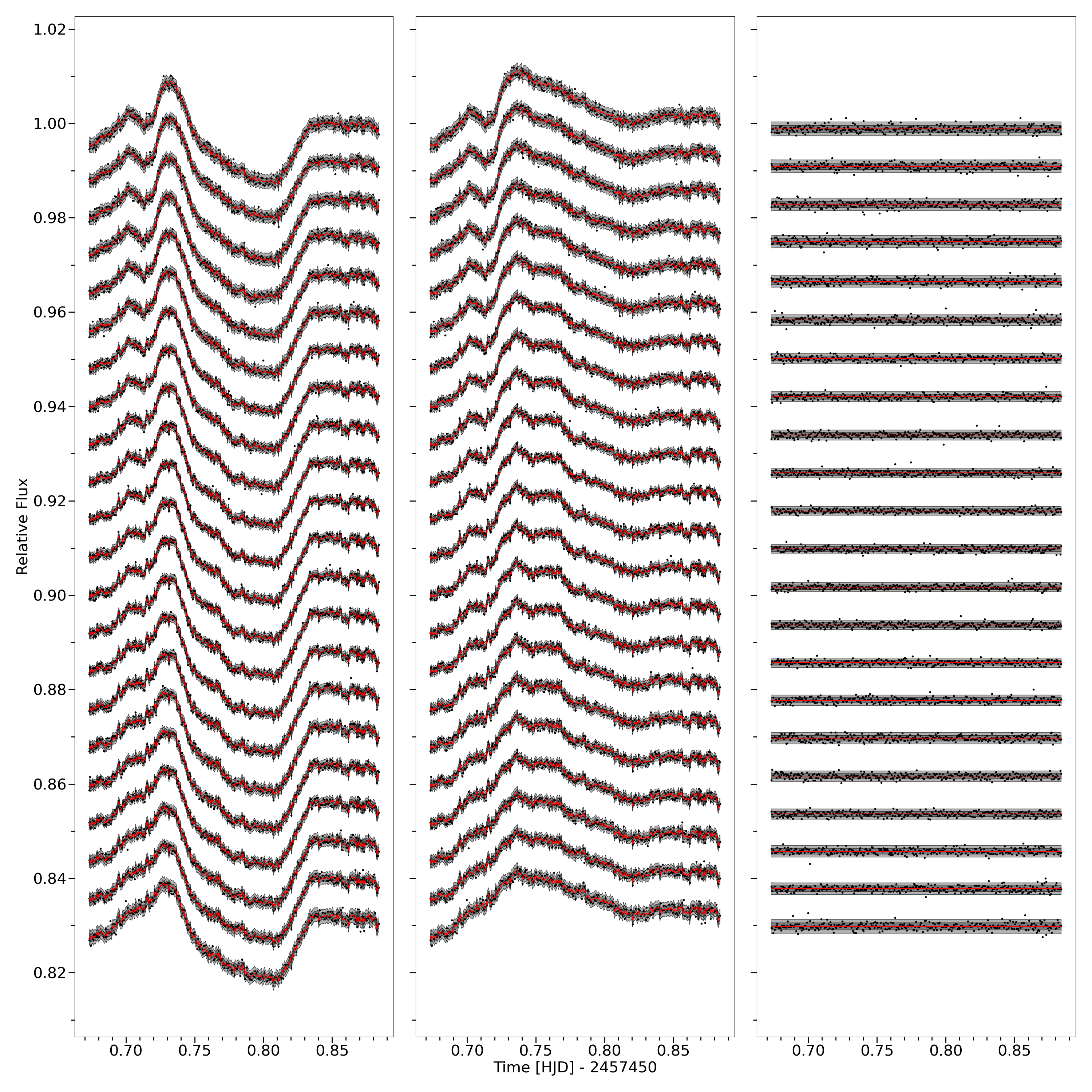}
        \caption{Same plot as Fig.~\ref{fig:blue residuals} but for 600RI data.}
        \label{fig:red residuals}
    \end{figure*}
    
    \begin{figure*}
    	\includegraphics[width=\textwidth]{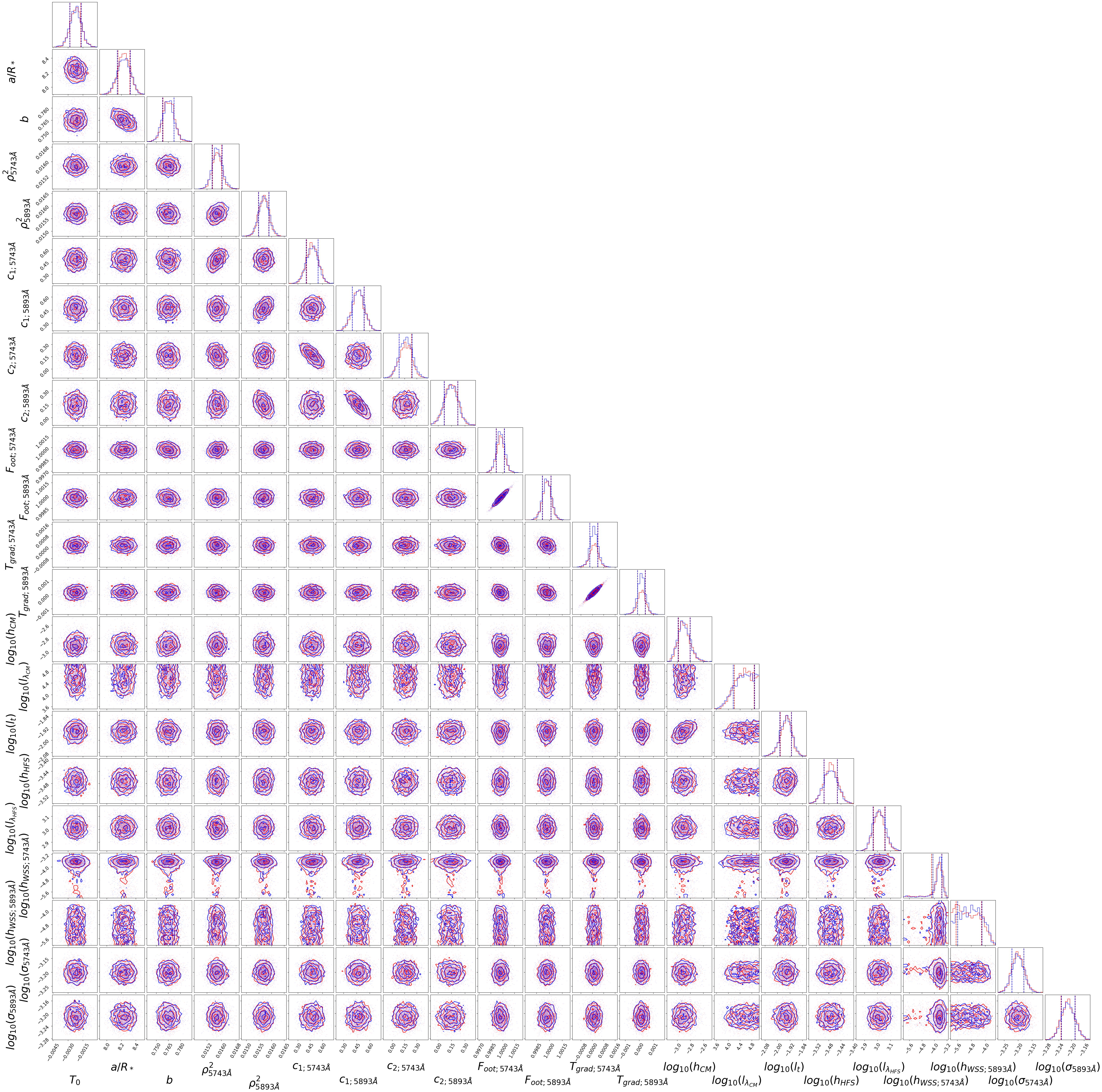}
        \caption{Corner plot for 600B data fit with 2D GP and marginalising over $T_0$, $a/R_*$ and $b$ (from Sect.~\ref{sec:Tab marginalisation}). Only the wavelength-dependent parameters from two of the light curves are included (from the wavelength bands immediately left of and centred on the Na feature) in addition to any parameters shared by all light curves. Note the wavelength length scale $l_\mathrm{\lambda_\mathrm{CM}}$ is consistent with the maximum prior limit, showing that these systematics may be constant in wavelength, while $l_\mathrm{\lambda_\mathrm{HFS}}$ is constrained to be within the prior bounds.}
        \label{fig:corner_blue}
    \end{figure*}
    
    \begin{figure*}
    	\includegraphics[width=\textwidth]{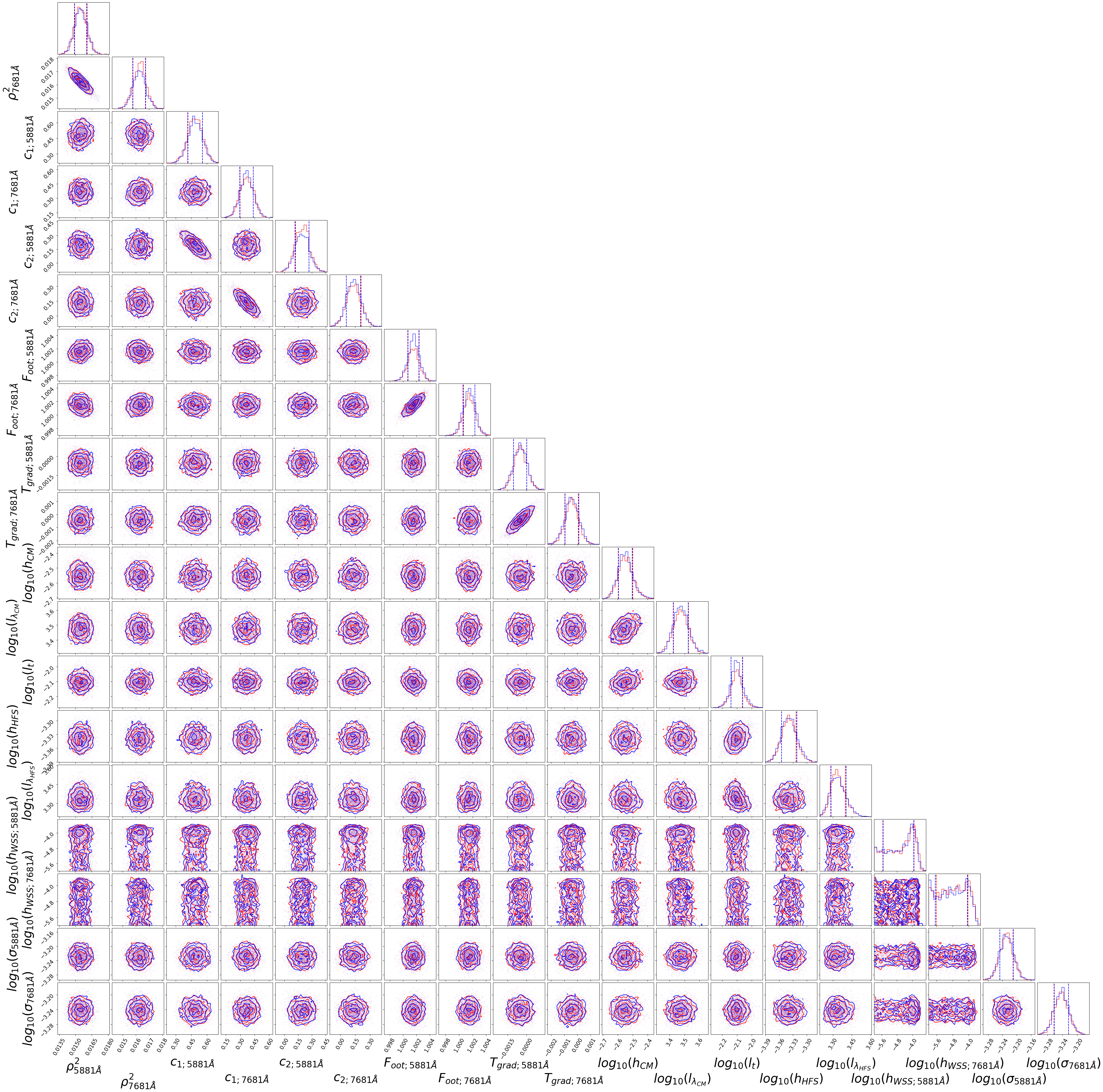}
        \caption{Corner plot for 600RI data for analogous 2D GP fit (from Sect.~\ref{sec:analogous_2D_GP}). The wavelength-dependent parameters included are from the light curves centred on the Na and K features. We note that both the wavelength length scales $l_\mathrm{\lambda_\mathrm{CM}}$ and $l_\mathrm{\lambda_\mathrm{HFS}}$ are constrained within the prior bounds for this dataset.}
        \label{fig:corner_red}
    \end{figure*}
    
\end{appendix}

\end{document}